\documentclass[%
reprint,
superscriptaddress,
amsmath,amssymb,
aps, 
prc
]{revtex4-2}

\usepackage{subfigure}
\usepackage{graphicx}
\usepackage{dcolumn}
\usepackage{bm}
\usepackage{threeparttable}
\usepackage{placeins}
\usepackage{hyperref}
\usepackage{xcolor} 

\usepackage{orcidlink}

\begin{document}

\title{Low-energy collective structure of $^{92,94}$Zr and $^{94}$Mo from ($e,e^{\prime}$) and ($p,p^{\prime}$) scattering
\newline I. Experimental results}

\author{C.~Walz}
\author{O.~Burda}
\affiliation{%
Institut f\"ur Kernphysik, Technische Universit\"at Darmstadt, 64289 Darmstadt, Germany
}%


\author{L.~M.~Donaldson,\orcidlink{0000-0001-8761-8257}}
\affiliation{%
School of Physics, University of the Witwatersrand, Johannesburg 2050, South Africa
}%
\affiliation{%
iThemba Laboratory for Accelerator Based Sciences, Somerset West 7129, South Africa
}%

\author{R.~W.~Fearick,\orcidlink{0000-0001-6778-1540}}
\affiliation{%
Department of Physics, University of Cape Town, Rondebosch 7700, South Africa
}%
\author{S.~V.~F\"ortsch,\orcidlink{0000-0003-0884-3283}}
\affiliation{%
iThemba Laboratory for Accelerator Based Sciences, Somerset West 7129, South Africa
}%
\author{H.~Fujita,\orcidlink{0000-0001-9711-7130}}
\affiliation{%
Research Center for Nuclear Physics, Osaka University, Ibaraki, Osaka 567-0047, Japan
}%

\author{\mbox{P.~von~Neumann-Cosel},\orcidlink{0000-0002-0256-5940}}
\email{Contact author: vnc@ikp.tu-darmstadt.de}
\affiliation{%
Institut f\"ur Kernphysik, Technische Universit\"at Darmstadt, 64289 Darmstadt, Germany
}%
\author{R.~Neveling,\orcidlink{0000-0001-5717-2725}}
\affiliation{%
iThemba Laboratory for Accelerator Based Sciences, Somerset West 7129, South Africa
}%
\author{N.~Pietralla,\orcidlink{0000-0002-4797-3032}}
\email{Email: pietralla@ikp.tu-darmstadt.de}
\author{A.~Richter,\orcidlink{0000-0001-6432-3405}}
\affiliation{%
Institut f\"ur Kernphysik, Technische Universit\"at Darmstadt, 64289 Darmstadt, Germany
}%
\author{F.~D.~Smit}
\affiliation{%
iThemba Laboratory for Accelerator Based Sciences, Somerset West 7129, South Africa
}%
\author{C.~Stahl}
\affiliation{%
Institut f\"ur Kernphysik, Technische Universit\"at Darmstadt, 64289 Darmstadt, Germany
}%
\author{J.~A.~Swartz,\orcidlink{0000-0002-5999-2791}}
\affiliation{%
iThemba Laboratory for Accelerator Based Sciences, Somerset West 7129, South Africa
}%
\affiliation{%
Department of Physics, University of Stellenbosch, Matieland 7602, South Africa
}%

\date{\today}

\begin{abstract}
This is the first of two papers discussing a new signature of quadrupole mixed-symmetry states in the vibrational nuclei $^{92,94}$Zr and $^{94}$Mo based on proton and neutron transition densities derived from the comparison of inelastic electron and proton scattering experiments.
Results of the $(e,e^\prime)$ and $(p,p^\prime)$ experiments on these nuclei as well as $(p,p^\prime)$ data for other potential cases $^{96}$Mo and $^{70}$Zn are presented.
Spin and parity quantum numbers of excited states are assigned based on angular distributions from the proton scattering data in comparison to calculations employing form factors from the collective model in distorted-wave Born approximation.
Possible candidates of one-phonon fully symmetric and mixed-symmetry states as well as members of two-phonon multiplets are identified.
\end{abstract}

\maketitle

\section{Introduction}

Isovector quadrupole valence-shell excitations are building blocks of low-energy nuclear structure.
They reflect the specific nature of the nuclear interaction as a two-component system consisting of protons and neutrons \cite{pietralla2008}.
A global description of such modes in heavy nuclei can be achieved in the Interacting Boson Model (IBM) \cite{iachello1987}.
The IBM is approximates a shell-model description of nuclei assuming dominance of pairing at low excitation energies and has been particularly useful in predicting collective excitations. 
The IBM-2 \cite{arima1977} extends this model to coupled neutron and proton pairs treated as distinct bosons implying new classes of collective states \cite{iachello1984}. 

The extra degree of freedom from the proton-neutron distinction in IBM-2 leads to a new quantum number called $F$ spin.
The lowest collective states or multiphonon states built on top of them are characterized by maximum $F$ spin, $F_{\mathrm{max}} = (N_{\pi} {+} N_{\nu})/2$, known as fully symmetric states (FSS).
Their existence is well established experimentally and was recognized early on \cite{otsuka1978}. 
States that have $F \leq F_{\mathrm{max}} - 1$ are called mixed symmetric states (MSS), and the first example experimentally identified was the so-called scissors mode observed in backward ($e,e^\prime$) experiments at low momentum transfers in Darmstadt \cite{bohle1984}. 
The properties of the scissors mode are closely linked to ground-state deformation and have been studied extensively \cite{heyde2010}.
In nuclei near shell closures, the lowest MSS is of a quadrupole nature.
This can be effectively described in the Q-phonon picture \cite{pietralla1994} where the FSS and MSS quadrupole vibrations are the one-phonon states. 
One can construct multi-phonon multiplets, and experimental evidence for their existence has been demonstrated in many cases \cite{pietralla2008}.

In order to test this picture, experimental signatures of MSS are crucial. 
They are typically based on electromagnetic transitions between FSS and MSS one- and two-phonon state candidates \cite{pietralla2008}. 
The Q-phonon scheme makes specific predictions like collective $E2$ transitions between one- and two-phonon states and, in particular, strong isovector $M1$ transitions between the quadrupole MSS and FSS. 
This scheme has been successfully tested for the case of $^{94}$Mo \cite{pietralla1999,pietralla2000,fransen2001} and serves as signature of MSS (see Refs.~\cite{stegmann2017,kern2019,kern2020,yaneva2020,stetz2025} for recent examples).
However, these states are always mixed to some degree with states outside the Q-phonon model space, which complicates the clear identification of MSS in many cases.

In nuclei with accessible model spaces, the above described features were shown to arise in shell-model calculations \cite{lisetskiy2000,werner2002,holt2007,sieja2009}. 
However, a model particularly suited for the microscopic description of low-energy collective modes and multiphonon structures built on it is the Quasiparticle-Phonon Model (QPM) \cite{soloviev1992}.
As demonstrated, e.g.\ in Refs.~\cite{pignanelli1993,ponomarev1999,ryezayeva2002,savran2018} and also in the present work, it provides a framework for an accurate description of low-energy states and transitions between them, which allows for conclusions to be drawn on their underlying structure.   
In Ref.~\cite{burda2007}, an alternative approach for the identification of MSSs in near-spherical nuclei was introduced and successfully tested for the case of $^{94}$Mo.
It is based on a comparison of proton and neutron transition densities derived from measurements of the momentum transfer dependence in ($e,e^\prime$) and ($p,p^\prime$) experiments. 
This should work particularly well in cases with one or two proton and neutron pairs outside shell or subshell closures.
The idea was further explored by the study of another case ($^{92}$Zr) and it was demonstrated that the collectivity of the MSS results from coupling to the isoscalar giant quadrupole resonance (ISGQR) \cite{walz2011}. 
The present work gives a full account of these ideas, provides an analysis for yet another case ($^{94}$Zr) and further investigates a possible extension to octupole and hexadecapole degrees of freedom.
While $3^-$ and $4^+$ FSS are easily identified, the possible existence of corresponding MSS is a longstanding controversial issue (see, e.g., Refs.~\cite{smirnova2000,fransen2003,scheck2010,hennig2015,gregor2017,thurauf2019} for $3^-$ and Refs.~\cite{fransen2003,fransen2005,hennig2015,casperson2013} for $4^+$ MSS). 

The extensive material is organized in two papers.
Paper I discusses the data base from
($e,e^\prime$) and ($p,p^\prime$) experiments and also  new ($p,p^\prime$) results for other cases of interest ($^{96}$Mo, $^{70}$Zn), where ($e,e^\prime$) data are not yet available. 
Paper II presents an extensive comparison of QPM calculations for many experimental observables in the nuclei of interest to demonstrate the level of predictive power. 
Possible candidates for octupole and hexadecapole MSS are discussed.
Then, a new signature of quadrupole MSS independent of electromagnetic decay properties based on proton and neutron transition densities is presented.
Finally, the role of high-lying 2-quasiparticle (2qp) states, in particular from giant resonances with the correct quantum numbers, in the generation of low-energy collectivity is investigated.

\section{Electron Scattering Experiments}
\label{sec:II}
\subsection{Experimental details}

The experiments were performed at the Darmstadt superconducting electron linear accelerator S-DALINAC \cite{pietralla2018}. 
The high-resolution spectrometer Lintott with its focal-plane detector system based on four single-sided silicon strip detectors, each providing 96 strips with a thickness of 500 $\mu$m and a pitch
of 650 $\mu$m \cite{lenhardt2006}, was used.
The experimental procedures and data analysis followed previous work with the same device (see, e.g., Refs.~\cite{burda2010,scheikh2013,scheikh2014,kremer2016}).

Data were taken at an incident electron beam energies $E_0 = 63$ MeV and scattering angles $\Theta_{\rm lab} = 69^\circ, 81^\circ, 93^\circ, 117^\circ, 165^\circ$ for $^{92}$Zr, $E_0 = 71$ MeV and $\Theta_{\rm lab} = 69^\circ, 81^\circ, 93^\circ, 165^\circ$ for $^{94}$Zr, and $E_0 = 70$ MeV and $\Theta_{lab} = 93^\circ, 117^\circ, 141^\circ, 165^\circ$ for $^{94}$Mo. 
The corresponding momentum transfers roughly cover the maximum region of $E2$ form factors. 
Beam currents typically varied between 0.5 and 2 $\mu$A.
Enriched self-supporting metallic foils 
were used as targets: $^{92}$Zr (94.6\%, 9.8 mg/cm$^2$),$^{94}$Zr (96.1\%, 10 mg/cm$^2$), and $^{94}$Mo (91.6\%, 9.7 mg/cm$^2$). 
Typical energy resolutions were 55 keV ($^{92}$Zr), 60 keV ($^{94}$Zr), and 30-45 keV ($^{94}$Mo) full width at half maximum (FWHM).

\subsection{Spectra}

\begin{figure}[tbh!]
\centering
\includegraphics[width=\columnwidth]{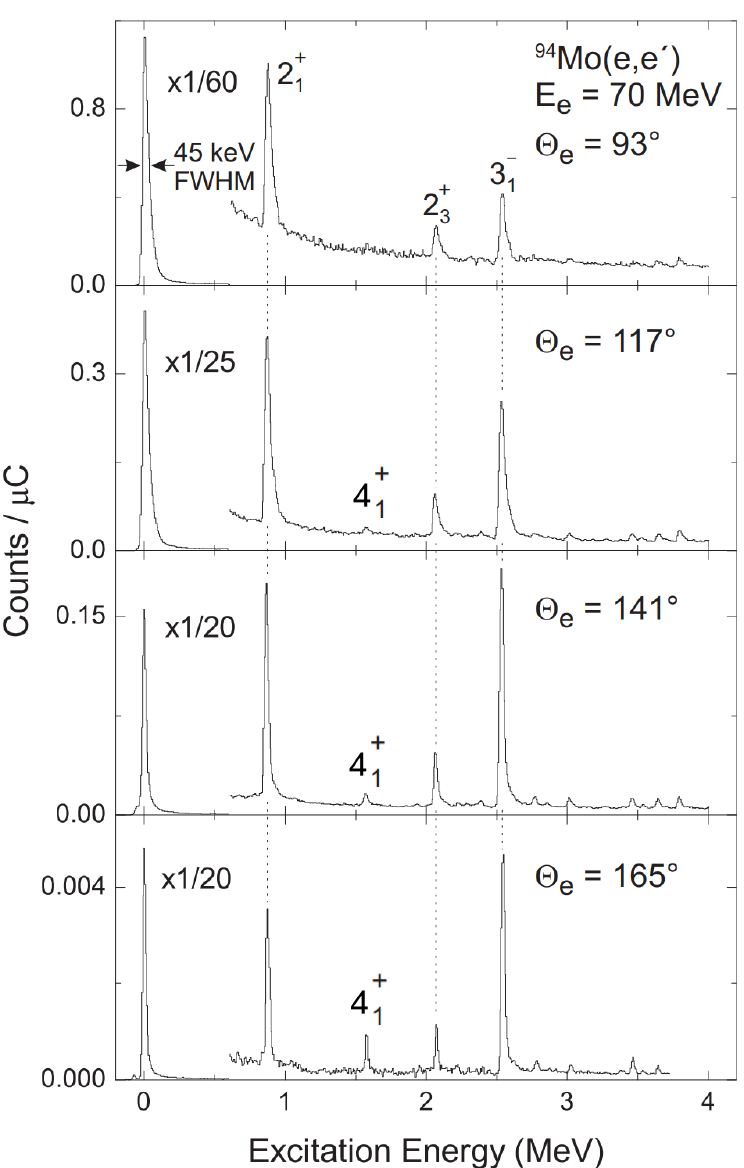}
\caption{Spectra of the $^{94}$Mo($e,e^{\prime}$) reaction at $E_0 = 70$ MeV and $\Theta_{\rm e} = 93^\circ, 117^\circ, 141^\circ, 165^\circ$.
The most prominent transitions are labeled with their spin and parity quantum numbers. 
The signals from elastic scattering are downscaled for visibility  by the factors given in the figure.
\label{Fig: Spectra 94Mo electron scattering} 
}
\end{figure}
The spectra obtained for $^{92}$Zr and $^{94}$Z can be found in Refs.~\cite{scheikh2013} and \cite{scheikh2014}, respectively.
Figure \ref{Fig: Spectra 94Mo electron scattering} presents the data for $^{94}$Mo.
A brief account of the experiment has been given in Ref.~\cite{burda2007}.
Signals from elastic scattering are downscaled for visibility of excited states by the factors given in the figure.
The most prominent signals of nuclear excitation in inelastic scattering reactions are labeled with the spin and parity quantum numbers of the corresponding excited states 
However, one can also identify a number of weaker transitions, cf.\ Fig.~1 in Ref.~\cite{burda2007}.
Many of them are known to be $J^\pi = 2^+$ states, but they also include candidates for octupole and hexadecapole one-phonon states.

\subsection{Form factors and spin-parity determination}
\label{sect: form  factors}

In $^{92,94}$Zr, further analysis is restricted to the transitions to the $2^+_{1,2}$ and $3^-_1$ states.
In contrast, the $^{94}$Mo measurement with better statistics and resolution shows candidates for further $2^+$ as well as $3^-$ and $4^+$ states.
The line contents were determined by simultaneous fits of the peaks in the spectra with the line shape given in Ref.~\cite{hofmann2002} and a bremsstrahlung background. 
Details of the extraction of the form factors are given in Ref.~\cite{burda2007a}.
The assignments of spin and parity quantum numbers of excited states is based on the comparison of the form factors to QPM calculations.
Accordingly, experimental and theoretical form factors are shown and discussed in paper II. 
The resulting spin-parity assignments are compared with the proton scattering results discussed below in Tab.~\ref{Tab: Results of 94Mo}.

\section{Proton Scattering}

\subsection{Experimental details}

The data presented in this paper were collected in two experimental campaigns performed at iThemba LABS using 200 MeV proton beams and the K600 magnetic spectrometer \cite{neveling2011} to analyze inelastically scattered protons. 
The magnetic fields in the spectrometer were always set to include the excitation-energy range $0 - 4$ MeV.
In the first experiment, proton beams with currents of $1 – 30$ nA (depending on angle) were used to bombard enriched self-supporting metallic foils of $^{92}$Zr (94.4\%, 1.3 mg/cm$^2$) and $^{94}$Mo (93.9\%, 1.2 mg/cm$^2$). 
A $^{24}$Mg target was used for calibration purposes.
Data were taken for spectrometer angles varying between $7^\circ$ and $26^\circ$ in $2^\circ$ steps.
The dispersion of the proton beamline was set to match the dispersion of the spectrometer. 
The energy resolution obtained with the dispersion-matched spectrometer \cite{neveling2011} was approximately 35 keV (FWHM).
 
Targets in the second experiment were $^{94}$Zr (96.1\%, 2.9 mg/cm$^2$), $^{96}$Mo (96.7\%, 3.0 mg/cm$^2$), and $^{70}$Zn (95.4\%, 4.9 mg/cm$^2$), along with a $^{24}$Mg target for calibration. 
Proton beam currents varied between 0.5 and 25 nA.
Data were taken for angles in the range 8$^\circ$ to 25$^\circ$ with a step size of 2$^\circ$ or 2.5$^\circ$. 
Energy resolutions of about $50 – 80$ keV (FWHM) for $^{94}$Zr, $30 – 45$ keV (FWHM) for $^{96}$Mo, and $30 – 50$ keV (FWHM) for $^{70}$Zn were achieved. 
There were accelerator problems while taking data on the $^{94}$Zr target, which adversely affected the resolution of the data set. 

\subsection{Data analysis}

\begin{figure}
\centering
\includegraphics[width=\columnwidth]{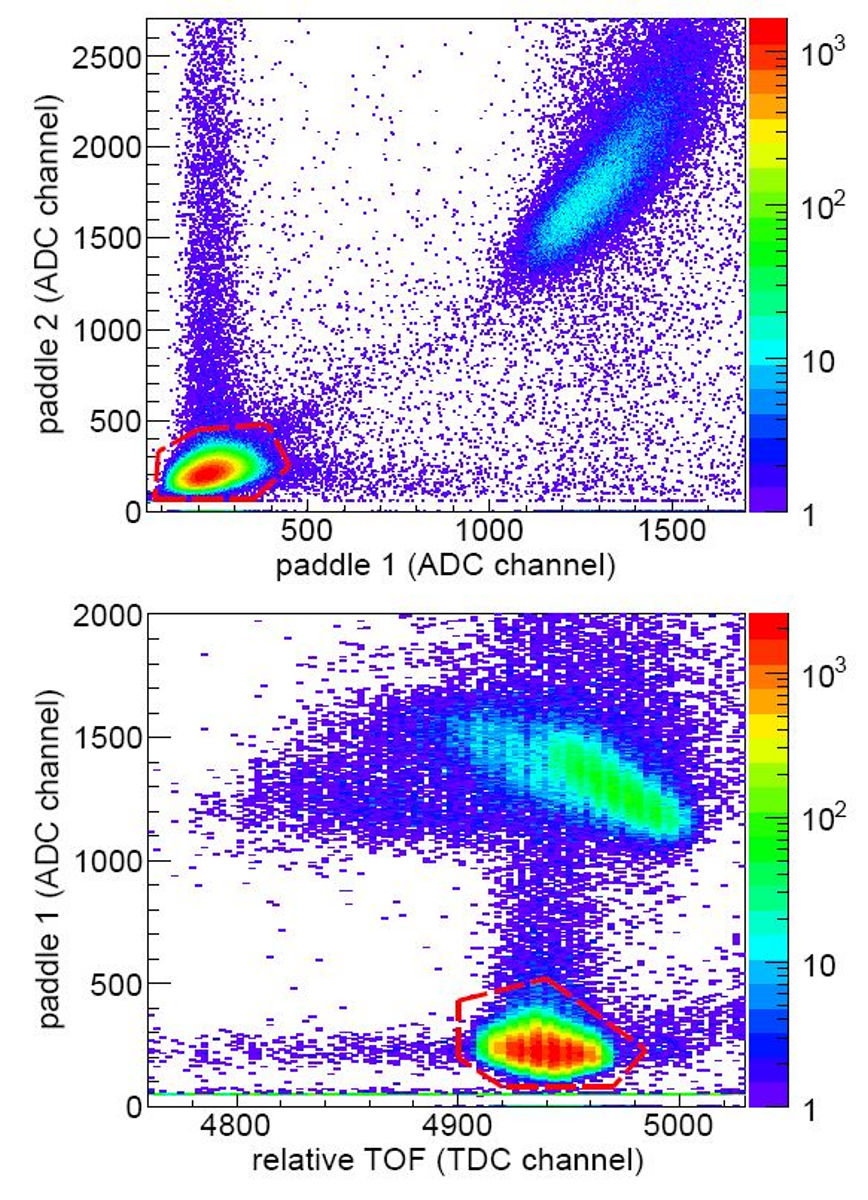}
\caption{
\label{Fig: 2D spectra}
Examples of $\Delta E_1$-$\Delta E_2$ (top) and $\Delta E_1$-TOF (bottom) spectra of the scintillator paddles used for particle identification. 
The energy losses are given in units of the ADC signals.
The software gates on scattered protons are marked in red. 
}
\end{figure}

The data were analyzed following the procedures described in Ref.~\cite{neveling2011}.
Particle identification was based on the correlation of energy loss ($\Delta E_{1,2}$) and time-of-flight (TOF) signals of two trigger scintillator detectors placed behind the vertical drift chambers. 
As can be seen in Fig.~\ref{Fig: 2D spectra}, 
proton scattering signals were well separated from background particles. 
The software gates marked in red were used to separate proton signals for further analysis.
The bump at $\Delta E \approx 1500$ is most likely due to $\alpha$ particles. 

The energies in the focal plane were determined from a fit to well-known transitions observed in the $^{24}$Mg($p,p^\prime$) reaction.
The accuracy of excitation energies deduced in the investigated targets is estimated to be $10 - 20$ keV.
Cross sections were determined from Gaussian fits to the peaks in the spectra.
The associated uncertainties have been calculated by combining the statistical and systematical errors in quadrature.
For further details, see Ref.~\cite{walz2014}. 

\subsection{Spectra}
\label{sect: Spectra}

\begin{figure*}
\centering
\includegraphics[width=0.7\textwidth]{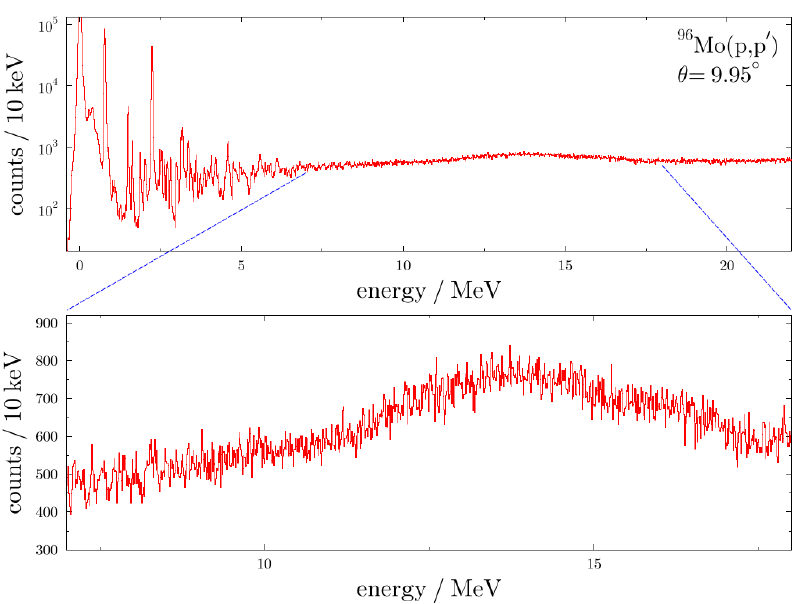}
\caption{
\label{Fig: Excitation spectra 96Mo(p,p') 1}
Top: Excitation-energy spectrum of the $^{94}$Mo($p,p^\prime$) reaction at $E_0 = 200$ MeV and $\theta_{\rm lab} = 9.95^\circ$ over the full momentum acceptance of the spectrometer. 
Bottom: Energy region of the ISGQR magnified.
}
\end{figure*}
The top part of Fig.~\ref{Fig: Excitation spectra 96Mo(p,p') 1}
presents an excitation-energy spectrum of the $^{94}$Mo($p,p^\prime$) reaction at $\theta_{\rm lab} =9.95^\circ$ over the full momentum acceptance of the spectrometer as an example.
Resolved transitions are observed up to an excitation energy of about 5 MeV and a (quasi)continuum is seen at higher excitation energies.
The bottom part magnifies the energy region between 7 and 18 MeV.
Up to the neutron threshold ($S_n = 9.15$ MeV), one still observes peaks, but on a background resulting from the superposition of many weak unresolved transitions due to the high level density \cite{martin2017} in this moderately deformed nucleus.
The scattering angle, for which the data are shown, roughly corresponds to the maximum cross section for $\Delta L = 2$ transitions.  
Thus, the broad bump peaking at about 14 MeV represents the ISGQR.
Owing to the good energy resolution, non-statistical fluctuations (i.e., fine structure) are visible in the giant resonance region, which can be interpreted in terms of the main decay mechanisms \cite{shevchenko2004,vonneumanncosel2019}.  

\begin{figure}
\centering
\includegraphics[width=\columnwidth]{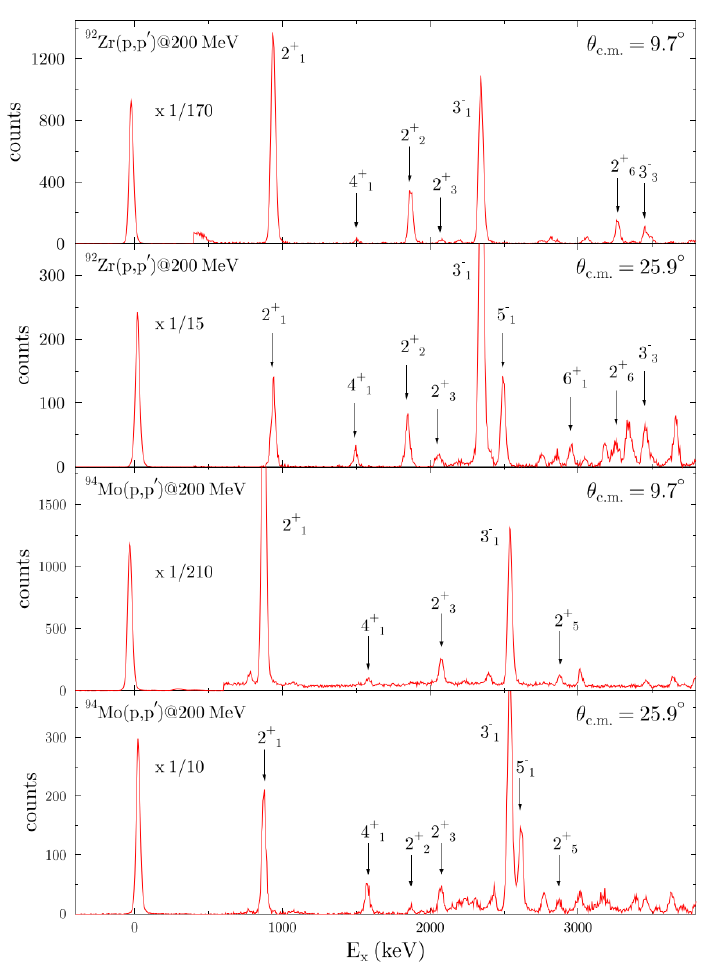}
\caption
{
\label{Fig: Represent spectra 92Zr 94Mo 2} Spectra of the $^{92}$Zr($p,p^{\prime}$) and $^{94}$Mo($p,p^{\prime}$) reactions measured at $E_p = 200$ MeV and $\theta_{\rm lab} = 9.7^\circ$ and $25.9^\circ$. 
Prominent transitions are labeled with their spin and parity quantum numbers. 
Signals from elastic scattering are downscaled by the factors indicated.
}
\end{figure}
Sample spectra of the $^{92}$Zr($p,p^{\prime}$) and $^{94}$Mo($p,p^{\prime}$) reactions at two different angles (the maximum angle for $\Delta L = 2$ transitions and the largest angle measured) are displayed in Fig.~\ref{Fig: Represent spectra 92Zr 94Mo 2} in the energy region up to 3.5 MeV.
Spin-parity assignments based on the analysis described below are indicated.  
Besides the dominant elastic scattering suppressed in the spectra by the factors indicated, at the smaller angle shown, one identifies states with spins $J = 2 - 4$  with a dominance of $J^\pi = 2^+$ states and the transition to the $J^\pi = 3^-$ FSS.
At the larger angle, one also finds states with $J = 4 -6$.

\begin{figure}
\centering
\includegraphics[width=\columnwidth]{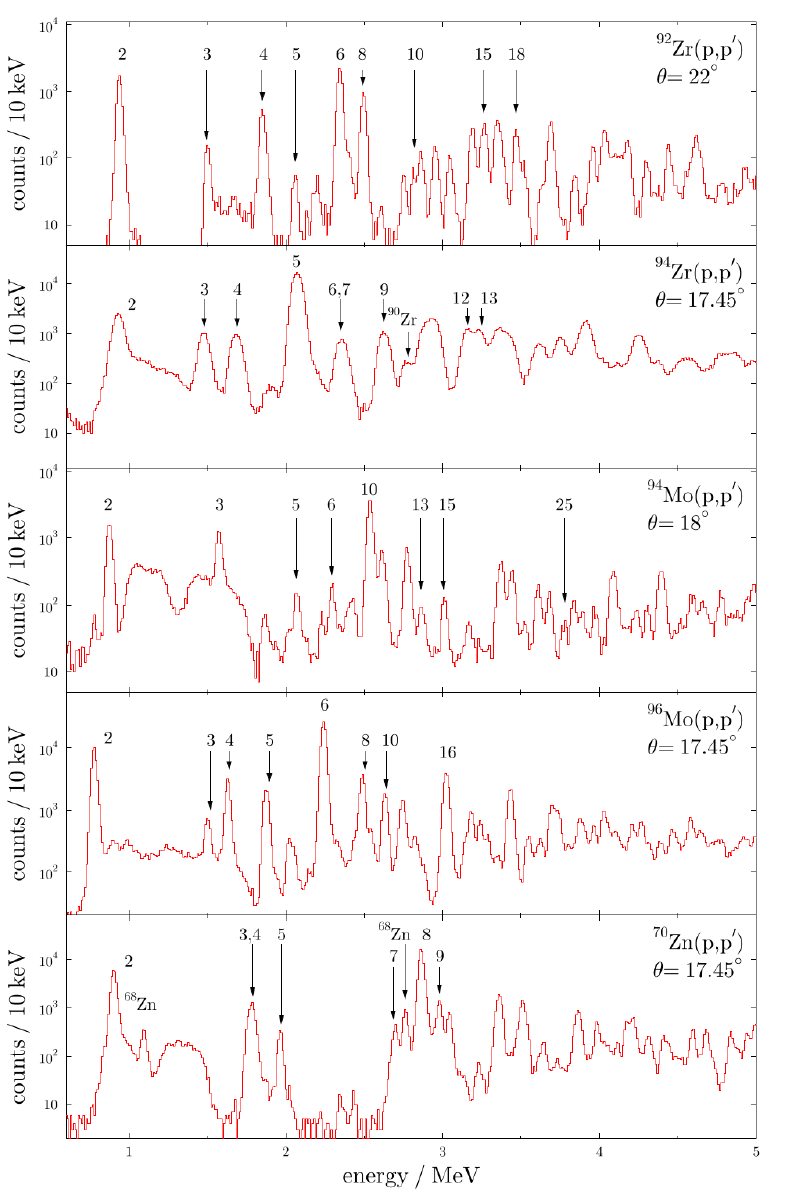}
\caption{
\label{Fig: Examples spectra Zr, Mo, Zn 3} Sample energy spectra obtained for $^{92,94}$Zr, $^{94,96}$Mo and $^{70}$Zn in the energy range $0.6 - 5$ MeV at the scattering angles indicated in the figure. 
Some states are labeled with numbers referring to the Tables in Sec.~\ref{sec:IV}. 
}
\end{figure}
Finally, Fig.~\ref{Fig: Examples spectra Zr, Mo, Zn 3} presents spectra of all measured targets for excitation energies of 0.6 - 5 MeV at a scattering angle of approximately $18^\circ$ (except $^{92}$Zr, where data are shown at $22^\circ$).
The $18^\circ$ spectra all show a broad bump in the excitation-energy region $1 -2$ MeV due to elastic scattering off hydrogen embedded in the targets.
The numbers assigned to peaks refer to the tables presented in Sec.~\ref{sec:IV}.
Comparing the $^{94}$Zr spectrum with the other data emphasizes the role of good energy resolution.
In the lighter nucleus, $^{70}$Zn, and in the semimagic $^{92}$Zr, one can essentially identify and resolve all natural-parity states with spins $J = 2 - 6$ up to about 3.5 MeV.
The excitation of unnatural-parity states, although possible in principle in $(p,p^\prime)$ scattering, is suppressed by the dominance of the isoscalar non-spinflip part of the effective proton-nucleus interaction for the kinematics of the experiments \cite{love1981}. 
Because of the onset of deformation with higher level densities, states in the $^{94,96}$Mo spectra can only be fully resolved up to about 3 MeV.

\subsection{Angular distributions and spin-parity determination}
\label{sect: angular distributions}

The experimental angular distributions presented in the next section are compared to distorted wave Born approximation (DWBA) calculations in order to assign spin and parity quantum numbers to the excited states.
These are based on the collective model assuming that the transition potential is given by the derivative of the optical potential.
The DWBA analysis was performed using the program CHUCK3 \cite{CHUCK3}.
The required optical model parameters were determined from fits to the elastic scattering cross sections using the global parameter set of Schwandt {\it et al.}~\cite{schwandt1982} as starting values.
Table \ref{tab: Optical model parameters} summarizes the resulting parameters for $^{92}$Zr as an example.
\begin{table}[tbh] 
\centering
\renewcommand*{\arraystretch}{1.3}
\caption{
Optical model parameters for 200 MeV elastic proton scattering off $^{92}$Zr deduced from the present experiment.
\label{tab: Optical model parameters}
}
\begin{threeparttable}[c]
\begin{tabular}{ccccccc}
\hline
\hline
  &\multicolumn{3}{c}{Wood-Saxon potential}&\multicolumn{3}{c}{$LS$ potential}\\
  & $V$ (MeV) & $r$ (fm) & $a$ (fm) & $V$ (MeV) & $r$ (fm) & $a$ (fm) \\
\hline
Re& 17.520  & 1.257  & 0.750  & -2.484  & 1.021  & 0.787  \\
Im& -10.980 & 1.253  & 0.822  &  1.853  & 1.020  & 0.592  \\
\hline
\hline
\end{tabular}
\end{threeparttable}
\end{table}
For the other target nuclei, slight adjustments of the parameters were performed by optimizing the reproduction of the angular distributions of the transitions to the $2^+_1$ and $3^-_1$ states, where the collective model is expected to be a very good approximation.  

Agreement between experimental angular distributions and theory over the entire covered angular range is required for an unambiguous spin-parity assignment in the present work.
Tentative spin-parity assignments are given in cases where the position of the first maximum is correctly reproduced. 
Otherwise, no quantum numbers are assigned.
No attempt was made to analyze the results quantitatively, e.g.\ in terms of deformation lengths.
A detailed quantitative comparison of the available experimental information for FSS and MSS candidates  with QPM predictions including the present data is given in paper II.  

\section{Results from the $(p,p^\prime$) reaction}
\label{sec:IV}

For each nucleus, experimental angular distributions and collective model predictions for all analyzed transitions are presented in a figure, and the results of the spin-parity assignments are summarized in a table. 
A brief discussion of further information and a comparison to other work for each excited state is provided. 

\subsection{ $^{92}$Zr }
\label{sect: $^{92}$Zr}

\begin{figure}
\centering
\includegraphics[width=\columnwidth]{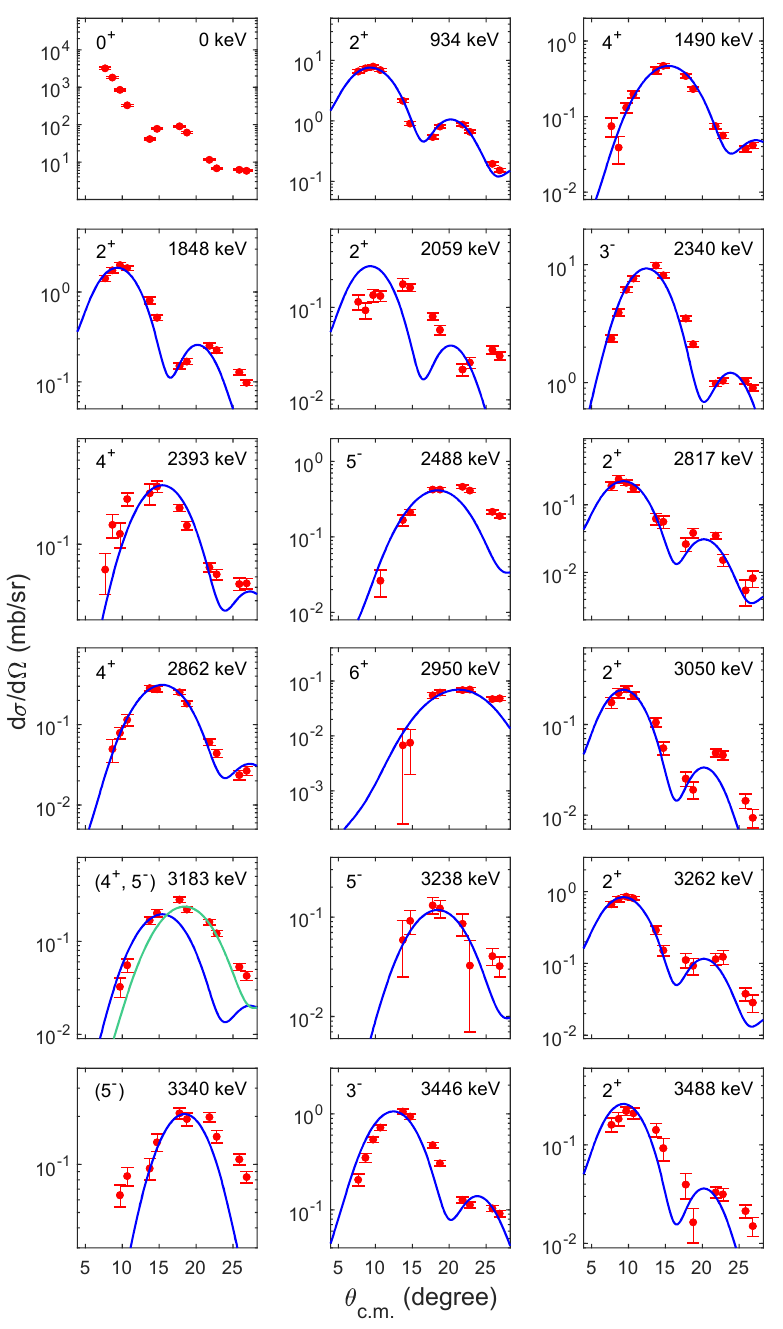}
\caption[Angular distributions of states excited in the $^{92}$Zr($p,p^{\prime}$) experiment.]{\label{Fig: Angular distributions of 92Zr 1} Angular distributions of states excited in the $^{92}$Zr($p,p^{\prime}$) reaction in comparison to DWBA calculations using a phenomenological optical potential and collective-model form factors for the transition densities.}
\end{figure}

The angular distribution of protons scattered inelastically off excited states in the $^{92}$Zr($,p,p^\prime$) reaction are shown in Fig.~\ref{Fig: Angular distributions of 92Zr 1} and compared with DWBA calculations. 
The information on the states extracted from the $^{92}$Zr($p,p^{\prime}$) data is summarized in  Tab.~\ref{Tab: Results of 92Zr}, where a comparison is given to results obtained with the $(\alpha,\alpha^\prime)$ \cite{aa94zr} and $(n,n^\prime\gamma)$ \cite{fransen2005} reactions.  

\begin{table}
\centering
\caption[Results of the $^{92}$Zr($p,p^{\prime}$) experiment.]{\label{Tab: Results of 92Zr} 
Excitation energies, spin and parity assignments of transitions excited in the $^{92}$Zr($p,p^{\prime}$) experiment and comparison to results obtained with the $(\alpha,\alpha^\prime)$ \cite{aa94zr} and $(n,n^\prime\gamma)$ \cite{fransen2005} reactions.
The peak numbers refer to Fig.~\ref{Fig: Examples spectra Zr, Mo, Zn 3}.} 

\begin{threeparttable}[c]
\begin{tabular}{ccccccc}
\hline
\hline
& $(p,p^\prime)$ & \cite{aa94zr} & \cite{fransen2005} 
     & $(p,p^\prime)$  & \cite{aa94zr} & \cite{fransen2005} \\
No. & $E_{\rm x}$ & $E_{\rm x}$  & $E_{\rm x}$ & $J^{\pi}$ & $J^{\pi}$ & $J^{\pi}$    \\
     \hline 
    & 0          & 0         & 0           & 0$^+$         & 0$^+$             & 0$^+$               \\
2    & 934(10)    & 935(10)   & 934.5(1)  & 2$^+$         & 2$^+$             & 2$^+$               \\
3    & 1490(10)   & 1495(10)  & 1495.5(1)   & 4$^+$         & 4$^+$             & 4$^+$               \\
4    & 1848(10)   & 1847(10)  & 1847.3(1)   & 2$^+$         & 2$^+$             & 2$^+$               \\
5    & 2059(10)   & 2053(10)  & 2066.6(1)   &          & 2$^+$             & 2$^+$               \\
6    & 2340(10)   & 2334(10)  & 2339.6(1)   & 3$^-$         & 3$^-$             & 3$^-$               \\
    & 2393(10)   & 2393(10)  & 2398.4(1 )  & 4$^+$         & 4$^+$             & 4$^+$               \\
8    & 2488(10)   & 2482(10)  & 2485.9(2)   & 5$^-$         & 5$^-$             & 5$^-$               \\
     & 2817(10)   & 2823(10)  & 2819.6(1)   & 2$^+$         & 2$^+$             & 2$^+$               \\
10   & 2862(10)   & 2869(10)  & 2864.7(2)   & 4$^+$         & 4$^+$             & 4$^+$               \\
   & 2950(10)   & 2963(10)  &             & 6$^+$         & (6$^+$)           &                     \\
   & 3050(10)   & 3055(10)  & 3057.5(3)   & 2$^+$         & 2$^+$             & 2$^+$               \\
   & 3183(10)   & 3187(10)  & 3178.3(2)   & (4$^+$+$5^-$)       & 4$^+$             & 4$^+$               \\
   &   &  & 3191.0(3)   &   &    & ($4^-$)                      \\
   & 3238(20)   & 3248(10)  &             & (5$^-$) & 4$^+$             &                     \\
15   & 3262(20)   & 3273(10)  & 3262.9(4)   & 2$^+$         & 2$^+$             & 2$^+$               \\
   & 3340(20)   & 3345(10)  &             & (5$^-$)         & 5$^-$             &                     \\
   & 3372(20)   & 3382(10)  &             &               & 3$^-$             &                     \\
18   & 3446(20)   & 3452(10)  & 3452.1(3)   & 3$^-$         & 3$^-$             & (2$^+$)             \\
   & 3488(20)   & 3491(10)  & 3500.1(3)   & 2$^+$         & (3$^-$)           & 2$^+$               \\
   & 3623(20)   & 3634(10)  & 3628.4(4)   &               &                   & (2,3)               \\
   & 3644(20)   &           & 3640.3(4)   &               &                   & (2$^+$)             \\
\hline
\hline
\end{tabular}
\end{threeparttable}
\end{table}

\textit{States at 934 keV, 1490 keV, 1848 keV, 2059 keV, 2340 keV, 2393 keV, 2488 keV, 2817 keV and 2862 keV:}
The angular distributions of the scattering cross sections are well described for all transitions by the collective model predictions. 
All spin and parity assignements agree with the results of Refs.~\cite{aa94zr,fransen2005}. 
The state at 1848 keV was discussed in the literature as the quadrupole one-phonon MSS of $^{92}$Zr \cite{scheikh2013}.
The transition to the 2$^+_3$ state at 2059~keV is an exception with poor agreement of the data with the theoretical angular distribution. 
This state is likely a member of the symmetric two-phonon $0^+,2^+,4^+$ triplet, where the dominance of one-step excitation underlying the present analysis is questionable. 
Fransen {\it et al.}~\cite{fransen2005} unambiguously identify it as 2$^+$ state and give upper limits of the $B(E2)$ strength to the ground state and to the 2$^+_1$ state of $<0.005$ Weisskopf units (W.u.) and $<16$ W.u., respectively. 
Both values are in agreement with the expectations for a two-phonon state.

\textit{$6^+$ state at 2950 keV:}
Based on the good agreement with the predicted angular distribution shape, $J^{\pi}=6^+$ is assigned to this state. 
Reference~\cite{fransen2005} does not find evidence for this state, while Ref.~\cite{aa94zr} makes a tentative $6^+$ assignment. 
The state has also been observed in the high-spin study of  Ref.~\cite{wang2014} with a $J^{\pi} = 6^+$ assignment.

\textit{$2^+$ state at 3050 keV:}
Despite some deviations from the DWBA calculation at the largest angles, $J^{\pi} = 2^+$ is assigned to this state in agreement with Refs.~\cite{fransen2005,aa94zr}.  

\textit{$(4^+ + 5^-)$ states at 3183 keV:}
For the 3183 keV state, there are deviations at large scattering angles from the behavior expected for a transition to a $4^+$ state.
A better description of the angular distribution is obtained assuming a superposition of $4^+$ and $5^-$ states. 
References\cite{fransen2005,aa94zr} both find a 4$^+$ state at this energy.
The former also assigns a state at 3191 keV with $J^\pi = (4^-)$.
Therefore, a tentative $J^{\pi}=(4^+ + 5^-)$ assignment is made from the present data. 

\textit{($5^-$) states at 3238 keV and 3340 keV:}
The first maximum of the angular distribution is consistent with a $J^\pi = 5^-$ assignment for both cases. 
The state at 3238 keV shows deviations at the largest angles, and Ref.~\cite{aa94zr} finds a compatible state with $J^\pi = 4^+$.
Deviations at small and large angles are observed for the 3340 keV state, but a corresponding $J^\pi = 5^-$ state is seen in Ref.~\cite{aa94zr}.
Therefore, only tentative $5^-$ spin-parity values are given for both.

\textit{$2^+$ state at 3262 keV:}
The angular distribution of the cross section is very well described assuming excitation of a $J^{\pi} = 2^+$ state consistent with Refs.~\cite{aa94zr,fransen2005}.

\textit{State at 3372 keV:}
It was not possible to extract a meaningful angular distribution for this state (not shown in Fig.~\ref{Fig: Angular distributions of 92Zr 1}).
Reference~\cite{aa94zr} finds evidence of a {$3^-$} state at a slightly higher energy of 3382 keV. 
Wang \textit{et al.}~\cite{wang2014} identify a $7^-$ state at 3379 keV and Ref.~\cite{fransen2005} finds a $1^-$ state at 3371 keV.

\textit{$3^-$ state at 3446 keV:}
This strongly excited state with a clear $\Delta L = 3$ angular distribution is also seen by Ref.~\cite{aa94zr}. 
However, Ref.~\cite{fransen2005} finds a tentative $J^{\pi}=2^+$ state at this energy.

\textit{2$^+$ state at 3488 keV:}
The $J^{\pi}=2^+$ assignment of Ref.~\cite{fransen2005} is confirmed. 
Reference~\cite{aa94zr} finds a tentative $J^{\pi}=3^-$ state at the same energy.

\textit{States at 3623 keV and 3644 keV:}
For both states (not shown in Fig.~\ref{Fig: Angular distributions of 92Zr 1}), no meaningful angular distribution could be obtained. 
Reference~\cite{aa94zr} finds \textcolor{blue}{a} $J=(2,3)$ state at 3628 keV and a tentative $J^{\pi}=2^+$ state at 3640 keV.
%


\FloatBarrier

\subsection{$^{94}$Zr }
\label{sect: $^{94}$Zr }

\begin{figure}
\centering
\includegraphics[width=\columnwidth]{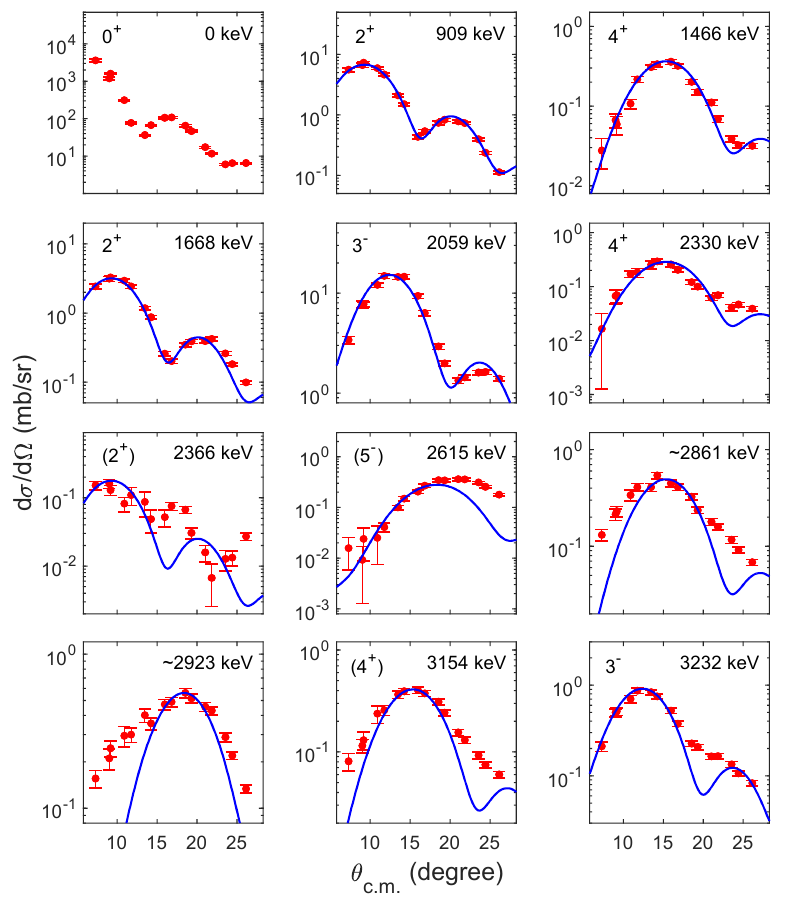}
\caption[Angular distributions of the $^{94}$Zr($p,p^{\prime}$) experiment.]{
\label{Fig: Angular distributions of Zr 1} Same as Fig.~\ref{Fig: Angular distributions of 92Zr 1} but for $^{94}$Zr($p,p^{\prime}$). }
\end{figure}

The information on the states extracted from the $^{94}$Zr($p,p^{\prime}$) data is summarized in Fig.~\ref {Fig: Angular distributions of Zr 1} and Tab.~\ref{Tab: Results of 94Zr}. 
The direct comparison is again limited to $(\alpha,\alpha^\prime)$ \cite{aa94zr} and $(n,n^\prime\gamma)$ \cite{elhami2008} reactions.

\begin{table}
\centering
\renewcommand*{\arraystretch}{1.3}
\caption[Results of the $^{94}$Zr($p,p^{\prime}$) experiment.]{\label{Tab: Results of 94Zr}
Excitation energies, spin and parity assignments of transitions excited in the $^{94}$Zr($p,p^{\prime}$) experiment and comparison to results obtained with the $(\alpha,\alpha^\prime)$ \cite{aa94zr} and $(n,n^\prime\gamma)$ \cite{elhami2008} reactions.
The peak numbers refer to Fig.~\ref{Fig: Examples spectra Zr, Mo, Zn 3}.} 
\begin{threeparttable}[c]
\begin{tabular}{ccccccc}
\hline
\hline
& $(p,p^\prime)$ & \cite{aa94zr} &  \cite{elhami2008}
     & $(p,p^\prime)$  & \cite{aa94zr} & \cite{elhami2008} \\
No. & $E_{\rm x}$  & $E_{\rm x}$  & $E_{\rm x}$  & $J^{\pi}$ & $J^{\pi}$ & $J^{\pi}$    \\
\hline
  & 0          & 0         & 0           &  0$^+$        & 0$^+$     & 0$^+$         \\
2  & 909(20)    & 919(10)   & 918.82(2)   &  2$^+$        & 2$^+$     & 2$^+$         \\
3  & 1466(10)   & 1469(10)  & 1469.70(2)  &  4$^+$        & 4$^+$     & 4$^+$         \\
4  & 1668(10)   & 1671(10)  & 1671.45(2)  &  2$^+$        & 2$^+$     & 2$^+$         \\
5  & 2059(10)   & 2057(10)  & 2057.87(2)  &  3$^-$        & 3$^-$     & 3$^-$         \\
6  &            &           & 2151.34(2)  &               &           & 2$^+$         \\
7  &            & 2336(10)  & 2329.97(2)  &  4$^+$        & 4$^+$     & 4$^+$         \\
  &            & 2372(10)  & 2366.34(2)  &  (2$^+$)      & 2$^+$     & 2$^+$         \\
9  & 2615(10)   & 2617(10)  & 2605.39(3)  &  (5$^-$)        & 5$^-$     & 5$^-$         \\
 & $\sim$2861 & 2881(10)  & 2873.65(3)  &               & 4$^+$     & (4$^+$)       \\
   &            &           & 2888.25(7)  &               &           & 4$^+$         \\
 & $\sim$2923 & 2940(10)  & 2908.04(2)  &               & 5$^-$     & 2$^+$         \\
   &            &           & 2927.50(5)  &               &           & 3$^-$         \\
   &            &           & 2945.33(5)  &               &           & 5$^-$         \\
12 & 3154(10)   & 3163(10)  & 3155.93(3)  &  ($4^+$)        & 4$^+$     & (4$^+$)       \\
13 & 3232(10)   & 3244(10)  & 3224.84(4)  &  3$^-$        & 3$^-$     & (4$^+$)       \\
\hline
\hline
\end{tabular}
\end{threeparttable}
\end{table}

\textit{$2^+$ states at 909 keV and 1668 keV:}
The experimental angular distributions are both described very well using $2^+$ collective model transition densities. 
This is consistent with the results of Refs.
\cite{aa94zr,elhami2008}.
The 1668 keV state is the candidate for the quadrupole one-phonon MSS of $^{94}$Zr \cite{scheikh2014}.
Recent literature discussed this state instead as originating from a cross-shell excitation over the $Z=40$ sub-shell closure \cite{gavrielov2022}. 

\textit{4$_1^+$ at 1466 keV:}
In agreement with Refs.~\cite{aa94zr,elhami2008}, the angular distribution can be well described assuming $J^{\pi}=4^+$.

\textit{3$^-$ state at 2059 keV:}
This strongly excited state is confirmed to be the octupole FSS in agreement with Refs.~\cite{aa94zr,elhami2008}.

\textit{$2^+$ state at 2151 keV:}
According to Ref.~\cite{elhami2008}, this well-known $2^+$ state has a small decay branch to the ground state with $B(E2)=0.019$ W.u. and a large transition strength to the $2^+_1$ state indicating significant two-phonon components in its wave function. 
Unfortunately, due to its proximity to the $3^-$ state, no meaningful angular distribution could be extracted from the present work.

\textit{$4^+$ state at 2330 keV and ($2^+$) state at 2366 keV:}
These two states form a doublet which cannot be resolved in the present data. 
Since the energies are precisely known \cite{elhami2008}, it was possible to fit the doublet by fixing the peak positions and to decompose the angular distributions. 
The results confirm the $J^{\pi}=4^+$ assignment of  Refs.~\cite{aa94zr,elhami2008} of the state at 2330 keV. 
For the state at 2366 keV, a tentative $J^{\pi}=2^+$ assignment is made consistent with the other work.

\textit{($5^-$) state at 2615 keV:}
The measured energy and the spin-parity assignment agree best with a 5$^-$ state as seen in Refs.~\cite{aa94zr,elhami2008}, but the deviations at larger angles make the assignment tentative only.

\textit{States at $\sim 2861$ keV and $\sim 2923$ keV:}
In this region one finds, according to Ref.~\cite{elhami2008}, five close-lying states that could not be resolved in the present work. 
Indeed, the measured angular distributions do not allow any spin assignments. 
It is possible, however, to draw some conclusions about the states most strongly excited in the present work. 
In Fig.~\ref{Fig: Angular distributions of Zr 1}, it can be seen that the structure at $\sim 2861$~keV is dominated by the transition to a $4^+$ state and the structure at $\sim 2923$~keV by the transition to a $5^-$ state.
Possible candidates have been seen in the literature. 

\textit{($4^+$) state at 3154 keV:}
A positive parity and a spin of $J=4$ is favored for this state in agreement with the findings of Refs.~\cite{aa94zr,elhami2008}.
The assignment is tentative only because of the deviations at larger scattering angles.
They might result from a tentative $6^+$ state found at 3142 keV in high-spin studies \cite{fotiades2002}.  

\textit{$3^-$ state at 3232 keV:}
In agreement with Ref.~\cite{aa94zr}, $J^{\pi}=3^-$ is assigned to this state.
Elhami {\it et al.}~\cite{elhami2008} find a tentative $4^+$ state close in energy.

\FloatBarrier

\subsection{$^{94}$Mo }
\label{sect: $^{94}$Mo }

\begin{figure}
\centering
\includegraphics[width=\columnwidth]{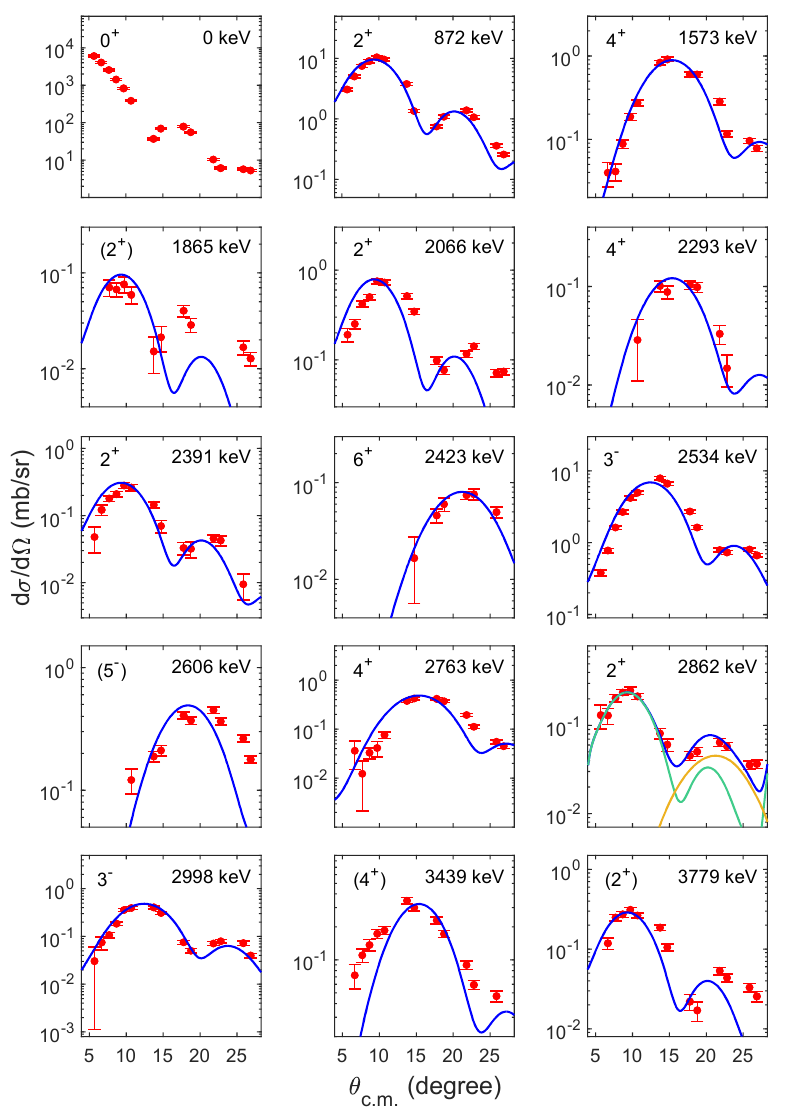}
\caption[Angular distributions of the $^{94}$Mo($p,p^{\prime}$) experiment.]{\label{Fig: Angular distributions of 94Mo 1}
Same as Fig.~\ref{Fig: Angular distributions of 92Zr 1} but for $^{94}$Mo($p,p^{\prime}$). }
\end{figure}

The angular distributions measured for the $^{94}$Mo($p,p^{\prime}$) reaction are displayed in Fig.~\ref{Fig: Angular distributions of 94Mo 1} and the resulting spin-parity assignments are given in Tab.~\ref{Tab: Results of 94Mo}.
The latter includes a comparison with results from the $(e,e^\prime)$ experiment described in Sec.~\ref{sec:II}, with the low-energy ($p,p^{\prime}$) and ($d,d^{\prime}$) experiments of Ref.~\cite{pp94mo}, and with the $(n,n^\prime\gamma)$ reaction \cite{fransen2003}.

\begin{table}
\renewcommand*{\arraystretch}{1.3}
\caption[Results of the $^{94}$Mo($p,p^{\prime}$) experiment.]{\label{Tab: Results of 94Mo}
Excitation energies (in keV), spin and parity assignments of the $^{94}$Mo($p,p^{\prime}$) and $(e,e^\prime)$ experiments and a comparison to data from Refs.~\cite{pp94mo} and \cite{fransen2003}.
The peak numbers (first column) refer to Fig.~\ref{Fig: Examples spectra Zr, Mo, Zn 3}.}
\setlength{\tabcolsep}{0.05mm}
\begin{tabular}{ccccccccc}
\hline
\hline
   & $(p,p^\prime)$ & $(e,e^\prime)$ & \cite{pp94mo} & \cite{fransen2003} & $(p,p^\prime)$ & $(e,e^\prime)$ & \cite{pp94mo} & \cite{fransen2003}  \\
No. & $E_{\rm x}$ & $E_{\rm x}$ & $E_{\rm x}$  & $E_{\rm x}$  & $J^{\pi}$ & $J^{\pi}$ & $J^{\pi}$ & $J^{\pi}$     \\
   \hline
    & 0          & 0         & 0   & 0       & 0$^+$         & 0$^+$             & 0$^+$    & 0$^+$        \\
2    & 872(10)   & 870(10)    & 871(2)    & 871.09(10)  & 2$^+$  & 2$^+$  & 2$^+$ & 2$^+$                 \\
3    & 1573(10) & 1579(10)   & 1573(2)   & 1573.72(14) & 4$^+$  & 4$^+$ & 4$^+$  & 4$^+$                 \\
    & 1865(10)  & 1857(10)  & 1864(2)   & 1864.3(1)   & (2$^+$) & 2$^+$  & 2$^+$ & 2$^+$                 \\
5    & 2066(10) & 2067(10)   & 2068(2)   & 2067.4(1)   & 2$^+$  & 2$^+$ & 2$^+$  & 2$^+$                 \\
6    & 2293(10) & 2297(10)  & 2295(2)   & 2294.7(2)   & 4$^+$  & 4$^+$  & 4$^+$  & 4$^+$                 \\
   & 2347(20) &  & 2322(2)   &             &               &           &        &                       \\
   & 2391(10) & 2390(10)   & 2393(2)   & 2393.1(1)   & 2$^+$  & 2$^+$  & 2$^+$ & 2$^+$                 \\
    & 2423(10)   &    & 2424(2)   & 2423.4(2)   & 6$^+$          &   & (6$^+$)     & 6$^+$    \\
10   & 2534(10) & 2531(10)   & 2534(5)   & 2533.8(3)   & 3$^-$  & 3$^-$ & 3$^-$  & 3$^-$                 \\
   & 2606(10) &  & 2611(5)   & 2610.5(2)   & (5$^-$)     &  & 5$^-$             & 5$^-$                 \\
   & 2763(10) & 2769(10)   & 2770(5)   & 2768.2(2)   & 4$^+$   & 4$^+$  & 4$^+$ & 4$^+$                 \\
13   & 2862(10) & 2862(10)   & 2854(5)   & 2870.02(2)  & 2$^+$ & 2$^+$ & (4$^+$)  & 2$^+$         \\
     &    &        &           & 2872.4(2)   &          &     &               & 6$^+$                 \\
   & 2946(10) &  & 2960(5)   &             &               & (4$^+$)   &        &                       \\
15     & 2998(10) & 3015(10)   & 3014(5)   & 3011.5(2)   & 3$^-$   & 3$^-$    & 3$^-$   & 3$^-$            \\
   & 3439(20)   &   &        &             & (4$^+$)       &         &          &                       \\
25   & 3779(20) & 3793(15)   &           &             & (2$^+$)   & 2$^+$   &     &                       \\
   & 3888(20)   & 3892(15)   &        &             &    & 2$^+$   &         &                       \\
\hline
\hline
\end{tabular}
\end{table}

\textit{$2_1^+$ state at 872 keV and 4$_1^+$ state at 1573 keV:}
All four studies agree on spins and parities $J^{\pi}=2^+$ and $4^+$, respectively, for these states.

\textit{$2_2^+$ state at 1865 keV:}
According to Ref.~\cite{fransen2003}, this state is a member of the symmetric two-phonon triplet. 
This is indicated by a large $B(E2)$ value  of 60$^{+20}_{-30}$ W.u.\ for the $E2$ decay to the 2$_1^+$ state. 
Indeed, the angular distribution of the cross section clearly differs from the expectation of a one-phonon 2$^+$ state at larger scattering angles giving a hint of two-step contributions to the excitation mechanism of the state. 
In the present data, there are indications that the 2$^+$ state at 1865 keV is a doublet with another close-lying state at $\sim$1900 keV. 
A possible candidate is the 3$_1^{-}$ state of $^{100}$Mo at an energy of 1908 keV. 
However, it is difficult to make this observation quantitative because of the limited statistics. 
One should, therefore, treat the measured cross sections with some reservation.

\textit{$2_3^+$ state at 2066 keV:}
It is supposed to be the one-phonon 2$^+$ mixed-symmetry state as indicated by a large $B(M1)$ value of 0.56(5) $\mu^2_N$ for the decay to the $2^+_1$ state \cite{fransen2003}. 
The maximum of the measured cross sections is clearly shifted to larger scattering angles with respect to the theoretical calculation. 
This can be explained by a smaller matter transition radius than assumed in the collective model due to the mixed-symmetric character of this state (see the discussion in paper II). 
Hence, a spin and parity of $J^{\pi}=2^+$ are assigned.

\textit{$4_2^+$ state at 2293 keV:}
Only 7 data points could be extracted for this state. 
However, they are well described by the collective model assuming spin and parity $J^{\pi}=4^+$. 
This assignment agrees with the other experiments.

\textit{State at 2347 keV:}
No meaningful angular distribution could be extracted for this state. Reference~\cite{pp94mo} reports a state at a similar energy but without additional information on quantum numbers.

\textit{$2_4^+$ state at 2391 keV:}
$J^{\pi}=2^+$ is assigned consistent with the other experiments.

\textit{$6_1^+$ state at 2423 keV:}
For the extraction of the angular distribution, the peak position was fixed to the known energy from $\gamma$ spectroscopy. 
Hence, no independent determination could be performed.
The angular distribution provides clear evidence for spin and parity $J^{\pi}=6^+$. 
Such a high-spin state could not be seen in the kinematics of the electron scattering experiments.
The result is in agreement with Ref.~\cite{fransen2003}, while Ref.~\cite{pp94mo} made a tentative $J^{\pi}=(5^-,6^+$) assignment. 

\textit{$3_1^-$ state at 2534 keV:}
$J^{\pi}=3^-$ is consistently assigned by all experiments.
The large cross sections in the proton and electron scattering experiments demonstrate the one-phonon character of this state.

\textit{($5_1^-$) state at 2606 keV:}
The position of the maximum points to a $5^-$ state. Due to deviations between experiment and the collective model results at larger angles, this assignment is tentative. 
References~\cite{pp94mo,fransen2003} find evidence of a $5^-$ state at consistent energies. 

\textit{$4_3^+$ state at 2763 keV:}
The state is clearly populated in all experiments and consistently interpreted to have $J^{\pi}=4^+$.

\textit{Doublet at 2862 keV:}
According to Ref.~\cite{fransen2003}, this state is a doublet consisting of a $2^+$ state and a $6^+$ state at energies of 2870 keV and 2872 keV, respectively. 
Due to the energy resolution of $30-40$ keV (FWHM), it was not possible to resolve the doublet in the present experiment. 
However, the angular distribution is well described using a superposition (blue curve in Fig.~\ref{Fig: Angular distributions of 94Mo 1}) of a $2^+$ (green curve) state and a $6^+$ state (yellow curve).
The $2^+$ cross sections dominate at smaller scattering angles while the $6^+$ cross sections are stronger at the larger scattering angles.

\textit{State at 2946 keV:}
No angular distribution could be obtained for this state. 
Reference~\cite{pp94mo} finds evidence of a tentative 4$^+$ state at a slightly higher excitation energy still compatible within uncertainties.

\textit{$3_2^-$ state at 2998 keV:}
The $\gamma$ spectroscopy results of Ref.~\cite{fransen2003} provide evidence for a $2^+$ state at 2993 keV and a $3^-$ state at 3011 keV. 
This is again a doublet which cannot be resolved in the present work. 
However, the cross section is very well described assuming a 3$^-$ cross section only.
The $(e,e^\prime)$ experiment observes a transition with consistent energy and an $E3$ form factor supporting a $J^\pi = 3^-$ assignment.

\textit{States at 3779 keV and 3888 keV:}
The first maximum of the angular distribution of the transition to the state at 3.779 keV is consistent with an interpretation as a $J^\pi = 2^+$ state. 
This is independently confirmed by the $(e,e^\prime)$ data.
Although the angular distribution populating the state at 3888 keV does not allow a unique assignment, the observation of an $E2$ transition to a state an energy of 3892 keV in the electron scattering data also indicates $J^\pi = 2^+$ for this state.

\textit{States at 3086 keV, 3158 keV, 3190 keV, 3243 keV, 3365 keV, 3439 keV, 3518 keV, 3617 keV, 3688 keV, 
3842 keV, and 3974 keV:}
The angular distributions of the cross sections suggest a tentative 4$^+$ state at 3439 keV.
No spin assignments were possible for the other states and their angular distributions are not shown in Fig.~\ref{Fig: Angular distributions of 94Mo 1}.
States with corresponding energies are found in some cases in the other experiments, but the data indicate that most or all of these transitions are multiplets.
Thus, no detailed comparison was attempted.



\FloatBarrier

\subsection{$^{96}$Mo }
\label{sect: $^{96}$Mo }

\begin{figure}[h!]
\centering
\includegraphics[width=\columnwidth]{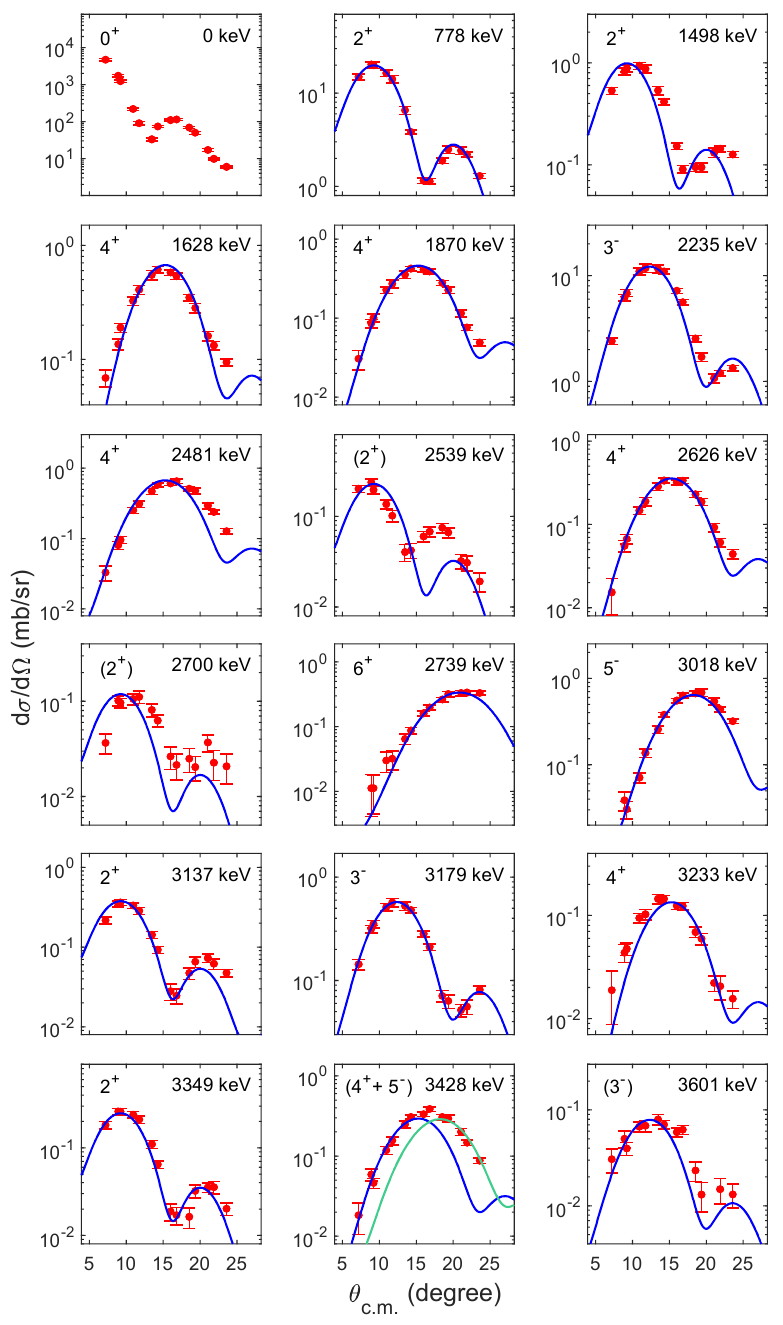}
\caption[Angular distributions some states of  $^{96}$Mo$(p,p^{\prime}$). Shown are only states where a meaningful angular distribution could be measured. Calculations are done using a phenomenological optical potential and a collective mode.]{\label{Fig: Angular distributions 96Mo1}
Same as Fig.~\ref{Fig: Angular distributions of 92Zr 1} but for $^{96}$Mo($p,p^{\prime}$). 
}
\end{figure}

Information on excited states in $^{96}$Mo analyzed in the present study is summarized in Fig.~\ref{Fig: Angular distributions 96Mo1} and Tab.~\ref{Tab: Results of 96Mo}. 
A direct comparison is made in Tab.~\ref{Tab: Results of 96Mo} with results from a $(p,p^\prime)$ experiment at low incident energies \cite{fretwurst1987} and the $(n,n^\prime\gamma)$ reaction \cite{lesher2007}.

\begin{table} \centering
\renewcommand*{\arraystretch}{1.3}
\caption[Results of the $^{96}$Mo($p,p^{\prime}$) experiment.]{\label{Tab: Results of 96Mo} 
Excitation energies (in keV), spin and parity assignments of the $^{96}$Mo($p,p^{\prime}$) experiment and comparison to 
a $(p,p^\prime)$ experiment at low incident energy \cite{fretwurst1987} and to the $(n,n^\prime\gamma)$ reaction \cite{lesher2007}.
The peak numbers (first column) refer to Fig.~\ref{Fig: Examples spectra Zr, Mo, Zn 3}.}
\begin{threeparttable}[c]
\setlength{\tabcolsep}{0.8mm}
\begin{tabular}{ccccccc}
\hline
\hline
 &$(p,p^\prime)$& \cite{fretwurst1987} & \cite{lesher2007} & $(p,p^\prime)$& \cite{fretwurst1987} & \cite{lesher2007}   \\
No. &$E_{\rm x}$ & $E_{\rm x}$ & $E_{\rm x}$  & $J^{\pi}$ & $J^{\pi}$ & $J^{\pi}$    \\
 \hline
& 0& 0& 0& 0$^+$& 0$^+$& 0$^+$         \\
2& 773(20)& 778(5)& 778.22(5)& 2$^+$& 2$^+$& 2$^+$         \\
3& 1498(10)& 1498(5)& 1497.76(5)& 2$^+$& 2$^+$& 2$^+$         \\
   &                &         &          1625.88(5)&                  &                 & 2$^+$         \\
4& 1628(10)& 1627(5)& 1628.20(5)& 4$^+$& 4$^+$& 4$^+$         \\
5& 1870(10)& 1870(5)& 1869.53(5)& 4$^+$& 4$^+$& 4$^+$         \\
6& 2235(10)& 2234(5)& 2234.63(6)& 3$^-$& 3$^-$& 3$^-$         \\
& 2428(10)& 2432(5)& 2426.07(6)&                   & 2$^+$& 2$^+$         \\
8& 2481(10)& 2481(5)& 2481.04(8)& 4$^+$& 4$^+$& 4$^+$         \\
& 2539(10)& 2542(5)& 2540.34(5)& (2$^+$)&                   & (2$^+$,3$^+$)         \\
10& 2626(10)& 2625(5)& 2625.23(8)& 4$^+$& 4$^+$& 4$^+$         \\
& 2700(10)&                   & 2700.08(8)& (2$^+$)&                   & 2$^+$         \\
& 2739(10)& 2734(5)& 2734.61(12)& 6$^+$& (5$^-$)& 4$^+$         \\
 &                  &                   &2755.12(30)&                   &                   & 6$^+$         \\
& 2809(10)& 2807(5)& 2818.53(10)& & (3$^-$)& 4$^+$         \\
& 2860(10)& 2875(5)&    &                &(4$^+$,6$^+$)&                         \\
& 2989(10)& 2981(5)& 2986.79(6)&                   & 1$^-$,2$^+$& 2$^+$         \\
16& 3018(20)& 3020(5)& 3024.47(6)& 5$^-$& 4$^+$& 2$^+$         \\
& 3137(20)& 3140(5)& 3134.50(7)& 2$^+$&                   & 2$^+$         \\
& 3179(20)& 3182(5)& 3178.74(7)& 3$^-$& 3$^-$& 3$^-$         \\
& 3233(20)& 3235(5)& 3232.45(10)& 4$^+$&                   & 3         \\
& 3283(20)& 3287(5)& 3284.89(17)&                   & 2$^-$,4$^+$& 2$^+$         \\
& 3349(20)& 3342(5)& 3352.01(12)& 2$^+$& 3$^-$,4$^+$& 2$^+$         \\
& 3428(20)& 3430(5)& 3433.58(25)& (4$^+$+5$^-$)& 4$^+$& 4$^+$         \\
 &                   &                   &3441.97(17)&                   &                   & 4$^+$         \\
& 3464(20)& & 3464.63(12)& &  & 3         \\
 & & 3474(5) & 3472.18(29)& &2$^+$,4$^+$,5$^-$ & 2$^+$         \\
& 3542(20)& 3549(5)& 3540.80(14)&                   & $3^-$ & 3         \\
& 3601(20)& 3597(5)& 3610.46(24)& (3$^-$)& 2$^+$& 2,3         \\
\hline
\hline
\end{tabular}
\end{threeparttable}
\end{table}

\textit{$2_1^+$ states  at 773 keV and 1498 keV:}
The two states are unanimously assigned $J^\pi = 2^+$ by all data, although the $(p,p^\prime)$ angular distribution of the 1498 keV state shows a small systematic shift to larger scattering angles indicating a smaller transition radius than predicted by the DWBA calculations. 

\textit{$4_{1,2}^+$ states  at 1628 keV and 1870 keV:}
Reference \cite{lesher2007} finds a $2^+$ state at 1625.9 keV and a $4^+$ state at 1628.2 keV.
These cannot be resolved in the present experiment. However, the cross sections are well described assuming a $J^\pi = 4^+$ state only, pointing toward a dominant
two-phonon character of the $2^+$ state with small cross sections in a direct one-step reaction.
$J^\pi = 4^+$ for the state at 1870 keV provides a very good description of the angular distribution consistent with findings in the other experiments.

\textit{$3_1^-$ state at 2235 keV:}
The angular distribution of the cross sections clearly indicates a 3$^-$ state in agreement with Refs. ~\cite{fretwurst1987,lesher2007}, and the large cross sections make it a candidate for the 1-phonon octupole FSS.
Reference \cite{lesher2007} finds further evidence for a $2^+$ state at 2096 keV and a $4^+$ state at 2219 keV. 
The former is assumed to be the $2^+$ MSS state. 
Due to the small energy difference from the strongly excited 3$^-$ state, it was not possible to resolve them in the present experiment.

\textit{State at 2428 keV:}
There is evidence for a state at 2428 keV, but no meaningful angular distribution of the cross section could be extracted. 
References~\cite{fretwurst1987,lesher2007} assign $J^\pi$ = 2$^+$ to this state.

\textit{$4^+$ states at 2481 keV and 2626 keV:}
Excitation energies and angular distributions measured in the present experiment agree with the 4$^+$ assignments of Refs.~\cite{fretwurst1987,lesher2007}.

\textit{($2^+$) states at 2539 keV and 2700 keV:}
The first maxima of the angular distributions indicate $J^\pi = 2^+$ for both states. 
However, at scattering angles larger than 15$^\circ$, the calculation fails to describe the data for the 2539 keV state. 
The deviations might be explained by considering contributions from the strongly excited 3$_1^-$ state in $^{94}$Mo observed at an energy of 2534 keV.
Reference \cite{lesher2007} makes a tentative $(2^+, 3^+)$ assignment. 
Taking the present data into consideration, $J^\pi = 2^+$ is more likely.
In case of the 2700 keV state, the description at larger scattering angles is rather poor.
Thus, only a tentative $J^\pi = 2^+$ assignment is made. 
Reference \cite{fretwurst1987} does not report this state, while Ref.~\cite{lesher2007} also finds a $2^+$ state at the same energy.

\textit{$6^+$ state at 2739 keV:}
The angular distribution is well described assuming spin and parity $J^\pi = 6^+$. 
Reference \cite{lesher2007} also reports a $6^+$ state but at a higher energy of 2755 keV slightly outside the respective error bars. 
Additionally, evidence for a $4^+$ state at 2735 keV is found. 
Reference \cite{fretwurst1987} reports a tentative $5^-$ state at an energy of 2734 keV.

\textit{States at 2809 keV, 2860 keV and 2989 keV:}
A meaningful angular distribution could not be extracted for any of these states. 
References \cite{fretwurst1987,lesher2007} find a number of  possible corresponding states to the 2809 keV and 2989 keV excitations with a variety of spin-parity assignments. 
A state fitting in energy  to the 2860 keV excitation with $(4^+,6^+)$ constraints on the spin-parity value is seen in Ref.~\cite{fretwurst1987}.

\textit{$5^-$ state at 3018 keV:}
In disagreement with Refs.~\cite{fretwurst1987} and \cite{lesher2007}, which report a $4^+$ state at the same energy and a $2^+$ state at an energy of 3024 keV, the present results are well described assuming $J^\pi = 5^-$. 

\textit{$2^+$ state at 3137 keV:}
The angular distribution of the cross sections clearly indicates a $2^+$ state in agreement with Ref.~\cite{lesher2007}. 
Reference \cite{fretwurst1987} also sees a state at 3140 keV, but without a spin or parity assignment.

\textit{$3^-$ state at 3179 keV:}
Previous $J^\pi = 3^-$ assignments \cite{fretwurst1987,lesher2007} are confirmed by the angular distribution data.

\textit{$4^+$ state at 3233 keV:}
The angular distribution data points toward a 4$^+$ state. 
References~\cite{fretwurst1987,lesher2007} find states at consistent energies with a $J = 3$ assignment by the former and no spin or parity information by the latter.

\textit{State at 3283 keV:}
No conclusions about spin and parity could be drawn for this state. 
Reference \cite{fretwurst1987} limits the possible quantum number of a corresponding state at 3287 keV to $2^-$ or $4^+$. 
In contrast, Ref.~\cite{lesher2007} finds a $2^+$ state at an energy of 3285 keV.

\textit{$2^+$ state at 3349 keV:}
The angular dependence of the cross sections provides clear evidence for a $2^+$ state in agreement with the findings of Ref.~\cite{lesher2007}. 
In contrast, Ref.~\cite{fretwurst1987} assigns $J^\pi = 3^-, 4^+$ to a state at 3342 keV.

\textit{($4^+$ {\rm +}$5^-$) state at 3428 keV:}
The angular distribution points to a 4$^+$ state at smaller angles and to a $5^-$ state at larger angles.
A satisfactory description can be achieved assuming a superposition of both.
Thus, a tentative assignment assuming a doublet is made. 
Reference \cite{fretwurst1987} assigns $J^\pi = 4^+$ to a state at 3430 keV, while Ref.~\cite{lesher2007} finds two $4^+$ states at energies of 3434 keV and 3442 keV.

\textit{States at 3464 keV and 3542 keV:}
No conclusion about the spin and parity could be drawn from the angular distributions of these states. 
Both Refs.~\cite{fretwurst1987} and \cite{lesher2007} find states at comparable energies with spin-parity suggestions summarized in Tab.~\ref{Tab: Results of 96Mo}.

\textit{($3^-$) state at 3601 keV:}
Because of deviations at scattering angles larger than $15^\circ$, only a tentative $J^\pi = 3^-$ assignment is given for this state. 
The result is consistent with the findings of Ref.~\cite{lesher2007}, which reports a $J = 2, 3$ state at 3610 keV, while Ref.~\cite{fretwurst1987} finds a $2^+$ state at 3597 keV.

\subsection{$^{70}$Zn }
\label{sect: $^{70}$Zn }

\begin{figure}
\centering
\includegraphics[width=\columnwidth]{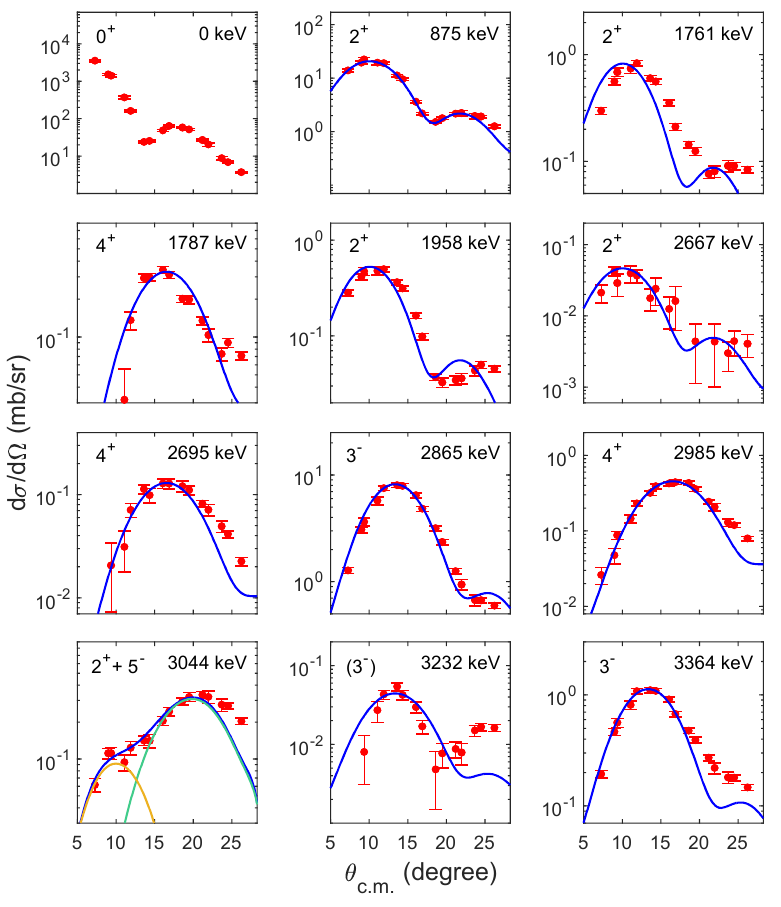}
\caption[Angular distributions some states of  $^{70}$Zn$(p,p^{\prime}$). Shown are only states where a meaningful angular distribution could be measured. Calculations are done using a phenomenological optical potential and a collective mode.]{\label{Fig: Angular distributions 70Zn}
Same as Fig.~\ref{Fig: Angular distributions of 92Zr 1} but for $^{70}$Zn($p,p^{\prime}$). 
}
\end{figure}

\begin{figure}
\centering
\includegraphics[width=\columnwidth]{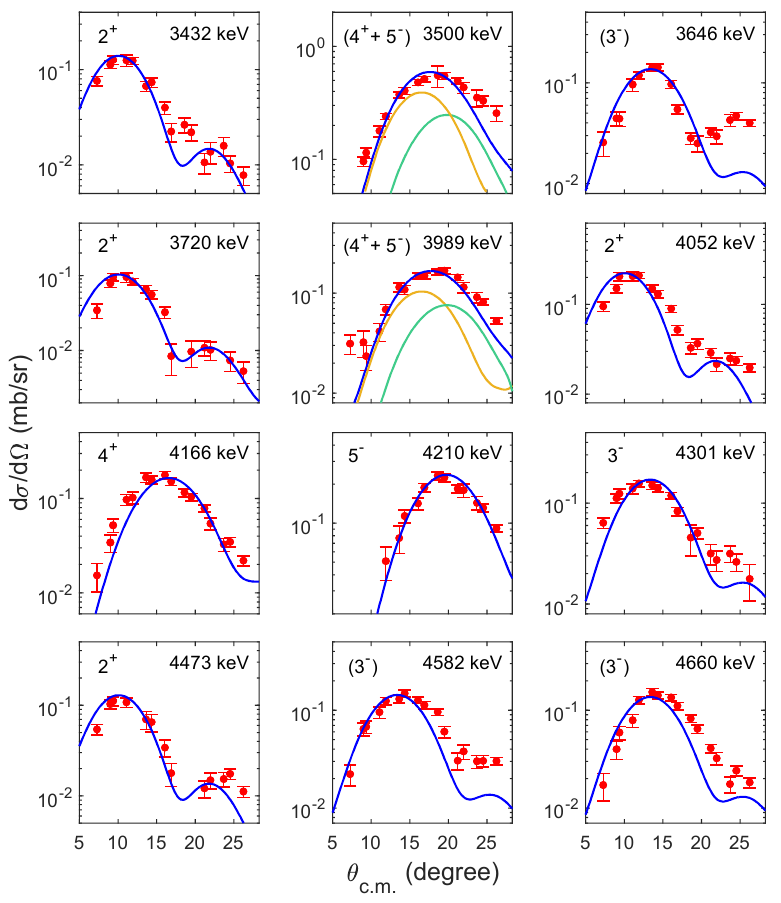}
\caption[Angular distributions some states of  $^{70}$Zn$(p,p^{\prime}$). Shown are only states where a meaningful angular distribution could be measured. Calculations are done using a phenomenological optical potential and a collective mode.]{\label{Fig: Angular distributions 70Zn 2}
Same as Fig.~\ref{Fig: Angular distributions of 92Zr 1} but for $^{70}$Zn($p,p^{\prime}$).
}
\end{figure}

Results for the study of the $^{70}$Zn$(p,p^\prime)$ reaction are presented in Figs.~\ref{Fig: Angular distributions 70Zn},\ref{Fig: Angular distributions 70Zn 2} and Tab.~\ref{Tab: Results of 70Zn}.
They are compared to low-energy $(p,p^\prime)$ \cite{jabbour1987} and $(t,p)$  experiments \cite{hudson1972} as well as a comprehensive study \cite{muecher2009} including $(\gamma,\gamma^\prime)$, $(n,n^\prime\gamma)$, and Coulomb excitation \cite{muecher2009a} measurements.

\begin{table*} 
\centering
\renewcommand*{\arraystretch}{1.3}
\caption[Results of the $^{70}$Zn($p,p^{\prime}$) experiment.]{\label{Tab: Results of 70Zn}
Excitation energies (in keV), spin and parity assignments in $^{70}$Zn from the present work and comparison to low-energy $(p,p^\prime)$ \cite{jabbour1987} and $(t,p)$  experiments \cite{hudson1972} as well as a comprehensive study \cite{muecher2009} including $(\gamma,\gamma^\prime)$, $(n,n^\prime\gamma)$, and Coulomb excitation \cite{muecher2009a} measurements.
No information on excitation uncertainties is available from Ref.~\cite{muecher2009}, but the errors should be of the order of 1 keV or less.
The peak numbers (first column) refer to Fig.~\ref{Fig: Examples spectra Zr, Mo, Zn 3}.}
\begin{threeparttable}[c]
\begin{tabular}{ccccccccc}
\hline
\hline
& $(p,p^\prime)$ & \cite{jabbour1987} & \cite{hudson1972} & \cite{muecher2009} & $(p,p^\prime)$ & \cite{jabbour1987} & \cite{hudson1972} & \cite{muecher2009}  \\
No. & $E_{\rm x}$ & $E_{\rm x}$ & $E_{\rm x}$  & $E_{\rm x}$  & $J^{\pi}$ & $J^{\pi}$ & $J^{\pi}$ & $J^{\pi}$     \\

\hline
& 0& 0& 0& 0& 0$^+$& 0$^+$& 0$^+$& 0$^+$       \\
2& 875(20)& 886(5)& 884(10)& 885.0& 2$^+$& 2$^+$& 2$^+$& 2$^+$      \\
3& 1761(10)& 1764(5)& 1767(10)& 1759.3& 2$^+$& 2$^+$ &   &  2$^+$  \\
4& 1787& 1790(5)& 1787(10)& 1786.8& 4$^+$& 4$^+$& 4$^+$& 4$^+$      \\
5& 1958(10)& 1960(5)& 1955(10)& 1957.2& 2$^+$& 2$^+$& 2$^+$& 2$^+$      \\
& 2667(10)& 2665(5)& 2661(10)& 2659.3& 2$^+$& 2$^+$& 2$^+$& 2$^+$      \\
7& 2695(10)& 2694(5)& 2690(10)& 2693.5& 4$^+$& (3$-$5)$^+$& 4$^+$  & 4$^+$    \\
8& 2865(10)& 2863(5)& 2856(10)& 2859.8& 3$^-$& 3$^-$& 3$^-$& 3$^-$      \\
5& 2930(20)& 2954(5)&                    &2949.3&                   &                   &                    &2$^+$,3$^+$     \\
9& 2985(10)& 2975(5)& 2971(10)& 2977.9& 4$^+$& 4$^+$& 4$^+$& 4$^+$      \\
& 3044(10)& 3042(5)& 3031(10)& 3038.2& 2$^+$+5$^-$& 5$^-$& (5$^-$)& 2,3      \\
& 3078(20)&                   &                   & 3083.1&                   &                   &                   & (1)      \\
& 3133(20)&                    &                   &3136.9&                   &                   &                   &                        \\
& 3232(10)& 3235(5)& 3232(10)& 3247.0& (3$^-$)& 4$^+$& (4$^+$)& (4$^-$)      \\
& 3364(10)&                    &3340(10)& 3341.8& 3$^-$&                   & (3$^-$)& 3$^-$      \\
& 3432(20)& 3419(5)& 3423(10)& 3422.0& 2$^+$& 3$^-$& (3$^-$)& 2$^+$      \\
& 3500(20)& 3464(5)& 3458(10)& 3454.3& 4$^+$+5$^-$& 4$^+$& 4$^+$&      \\
 & 3500(20)  &3506(5)& 3502(10)& 3520.7&  4$^+$+5$^-$                 & 5$^-$& 5$^-$& $J$$\leq$4    \\
& 3646(10)& 3635(5)& 3631(10)& 3634.7& (3$^-$)& 2$^+$& 2$^+$& 2$^+$      \\
& 3720(10)& 3712(5)& 3707(10)& 3723.9& 2$^+$& 2$^+$& 2$^+$& 2$^+$      \\
& 3741(20)& 3750(5)& 3746(10)&                   &                   &                   &                   &                         \\
& 3803(20)& 3813(5)& 3802(10)&                   &                   &                   &                   &                         \\
& 3844(10)& 3844(5)& 3839(10)& 3848.3&                   &                   & 1$^-$&                        \\
& 3908(10)& 3888(5)& 3914(10)& 3904.9&                    &(3$-$5)$^+$&                   &                        \\
& 3989(10)&                    &3999(10)&                   &                   & 4$^+$+5$^-$& 2$^+$&                         \\
& 4052(10)& 4066(10)& 4060(10)& 4061.4 & 2$^+$& 4$^+$& 4$^+$& 4$^+$      \\
& 4107(20)& 4136(10)& 4140(10)&                   &                   & (1$^-$3)$^+$&                   &                        \\
& 4166(10)& 4172(10)&                   &                   & 4$^+$& 5$^-$&                   &                         \\
& 4210(10)&                   &                   &                   & 5$^-$&                   &                   &                        \\
& 4259(20)& 4284(10)& 4297(10)& 4264.6&                   & 2$^+$& 2$^+$&                         \\
& 4301(10)& 4309(10)&                   & 4309.0 & 3$^-$&                   &                   &                       \\
& 4332(20)&                   &                   &                   &                   &                   &                    &                      \\
& 4395(20)& 4367(10)&                    &                   &(3$^-$5)$^+$&                   &                       \\
& 4449(20)& 4444(10)&                   &                   &                   & $(3-5)^+$&                   &                      \\
& 4473(10)&                   &                   & 4463.8 &2$^+$&                   &                   &                       \\
& 4515(20)&                   &                   & 4514.4&                   &                   &                   &                        \\
& 4582(10)&                   &                   & 4588.4&  (3$^-$)&                   &                   &                        \\
& 4616(20)&                   &                   &                   &                   &                   &                   &                        \\
& 4660(10)&                    &                   &                   &(3$^-$)&                   &                   &                        \\
& 4720(20)&                   &                   & 4710.1&                   &                   &                   &                        \\
\hline
\hline
\end{tabular}
\end{threeparttable}
\end{table*}

\textit{$2^+$ states at 875 keV, 1958 keV and 2667 keV:}
The angular distributions from the present work as well as all other experimental results agree on $J^\pi =2^+$ for all three states.
The states 875 keV and 1958 keV are candidates for the one-phonon FSS and MSS.
The assignment of the latter is based on the large $B(M1)$ strength of 0.231(27) $\mu_N^2$ in the decay to the 2$_1^+$ state \cite{muecher2009}. 
Its $B(E2)$ strength to the g.s.\ is at least a factor of $\sim330$ smaller than the one of the 2$_1^+$ state \cite{muecher2009}. 
This is in contrast to the proton scattering cross sections where the first maxima differ only by a factor of $\sim 40$. 
This points to a dominance of neutron components in the wave function of the $2_3^+$state.

\textit{2$_2^+$ and 4$_1^+$ states at 1761 and 1787 keV:}
The two states form a doublet, which cannot be resolved in the present experiment. 
To nevertheless get separate angular distributions, the positions of the centroids were fixed in the fit using the known energies \cite{tuli2004}.
At low scattering angles, the contribution of the 4$_1^+$ state to the doublet is negligible and, therefore, the energy of the $2_2^+$ state could be determined independently. 
The spin and parity assignments of Refs.~\cite{jabbour1987,hudson1972,muecher2009}
are confirmed. 
A clear shift of the angular distribution to higher scattering angles compared to the model calculations can be seen for the 2$_2^+$ state, indicating a smaller matter transition radius than predicted by the collective model (cf.\ paper II). 
The 2$_2^+$ state has a large two-phonon component in its wave function as indicated by the large $B(E2)$ value of 22.5(39) W.u.\ to the 2$_1^+$ state \cite{muecher2009}. 
However, the shape of the angular distribution is consistent with a one-step excitation mechanism also indicating a sizable one-phonon component in the wave function of the state.

\textit{Signals at 2340 and 2410 keV:}
These states have not been seen in Refs.~\cite{hudson1972,muecher2009,tuli2004}, but Ref.~\cite{jabbour1987} finds a 2$^+$ state at 2375(5) keV. 
It was not possible to extract a meaningful angular distribution for the 2410 keV state, but the 2340 keV state clearly has $J^\pi = 2^+$. 
Due to the proximity in energy of those two states to a 2$^+$ state and a 4$^+$ state in $^{68}$Zn (2338 keV and 2417 keV, respectively), these states are interpreted as excitations of $^{68}$Zn.

\textit{$4_{2,3}^+$ states at 2695 and 2985 keV, and $3_1^-$ state 2865 keV:}
All states were also seen in Refs.~\cite{jabbour1987,hudson1972,muecher2009}.  
The angular distributions are well described assuming either spin and parity of $4^+$ or $3^-$, consistent with the findings in the other experiments.

\textit{State at 2930 keV:}
No conclusions about spin and parity could be drawn from the angular distribution of this state. 
Reference \cite{muecher2009} limits the possible quantum number to $2^+$ or $3^+$. 

\textit{2$^+$ + 5$_1^-$ doublet at 3044 keV:}
All other experiments identify a 5$^-$ state consistent in energy. 
The angular distribution obtained in the present work is best described assuming a doublet of a 2$^+$ and a 5$^-$ state. 
This is in agreement with Ref.~\cite{muecher2009}, which found a corresponding state with $J = (2, 3)$.  From the present results, $J^\pi = 2^+$ can be assigned to the second state.

\textit{States at 3078 and 3133 keV:}
No spins could be assigned to these states based on the angular distributions. 
Only Ref.~\cite{muecher2009} finds two states at compatible energies with a tentative spin assignment of $J = 1$ for the level at 3078 keV. 

\textit{(3$^-$) state at 3232 keV:}
The position of the maximum of the angular distribution points to $J^\pi = 3^-$. 
However, there are deviations from the expected angular distribution of a 3$^-$ state at small and large angles. 
Thus, only a tentative $3^-$ assignment is made. 
References \cite{jabbour1987,hudson1972} both find a state corresponding in energy assigned $J^\pi$ = 4$^+$. Reference \cite{muecher2009} reports two close-lying states at 3224 keV and 3247 keV with quantum numbers $J = 1$ and $J^\pi = (4^-)$, respectively. 
The relation to the state(s) observed in the present experiment remains unclear.

\textit{$3_2^-$  state at 3364 keV:}
The angular distributions of the cross section clearly indicates a $3^-$ state. 
This strongly excited state was not observed in Ref.~\cite{jabbour1987}. 
References \cite{hudson1972,tuli2004} quote a $3^-$ state at an energy of 3342 keV. 
It is not clear if this state corresponds to the one seen at 3364 keV or rather to a state at 3357 keV with possible quantum numbers $J = 1,2,3$.
Furthermore, Ref.~\cite{muecher2009} observes a state at 3357 keV with decays to lower-lying $2^+$ states but not to the ground state.
Such a decay behavior is at least compatible with a $3^-$ assignment.

\textit{2$_5^+$ state at 3432 keV:}
The good agreement between the model predictions and experimental data gives evidence for $J^\pi = 2^+$. 
This is in agreement with Ref.~\cite{muecher2009}. 
On the other hand, Refs.~\cite{jabbour1987,hudson1972} report $J^\pi = 3^-$.

\textit{Doublets at 3500 keV and 3989 keV:}
The measured angular distributions are best described in both cases assuming a doublet with $J^\pi = 4^+$ and $5^-$. 
The variation of the asymmetry of the line shape with angle for the 3500 keV state suggests that the $4^+$ state is at a higher excitation energy. 
References \cite{jabbour1987,hudson1972} find a $5^-$ and a $4^+$ state close to 3500 keV. 
However, in these two experiments the $5^-$ state is at a higher energy.
A state with comparable excitation energy to the 3989 keV doublet was only seen in Ref.~\cite{hudson1972}, but with a $J^\pi = 2^+$ assignment.

\textit{(3$^-$) state at 3646 keV:}
In disagreement with Refs.~\cite{jabbour1987,hudson1972,muecher2009} all favoring $J^\pi = 2^+$, the position of the angular distribution maximum is consistent with $J^\pi = 3^-$. Because of deviations at larger scattering angles, only a tentative $3^-$ assignment is made. 
However, there is no hint of a $2^+$ state as seen in the other experiments.

\textit{2$_6^+$ state at 3720 keV:}
$J^\pi$ = 2$^+$ assigned in Refs.~\cite{jabbour1987,hudson1972}
is confirmed by the angular distribution data.

\textit{States at 3741, 3803, 3844 and 3908 keV:}
No conclusions about spin and parity could be drawn for these states. 
References \cite{jabbour1987,hudson1972,tuli2004} find states at similar energies. 
A $J^\pi$ = 1$^-$ assignment is made for the state at 3844 keV \cite{hudson1972}, but such a transition  
would be only weakly excited in the present experiment.

\textit{($2^+$) state at 4052 keV:}
In contrast to the other experimental results, the angular distribution favors a 2$^+$ state, although the data are systematically shifted somewhat to larger angles than the collective model predicts. 
All other experiments assign $J^\pi =4^+$ to this state. 

Up to this point, an attempt has been made to make a one-to-one comparison between the data in this paper and the states seen in the various experiments.
This is very difficult for states above 4 MeV because of the high level density, i.e., the comparison given in Tab.~\ref{Tab: Results of 70Zn} should be treated with caution. 
There are usually several candidates in the other references overlapping in energy with the states seen in this experiment. 
Furthermore, with an energy resolution of $30 - 50$ keV (FWHM), the assumption of exciting resolved single states becomes increasingly questionable.
However, based on the experimental angular distributions shown in Fig.~\ref{Fig: Angular distributions 70Zn} and in combination with the available experimental data it is still possible to make some tentative assignments. 

\section{Summary}

We report measurements of proton-scattering reactions off the nuclides $^{70}$Zn, $^{92,94}$Zr, and $^{94,96}$Mo. 
All of them are cases with one or two proton and neutron pairs outside shell or subshell closures.
Absolute cross sections were determined for elastic and inelastic scattering as a function of scattering angles covering a range $8^\circ -25^\circ$ in the laboratory frame of reference. 
These angular distributions are compared to DWBA calculations based on collective-model form factors. The comparison allows for several unambiguous and independent spin-parity assignments for states excited by the $(p,p^\prime)$ reaction in the excitation-energy range up to approximately 3.5 MeV. 
Pre-existing assignments from the literature are mostly confirmed. 
In a few cases, new assignments are possible or ambiguities are clarified. 

These results provide insight into the low-energy nuclear structure in terms of its collective building blocks, such as one-phonon FSS and MSS, as well as candidates for multiphonon excitations built from them. They are used in the subsequent paper II to investigate a new signature of quadrupole MSS based on transition densities and a possible existence of octupole or hexadecapole MSS.

\section*{Acknowledgements}

We wish to thank J.~L.~Conradie and the accelerator team at iThemba LABS for their support in the experiments. 
We are indebted to J.~Carter for help in the data taking and to V.~Yu.~Ponomarev for help with the DWBA calculations.
This work was supported by the South African National Research Foundation (NRF) and by the Deutsche Forschungsgemeinschaft (DFG, German Research Foundation) under Grant No.\ SFB 1245 (project ID 279384907). 

\section*{Data Availability}

The data that support the findings of this article are not publicly available. 
They are available from the authors upon reasonable request.

\bibliography{MSSPaper1}

\begin{thebibliography}{64}%
\makeatletter
\providecommand \@ifxundefined [1]{%
 \@ifx{#1\undefined}
}%
\providecommand \@ifnum [1]{%
 \ifnum #1\expandafter \@firstoftwo
 \else \expandafter \@secondoftwo
 \fi
}%
\providecommand \@ifx [1]{%
 \ifx #1\expandafter \@firstoftwo
 \else \expandafter \@secondoftwo
 \fi
}%
\providecommand \natexlab [1]{#1}%
\providecommand \enquote  [1]{``#1''}%
\providecommand \bibnamefont  [1]{#1}%
\providecommand \bibfnamefont [1]{#1}%
\providecommand \citenamefont [1]{#1}%
\providecommand \href@noop [0]{\@secondoftwo}%
\providecommand \href [0]{\begingroup \@sanitize@url \@href}%
\providecommand \@href[1]{\@@startlink{#1}\@@href}%
\providecommand \@@href[1]{\endgroup#1\@@endlink}%
\providecommand \@sanitize@url [0]{\catcode `\\12\catcode `\$12\catcode
  `\&12\catcode `\#12\catcode `\^12\catcode `\_12\catcode `\%12\relax}%
\providecommand \@@startlink[1]{}%
\providecommand \@@endlink[0]{}%
\providecommand \url  [0]{\begingroup\@sanitize@url \@url }%
\providecommand \@url [1]{\endgroup\@href {#1}{\urlprefix }}%
\providecommand \urlprefix  [0]{URL }%
\providecommand \Eprint [0]{\href }%
\providecommand \doibase [0]{https://doi.org/}%
\providecommand \selectlanguage [0]{\@gobble}%
\providecommand \bibinfo  [0]{\@secondoftwo}%
\providecommand \bibfield  [0]{\@secondoftwo}%
\providecommand \translation [1]{[#1]}%
\providecommand \BibitemOpen [0]{}%
\providecommand \bibitemStop [0]{}%
\providecommand \bibitemNoStop [0]{.\EOS\space}%
\providecommand \EOS [0]{\spacefactor3000\relax}%
\providecommand \BibitemShut  [1]{\csname bibitem#1\endcsname}%
\let\auto@bib@innerbib\@empty
\bibitem [{\citenamefont {Pietralla}\ \emph {et~al.}(2008)\citenamefont
  {Pietralla}, \citenamefont {{von Brentano}},\ and\ \citenamefont
  {Lisetskiy}}]{pietralla2008}%
  \BibitemOpen
  \bibfield  {author} {\bibinfo {author} {\bibfnamefont {N.}~\bibnamefont
  {Pietralla}}, \bibinfo {author} {\bibfnamefont {P.}~\bibnamefont {{von
  Brentano}}},\ and\ \bibinfo {author} {\bibfnamefont {A.~F.}\ \bibnamefont
  {Lisetskiy}},\ }\bibfield  {title} {\bibinfo {title} {Experiments on
  multiphonon states with proton–neutron mixed symmetry in vibrational
  nuclei},\ }\href {https://doi.org/https://doi.org/10.1016/j.ppnp.2007.08.002}
  {\bibfield  {journal} {\bibinfo  {journal} {Prog. Part. Nucl. Phys.}\
  }\textbf {\bibinfo {volume} {60}},\ \bibinfo {pages} {225} (\bibinfo {year}
  {2008})}\BibitemShut {NoStop}%
\bibitem [{\citenamefont {Iachello}\ and\ \citenamefont
  {Arima}(1987)}]{iachello1987}%
  \BibitemOpen
  \bibfield  {author} {\bibinfo {author} {\bibfnamefont {A.}~\bibnamefont
  {Iachello}}\ and\ \bibinfo {author} {\bibfnamefont {A.}~\bibnamefont
  {Arima}},\ }\href@noop {} {\emph {\bibinfo {title} {The Interacting Boson
  Model}}}\ (\bibinfo  {publisher} {Cambridge University Press},\ \bibinfo
  {address} {Cambridge},\ \bibinfo {year} {1987})\BibitemShut {NoStop}%
\bibitem [{\citenamefont {Arima}\ \emph {et~al.}(1977)\citenamefont {Arima},
  \citenamefont {Otsuka}, \citenamefont {Iachello},\ and\ \citenamefont
  {Talmi}}]{arima1977}%
  \BibitemOpen
  \bibfield  {author} {\bibinfo {author} {\bibfnamefont {A.}~\bibnamefont
  {Arima}}, \bibinfo {author} {\bibfnamefont {T.}~\bibnamefont {Otsuka}},
  \bibinfo {author} {\bibfnamefont {F.}~\bibnamefont {Iachello}},\ and\
  \bibinfo {author} {\bibfnamefont {I.}~\bibnamefont {Talmi}},\ }\bibfield
  {title} {\bibinfo {title} {{C}ollective nuclear states as symmetric couplings
  of proton and neutron excitations},\ }\href
  {https://doi.org/https://doi.org/10.1016/0370-2693(77)90860-7} {\bibfield
  {journal} {\bibinfo  {journal} {Phys. Lett. B}\ }\textbf {\bibinfo {volume}
  {66}},\ \bibinfo {pages} {205} (\bibinfo {year} {1977})}\BibitemShut
  {NoStop}%
\bibitem [{\citenamefont {Iachello}(1984)}]{iachello1984}%
  \BibitemOpen
  \bibfield  {author} {\bibinfo {author} {\bibfnamefont {F.}~\bibnamefont
  {Iachello}},\ }\bibfield  {title} {\bibinfo {title} {New class of low-lying
  collective modes in nuclei},\ }\href
  {https://doi.org/10.1103/PhysRevLett.53.1427} {\bibfield  {journal} {\bibinfo
   {journal} {Phys. Rev. Lett.}\ }\textbf {\bibinfo {volume} {53}},\ \bibinfo
  {pages} {1427} (\bibinfo {year} {1984})}\BibitemShut {NoStop}%
\bibitem [{\citenamefont {Otsuka}\ \emph {et~al.}(1978)\citenamefont {Otsuka},
  \citenamefont {Arima},\ and\ \citenamefont {Iachello}}]{otsuka1978}%
  \BibitemOpen
  \bibfield  {author} {\bibinfo {author} {\bibfnamefont {T.}~\bibnamefont
  {Otsuka}}, \bibinfo {author} {\bibfnamefont {A.}~\bibnamefont {Arima}},\ and\
  \bibinfo {author} {\bibfnamefont {F.}~\bibnamefont {Iachello}},\ }\bibfield
  {title} {\bibinfo {title} {Nuclear shell model and interacting bosons},\
  }\href {https://doi.org/https://doi.org/10.1016/0375-9474(78)90532-8}
  {\bibfield  {journal} {\bibinfo  {journal} {Nucl. Phys. A}\ }\textbf
  {\bibinfo {volume} {309}},\ \bibinfo {pages} {1} (\bibinfo {year}
  {1978})}\BibitemShut {NoStop}%
\bibitem [{\citenamefont {Bohle}\ \emph {et~al.}(1984)\citenamefont {Bohle},
  \citenamefont {Richter}, \citenamefont {Steffen}, \citenamefont {Dieperink},
  \citenamefont {{Lo Iudice}}, \citenamefont {Palumbo},\ and\ \citenamefont
  {Scholten}}]{bohle1984}%
  \BibitemOpen
  \bibfield  {author} {\bibinfo {author} {\bibfnamefont {D.}~\bibnamefont
  {Bohle}}, \bibinfo {author} {\bibfnamefont {A.}~\bibnamefont {Richter}},
  \bibinfo {author} {\bibfnamefont {W.}~\bibnamefont {Steffen}}, \bibinfo
  {author} {\bibfnamefont {A.}~\bibnamefont {Dieperink}}, \bibinfo {author}
  {\bibfnamefont {N.}~\bibnamefont {{Lo Iudice}}}, \bibinfo {author}
  {\bibfnamefont {F.}~\bibnamefont {Palumbo}},\ and\ \bibinfo {author}
  {\bibfnamefont {O.}~\bibnamefont {Scholten}},\ }\bibfield  {title} {\bibinfo
  {title} {New magnetic dipole excitation mode studied in the heavy deformed
  nucleus $^{156}${G}d by inelastic electron scattering},\ }\href
  {https://doi.org/https://doi.org/10.1016/0370-2693(84)91099-2} {\bibfield
  {journal} {\bibinfo  {journal} {Phys. Lett. B}\ }\textbf {\bibinfo {volume}
  {137}},\ \bibinfo {pages} {27} (\bibinfo {year} {1984})}\BibitemShut
  {NoStop}%
\bibitem [{\citenamefont {Heyde}\ \emph {et~al.}(2010)\citenamefont {Heyde},
  \citenamefont {von Neumann-Cosel},\ and\ \citenamefont
  {Richter}}]{heyde2010}%
  \BibitemOpen
  \bibfield  {author} {\bibinfo {author} {\bibfnamefont {K.}~\bibnamefont
  {Heyde}}, \bibinfo {author} {\bibfnamefont {P.}~\bibnamefont {von
  Neumann-Cosel}},\ and\ \bibinfo {author} {\bibfnamefont {A.}~\bibnamefont
  {Richter}},\ }\bibfield  {title} {\bibinfo {title} {Magnetic dipole
  excitations in nuclei: Elementary modes of nucleonic motion},\ }\href
  {https://doi.org/10.1103/RevModPhys.82.2365} {\bibfield  {journal} {\bibinfo
  {journal} {Rev. Mod. Phys.}\ }\textbf {\bibinfo {volume} {82}},\ \bibinfo
  {pages} {2365} (\bibinfo {year} {2010})}\BibitemShut {NoStop}%
\bibitem [{\citenamefont {Pietralla}\ \emph {et~al.}(1994)\citenamefont
  {Pietralla}, \citenamefont {von Brentano}, \citenamefont {Casten},
  \citenamefont {Otsuka},\ and\ \citenamefont {Zamfir}}]{pietralla1994}%
  \BibitemOpen
  \bibfield  {author} {\bibinfo {author} {\bibfnamefont {N.}~\bibnamefont
  {Pietralla}}, \bibinfo {author} {\bibfnamefont {P.}~\bibnamefont {von
  Brentano}}, \bibinfo {author} {\bibfnamefont {R.~F.}\ \bibnamefont {Casten}},
  \bibinfo {author} {\bibfnamefont {T.}~\bibnamefont {Otsuka}},\ and\ \bibinfo
  {author} {\bibfnamefont {N.~V.}\ \bibnamefont {Zamfir}},\ }\bibfield  {title}
  {\bibinfo {title} {Distribution of low-lying quadrupole phonon strength in
  nuclei},\ }\href {https://doi.org/10.1103/PhysRevLett.73.2962} {\bibfield
  {journal} {\bibinfo  {journal} {Phys. Rev. Lett.}\ }\textbf {\bibinfo
  {volume} {73}},\ \bibinfo {pages} {2962} (\bibinfo {year}
  {1994})}\BibitemShut {NoStop}%
\bibitem [{\citenamefont {Pietralla}\ \emph {et~al.}(1999)\citenamefont
  {Pietralla}, \citenamefont {Fransen}, \citenamefont {Belic}, \citenamefont
  {von Brentano}, \citenamefont {Frie\ss{}ner}, \citenamefont {Kneissl},
  \citenamefont {Linnemann}, \citenamefont {Nord}, \citenamefont {Pitz},
  \citenamefont {Otsuka}, \citenamefont {Schneider}, \citenamefont {Werner},\
  and\ \citenamefont {Wiedenh\"over}}]{pietralla1999}%
  \BibitemOpen
  \bibfield  {author} {\bibinfo {author} {\bibfnamefont {N.}~\bibnamefont
  {Pietralla}}, \bibinfo {author} {\bibfnamefont {C.}~\bibnamefont {Fransen}},
  \bibinfo {author} {\bibfnamefont {D.}~\bibnamefont {Belic}}, \bibinfo
  {author} {\bibfnamefont {P.}~\bibnamefont {von Brentano}}, \bibinfo {author}
  {\bibfnamefont {C.}~\bibnamefont {Frie\ss{}ner}}, \bibinfo {author}
  {\bibfnamefont {U.}~\bibnamefont {Kneissl}}, \bibinfo {author} {\bibfnamefont
  {A.}~\bibnamefont {Linnemann}}, \bibinfo {author} {\bibfnamefont
  {A.}~\bibnamefont {Nord}}, \bibinfo {author} {\bibfnamefont {H.~H.}\
  \bibnamefont {Pitz}}, \bibinfo {author} {\bibfnamefont {T.}~\bibnamefont
  {Otsuka}}, \bibinfo {author} {\bibfnamefont {I.}~\bibnamefont {Schneider}},
  \bibinfo {author} {\bibfnamefont {V.}~\bibnamefont {Werner}},\ and\ \bibinfo
  {author} {\bibfnamefont {I.}~\bibnamefont {Wiedenh\"over}},\ }\bibfield
  {title} {\bibinfo {title} {Transition rates between mixed symmetry states:
  First measurement in ${}^{94}\mathrm{Mo}$},\ }\href
  {https://doi.org/10.1103/PhysRevLett.83.1303} {\bibfield  {journal} {\bibinfo
   {journal} {Phys. Rev. Lett.}\ }\textbf {\bibinfo {volume} {83}},\ \bibinfo
  {pages} {1303} (\bibinfo {year} {1999})}\BibitemShut {NoStop}%
\bibitem [{\citenamefont {Pietralla}\ \emph {et~al.}(2000)\citenamefont
  {Pietralla}, \citenamefont {Fransen}, \citenamefont {von Brentano},
  \citenamefont {Dewald}, \citenamefont {Fitzler}, \citenamefont
  {Frie\ss{}ner},\ and\ \citenamefont {Gableske}}]{pietralla2000}%
  \BibitemOpen
  \bibfield  {author} {\bibinfo {author} {\bibfnamefont {N.}~\bibnamefont
  {Pietralla}}, \bibinfo {author} {\bibfnamefont {C.}~\bibnamefont {Fransen}},
  \bibinfo {author} {\bibfnamefont {P.}~\bibnamefont {von Brentano}}, \bibinfo
  {author} {\bibfnamefont {A.}~\bibnamefont {Dewald}}, \bibinfo {author}
  {\bibfnamefont {A.}~\bibnamefont {Fitzler}}, \bibinfo {author} {\bibfnamefont
  {C.}~\bibnamefont {Frie\ss{}ner}},\ and\ \bibinfo {author} {\bibfnamefont
  {J.}~\bibnamefont {Gableske}},\ }\bibfield  {title} {\bibinfo {title}
  {Proton-neutron mixed-symmetry ${3}_{\mathrm{ms}}^{+}$ state in
  $^{94}${M}o},\ }\href {https://doi.org/10.1103/PhysRevLett.84.3775}
  {\bibfield  {journal} {\bibinfo  {journal} {Phys. Rev. Lett.}\ }\textbf
  {\bibinfo {volume} {84}},\ \bibinfo {pages} {3775} (\bibinfo {year}
  {2000})}\BibitemShut {NoStop}%
\bibitem [{\citenamefont {Fransen}\ \emph {et~al.}(2001)\citenamefont
  {Fransen}, \citenamefont {Pietralla}, \citenamefont {{von Brentano}},
  \citenamefont {Dewald}, \citenamefont {Gableske}, \citenamefont {Gade},
  \citenamefont {Lisetskiy},\ and\ \citenamefont {Werner}}]{fransen2001}%
  \BibitemOpen
  \bibfield  {author} {\bibinfo {author} {\bibfnamefont {C.}~\bibnamefont
  {Fransen}}, \bibinfo {author} {\bibfnamefont {N.}~\bibnamefont {Pietralla}},
  \bibinfo {author} {\bibfnamefont {P.}~\bibnamefont {{von Brentano}}},
  \bibinfo {author} {\bibfnamefont {A.}~\bibnamefont {Dewald}}, \bibinfo
  {author} {\bibfnamefont {J.}~\bibnamefont {Gableske}}, \bibinfo {author}
  {\bibfnamefont {A.}~\bibnamefont {Gade}}, \bibinfo {author} {\bibfnamefont
  {A.}~\bibnamefont {Lisetskiy}},\ and\ \bibinfo {author} {\bibfnamefont
  {V.}~\bibnamefont {Werner}},\ }\bibfield  {title} {\bibinfo {title} {First
  observation of a mixed-symmetry two-{Q}-phonon $2_{2,{ \rm ms}}^+$ state in
  $^{94}${M}o},\ }\href
  {https://doi.org/https://doi.org/10.1016/S0370-2693(01)00345-8} {\bibfield
  {journal} {\bibinfo  {journal} {Phys. Lett. B}\ }\textbf {\bibinfo {volume}
  {508}},\ \bibinfo {pages} {219} (\bibinfo {year} {2001})}\BibitemShut
  {NoStop}%
\bibitem [{\citenamefont {Stegmann}\ \emph {et~al.}(2017)\citenamefont
  {Stegmann}, \citenamefont {Stahl}, \citenamefont {Rainovski}, \citenamefont
  {Pietralla}, \citenamefont {Stoyanov}, \citenamefont {Carpenter},
  \citenamefont {Janssens}, \citenamefont {Lettmann}, \citenamefont {Möller},
  \citenamefont {Möller}, \citenamefont {Werner},\ and\ \citenamefont
  {Zhu}}]{stegmann2017}%
  \BibitemOpen
  \bibfield  {author} {\bibinfo {author} {\bibfnamefont {R.}~\bibnamefont
  {Stegmann}}, \bibinfo {author} {\bibfnamefont {C.}~\bibnamefont {Stahl}},
  \bibinfo {author} {\bibfnamefont {G.}~\bibnamefont {Rainovski}}, \bibinfo
  {author} {\bibfnamefont {N.}~\bibnamefont {Pietralla}}, \bibinfo {author}
  {\bibfnamefont {C.}~\bibnamefont {Stoyanov}}, \bibinfo {author}
  {\bibfnamefont {M.}~\bibnamefont {Carpenter}}, \bibinfo {author}
  {\bibfnamefont {R.}~\bibnamefont {Janssens}}, \bibinfo {author}
  {\bibfnamefont {M.}~\bibnamefont {Lettmann}}, \bibinfo {author}
  {\bibfnamefont {T.}~\bibnamefont {Möller}}, \bibinfo {author} {\bibfnamefont
  {O.}~\bibnamefont {Möller}}, \bibinfo {author} {\bibfnamefont
  {V.}~\bibnamefont {Werner}},\ and\ \bibinfo {author} {\bibfnamefont
  {S.}~\bibnamefont {Zhu}},\ }\bibfield  {title} {\bibinfo {title}
  {Identification of the one-quadrupole phonon $2^+_{1,{\rm ms}}$ state of
  $^{204}${H}g},\ }\href
  {https://doi.org/https://doi.org/10.1016/j.physletb.2017.04.032} {\bibfield
  {journal} {\bibinfo  {journal} {Phys. Lett. B}\ }\textbf {\bibinfo {volume}
  {770}},\ \bibinfo {pages} {77} (\bibinfo {year} {2017})}\BibitemShut
  {NoStop}%
\bibitem [{\citenamefont {Kern}\ \emph {et~al.}(2019)\citenamefont {Kern},
  \citenamefont {Stegmann}, \citenamefont {Pietralla}, \citenamefont
  {Rainovski}, \citenamefont {Carpenter}, \citenamefont {Janssens},
  \citenamefont {Lettmann}, \citenamefont {M\"oller}, \citenamefont {M\"oller},
  \citenamefont {Stahl}, \citenamefont {Werner},\ and\ \citenamefont
  {Zhu}}]{kern2019}%
  \BibitemOpen
  \bibfield  {author} {\bibinfo {author} {\bibfnamefont {R.}~\bibnamefont
  {Kern}}, \bibinfo {author} {\bibfnamefont {R.}~\bibnamefont {Stegmann}},
  \bibinfo {author} {\bibfnamefont {N.}~\bibnamefont {Pietralla}}, \bibinfo
  {author} {\bibfnamefont {G.}~\bibnamefont {Rainovski}}, \bibinfo {author}
  {\bibfnamefont {M.~P.}\ \bibnamefont {Carpenter}}, \bibinfo {author}
  {\bibfnamefont {R.~V.~F.}\ \bibnamefont {Janssens}}, \bibinfo {author}
  {\bibfnamefont {M.}~\bibnamefont {Lettmann}}, \bibinfo {author}
  {\bibfnamefont {O.}~\bibnamefont {M\"oller}}, \bibinfo {author}
  {\bibfnamefont {T.}~\bibnamefont {M\"oller}}, \bibinfo {author}
  {\bibfnamefont {C.}~\bibnamefont {Stahl}}, \bibinfo {author} {\bibfnamefont
  {V.}~\bibnamefont {Werner}},\ and\ \bibinfo {author} {\bibfnamefont
  {S.}~\bibnamefont {Zhu}},\ }\bibfield  {title} {\bibinfo {title} {Nuclear
  isovector valence-shell excitation of $^{202}\mathrm{Hg}$},\ }\href
  {https://doi.org/10.1103/PhysRevC.99.011303} {\bibfield  {journal} {\bibinfo
  {journal} {Phys. Rev. C}\ }\textbf {\bibinfo {volume} {99}},\ \bibinfo
  {pages} {011303(R)} (\bibinfo {year} {2019})}\BibitemShut {NoStop}%
\bibitem [{\citenamefont {Kern}\ \emph {et~al.}(2020)\citenamefont {Kern},
  \citenamefont {Zidarova}, \citenamefont {Pietralla}, \citenamefont
  {Rainovski}, \citenamefont {Stegmann}, \citenamefont {Blazhev}, \citenamefont
  {Boukhari}, \citenamefont {Cederk\"all}, \citenamefont {Cubiss},
  \citenamefont {Djongolov}, \citenamefont {Fransen}, \citenamefont {Gaffney},
  \citenamefont {Gladnishki}, \citenamefont {Giannopoulos}, \citenamefont
  {Hess}, \citenamefont {Jolie}, \citenamefont {Karayonchev}, \citenamefont
  {Kaya}, \citenamefont {Keatings}, \citenamefont {Kocheva}, \citenamefont
  {Kr\"oll}, \citenamefont {M\"oller}, \citenamefont {O'Neill}, \citenamefont
  {Pakarinen}, \citenamefont {Reiter}, \citenamefont {Rosiak}, \citenamefont
  {Scheck}, \citenamefont {Snall}, \citenamefont {S\"oderstr\"om},
  \citenamefont {Spagnoletti}, \citenamefont {Stoyanova}, \citenamefont
  {Thiel}, \citenamefont {Vogt}, \citenamefont {Warr}, \citenamefont {Welker},
  \citenamefont {Werner}, \citenamefont {Wiederhold},\ and\ \citenamefont
  {De~Witte}}]{kern2020}%
  \BibitemOpen
  \bibfield  {author} {\bibinfo {author} {\bibfnamefont {R.}~\bibnamefont
  {Kern}}, \bibinfo {author} {\bibfnamefont {R.}~\bibnamefont {Zidarova}},
  \bibinfo {author} {\bibfnamefont {N.}~\bibnamefont {Pietralla}}, \bibinfo
  {author} {\bibfnamefont {G.}~\bibnamefont {Rainovski}}, \bibinfo {author}
  {\bibfnamefont {R.}~\bibnamefont {Stegmann}}, \bibinfo {author}
  {\bibfnamefont {A.}~\bibnamefont {Blazhev}}, \bibinfo {author} {\bibfnamefont
  {A.}~\bibnamefont {Boukhari}}, \bibinfo {author} {\bibfnamefont
  {J.}~\bibnamefont {Cederk\"all}}, \bibinfo {author} {\bibfnamefont {J.~G.}\
  \bibnamefont {Cubiss}}, \bibinfo {author} {\bibfnamefont {M.}~\bibnamefont
  {Djongolov}}, \bibinfo {author} {\bibfnamefont {C.}~\bibnamefont {Fransen}},
  \bibinfo {author} {\bibfnamefont {L.~P.}\ \bibnamefont {Gaffney}}, \bibinfo
  {author} {\bibfnamefont {K.}~\bibnamefont {Gladnishki}}, \bibinfo {author}
  {\bibfnamefont {E.}~\bibnamefont {Giannopoulos}}, \bibinfo {author}
  {\bibfnamefont {H.}~\bibnamefont {Hess}}, \bibinfo {author} {\bibfnamefont
  {J.}~\bibnamefont {Jolie}}, \bibinfo {author} {\bibfnamefont
  {V.}~\bibnamefont {Karayonchev}}, \bibinfo {author} {\bibfnamefont
  {L.}~\bibnamefont {Kaya}}, \bibinfo {author} {\bibfnamefont {J.~M.}\
  \bibnamefont {Keatings}}, \bibinfo {author} {\bibfnamefont {D.}~\bibnamefont
  {Kocheva}}, \bibinfo {author} {\bibfnamefont {T.}~\bibnamefont {Kr\"oll}},
  \bibinfo {author} {\bibfnamefont {O.}~\bibnamefont {M\"oller}}, \bibinfo
  {author} {\bibfnamefont {G.~G.}\ \bibnamefont {O'Neill}}, \bibinfo {author}
  {\bibfnamefont {J.}~\bibnamefont {Pakarinen}}, \bibinfo {author}
  {\bibfnamefont {P.}~\bibnamefont {Reiter}}, \bibinfo {author} {\bibfnamefont
  {D.}~\bibnamefont {Rosiak}}, \bibinfo {author} {\bibfnamefont
  {M.}~\bibnamefont {Scheck}}, \bibinfo {author} {\bibfnamefont
  {J.}~\bibnamefont {Snall}}, \bibinfo {author} {\bibfnamefont {P.-A.}\
  \bibnamefont {S\"oderstr\"om}}, \bibinfo {author} {\bibfnamefont
  {P.}~\bibnamefont {Spagnoletti}}, \bibinfo {author} {\bibfnamefont
  {M.}~\bibnamefont {Stoyanova}}, \bibinfo {author} {\bibfnamefont
  {S.}~\bibnamefont {Thiel}}, \bibinfo {author} {\bibfnamefont
  {A.}~\bibnamefont {Vogt}}, \bibinfo {author} {\bibfnamefont {N.}~\bibnamefont
  {Warr}}, \bibinfo {author} {\bibfnamefont {A.}~\bibnamefont {Welker}},
  \bibinfo {author} {\bibfnamefont {V.}~\bibnamefont {Werner}}, \bibinfo
  {author} {\bibfnamefont {J.}~\bibnamefont {Wiederhold}},\ and\ \bibinfo
  {author} {\bibfnamefont {H.}~\bibnamefont {De~Witte}},\ }\bibfield  {title}
  {\bibinfo {title} {Restoring the valence-shell stabilization in
  $^{140}\mathrm{Nd}$},\ }\href {https://doi.org/10.1103/PhysRevC.102.041304}
  {\bibfield  {journal} {\bibinfo  {journal} {Phys. Rev. C}\ }\textbf {\bibinfo
  {volume} {102}},\ \bibinfo {pages} {041304(R)} (\bibinfo {year}
  {2020})}\BibitemShut {NoStop}%
\bibitem [{\citenamefont {Yaneva}\ \emph {et~al.}(2020)\citenamefont {Yaneva},
  \citenamefont {Kocheva}, \citenamefont {Rainovski}, \citenamefont {Jolie},
  \citenamefont {Pietralla}, \citenamefont {Blazhev}, \citenamefont {Dewald},
  \citenamefont {Djongolov}, \citenamefont {Fransen}, \citenamefont
  {Gladnishki}, \citenamefont {Henrich}, \citenamefont {Homm}, \citenamefont
  {Ide}, \citenamefont {John}, \citenamefont {Kalaydjieva}, \citenamefont
  {Karayonchev}, \citenamefont {Kern}, \citenamefont {Kleemann}, \citenamefont
  {Kr{\"o}ll}, \citenamefont {M{\"u}ller-Gatermann}, \citenamefont {Scheck},
  \citenamefont {Spagnoletti}, \citenamefont {Stoyanova},\ and\ \citenamefont
  {Werner}}]{yaneva2020}%
  \BibitemOpen
  \bibfield  {author} {\bibinfo {author} {\bibfnamefont {A.}~\bibnamefont
  {Yaneva}}, \bibinfo {author} {\bibfnamefont {D.}~\bibnamefont {Kocheva}},
  \bibinfo {author} {\bibfnamefont {G.}~\bibnamefont {Rainovski}}, \bibinfo
  {author} {\bibfnamefont {J.}~\bibnamefont {Jolie}}, \bibinfo {author}
  {\bibfnamefont {N.}~\bibnamefont {Pietralla}}, \bibinfo {author}
  {\bibfnamefont {A.}~\bibnamefont {Blazhev}}, \bibinfo {author} {\bibfnamefont
  {A.}~\bibnamefont {Dewald}}, \bibinfo {author} {\bibfnamefont
  {M.}~\bibnamefont {Djongolov}}, \bibinfo {author} {\bibfnamefont
  {C.}~\bibnamefont {Fransen}}, \bibinfo {author} {\bibfnamefont {K.~A.}\
  \bibnamefont {Gladnishki}}, \bibinfo {author} {\bibfnamefont
  {C.}~\bibnamefont {Henrich}}, \bibinfo {author} {\bibfnamefont
  {I.}~\bibnamefont {Homm}}, \bibinfo {author} {\bibfnamefont {K.~E.}\
  \bibnamefont {Ide}}, \bibinfo {author} {\bibfnamefont {P.~R.}\ \bibnamefont
  {John}}, \bibinfo {author} {\bibfnamefont {D.}~\bibnamefont {Kalaydjieva}},
  \bibinfo {author} {\bibfnamefont {V.}~\bibnamefont {Karayonchev}}, \bibinfo
  {author} {\bibfnamefont {R.}~\bibnamefont {Kern}}, \bibinfo {author}
  {\bibfnamefont {J.}~\bibnamefont {Kleemann}}, \bibinfo {author}
  {\bibfnamefont {T.}~\bibnamefont {Kr{\"o}ll}}, \bibinfo {author}
  {\bibfnamefont {C.}~\bibnamefont {M{\"u}ller-Gatermann}}, \bibinfo {author}
  {\bibfnamefont {M.}~\bibnamefont {Scheck}}, \bibinfo {author} {\bibfnamefont
  {P.}~\bibnamefont {Spagnoletti}}, \bibinfo {author} {\bibfnamefont
  {M.}~\bibnamefont {Stoyanova}},\ and\ \bibinfo {author} {\bibfnamefont
  {V.}~\bibnamefont {Werner}},\ }\bibfield  {title} {\bibinfo {title}
  {Experimental evidence for low-lying quadrupole isovector excitation of
  $^{208}${P}o},\ }\href {https://doi.org/10.1140/epja/s10050-020-00259-w}
  {\bibfield  {journal} {\bibinfo  {journal} {Eur. Phys. J. A}\ }\textbf
  {\bibinfo {volume} {56}},\ \bibinfo {pages} {246} (\bibinfo {year}
  {2020})}\BibitemShut {NoStop}%
\bibitem [{\citenamefont {Stetz}\ \emph {et~al.}(2025)\citenamefont {Stetz},
  \citenamefont {Mayr}, \citenamefont {Werner}, \citenamefont {Pietralla},
  \citenamefont {Tsunoda}, \citenamefont {Otsuka}, \citenamefont {Rainovski},
  \citenamefont {Beck}, \citenamefont {Borcea}, \citenamefont {Calinescu},
  \citenamefont {Costache}, \citenamefont {Dinescu}, \citenamefont {Florea},
  \citenamefont {Gladnishki}, \citenamefont {Ide}, \citenamefont {Ionescu},
  \citenamefont {Kocheva}, \citenamefont {Koseoglou}, \citenamefont {Lica},
  \citenamefont {M\ifmmode~\u{a}\else \u{a}\fi{}rginean}, \citenamefont
  {M\ifmmode~\u{a}\else \u{a}\fi{}rginean}, \citenamefont {Mihai},
  \citenamefont {Mihai}, \citenamefont {Mitu}, \citenamefont {Nickel},
  \citenamefont {Nita}, \citenamefont {Pascu}, \citenamefont {Stan},
  \citenamefont {Toma}, \citenamefont {Turturic\ifmmode~\u{a}\else
  \u{a}\fi{}},\ and\ \citenamefont {Zidarova}}]{stetz2025}%
  \BibitemOpen
  \bibfield  {author} {\bibinfo {author} {\bibfnamefont {T.}~\bibnamefont
  {Stetz}}, \bibinfo {author} {\bibfnamefont {H.}~\bibnamefont {Mayr}},
  \bibinfo {author} {\bibfnamefont {V.}~\bibnamefont {Werner}}, \bibinfo
  {author} {\bibfnamefont {N.}~\bibnamefont {Pietralla}}, \bibinfo {author}
  {\bibfnamefont {Y.}~\bibnamefont {Tsunoda}}, \bibinfo {author} {\bibfnamefont
  {T.}~\bibnamefont {Otsuka}}, \bibinfo {author} {\bibfnamefont
  {G.}~\bibnamefont {Rainovski}}, \bibinfo {author} {\bibfnamefont
  {T.}~\bibnamefont {Beck}}, \bibinfo {author} {\bibfnamefont {R.}~\bibnamefont
  {Borcea}}, \bibinfo {author} {\bibfnamefont {S.}~\bibnamefont {Calinescu}},
  \bibinfo {author} {\bibfnamefont {C.}~\bibnamefont {Costache}}, \bibinfo
  {author} {\bibfnamefont {I.~E.}\ \bibnamefont {Dinescu}}, \bibinfo {author}
  {\bibfnamefont {N.~M.}\ \bibnamefont {Florea}}, \bibinfo {author}
  {\bibfnamefont {K.}~\bibnamefont {Gladnishki}}, \bibinfo {author}
  {\bibfnamefont {K.~E.}\ \bibnamefont {Ide}}, \bibinfo {author} {\bibfnamefont
  {A.~N.}\ \bibnamefont {Ionescu}}, \bibinfo {author} {\bibfnamefont
  {D.}~\bibnamefont {Kocheva}}, \bibinfo {author} {\bibfnamefont
  {P.}~\bibnamefont {Koseoglou}}, \bibinfo {author} {\bibfnamefont
  {R.}~\bibnamefont {Lica}}, \bibinfo {author} {\bibfnamefont {N.}~\bibnamefont
  {M\ifmmode~\u{a}\else \u{a}\fi{}rginean}}, \bibinfo {author} {\bibfnamefont
  {R.}~\bibnamefont {M\ifmmode~\u{a}\else \u{a}\fi{}rginean}}, \bibinfo
  {author} {\bibfnamefont {C.}~\bibnamefont {Mihai}}, \bibinfo {author}
  {\bibfnamefont {R.~E.}\ \bibnamefont {Mihai}}, \bibinfo {author}
  {\bibfnamefont {A.}~\bibnamefont {Mitu}}, \bibinfo {author} {\bibfnamefont
  {C.~M.}\ \bibnamefont {Nickel}}, \bibinfo {author} {\bibfnamefont {C.~R.}\
  \bibnamefont {Nita}}, \bibinfo {author} {\bibfnamefont {S.}~\bibnamefont
  {Pascu}}, \bibinfo {author} {\bibfnamefont {L.}~\bibnamefont {Stan}},
  \bibinfo {author} {\bibfnamefont {S.}~\bibnamefont {Toma}}, \bibinfo {author}
  {\bibfnamefont {A.}~\bibnamefont {Turturic\ifmmode~\u{a}\else \u{a}\fi{}}},\
  and\ \bibinfo {author} {\bibfnamefont {R.}~\bibnamefont {Zidarova}},\
  }\bibfield  {title} {\bibinfo {title} {Isolated mixed-symmetry ${2}^{+}$
  state of the radioactive neutron-rich nuclide $^{132}\mathrm{Te}$},\ }\href
  {https://doi.org/10.1103/l6dj-yfz7} {\bibfield  {journal} {\bibinfo
  {journal} {Phys. Rev. C}\ }\textbf {\bibinfo {volume} {112}},\ \bibinfo
  {pages} {034325} (\bibinfo {year} {2025})}\BibitemShut {NoStop}%
\bibitem [{\citenamefont {Lisetskiy}\ \emph {et~al.}(2000)\citenamefont
  {Lisetskiy}, \citenamefont {Pietralla}, \citenamefont {Fransen},
  \citenamefont {Jolos},\ and\ \citenamefont {{von Brentano}}}]{lisetskiy2000}%
  \BibitemOpen
  \bibfield  {author} {\bibinfo {author} {\bibfnamefont {A.~F.}\ \bibnamefont
  {Lisetskiy}}, \bibinfo {author} {\bibfnamefont {N.}~\bibnamefont
  {Pietralla}}, \bibinfo {author} {\bibfnamefont {C.}~\bibnamefont {Fransen}},
  \bibinfo {author} {\bibfnamefont {R.~V.}\ \bibnamefont {Jolos}},\ and\
  \bibinfo {author} {\bibfnamefont {P.}~\bibnamefont {{von Brentano}}},\
  }\bibfield  {title} {\bibinfo {title} {Shell model description of
  “mixed-symmetry” states in $^{94}${M}o},\ }\href
  {https://doi.org/https://doi.org/10.1016/S0375-9474(00)00255-4} {\bibfield
  {journal} {\bibinfo  {journal} {Nucl. Phys. A}\ }\textbf {\bibinfo {volume}
  {677}},\ \bibinfo {pages} {100} (\bibinfo {year} {2000})}\BibitemShut
  {NoStop}%
\bibitem [{\citenamefont {Werner}\ \emph {et~al.}(2002)\citenamefont {Werner},
  \citenamefont {Belic}, \citenamefont {{von Brentano}}, \citenamefont
  {Fransen}, \citenamefont {Gade}, \citenamefont {{von Garrel}}, \citenamefont
  {Jolie}, \citenamefont {Kneissl}, \citenamefont {Kohstall}, \citenamefont
  {Linnemann}, \citenamefont {Lisetskiy}, \citenamefont {Pietralla},
  \citenamefont {Pitz}, \citenamefont {Scheck}, \citenamefont {Speidel},
  \citenamefont {Stedile},\ and\ \citenamefont {Yates}}]{werner2002}%
  \BibitemOpen
  \bibfield  {author} {\bibinfo {author} {\bibfnamefont {V.}~\bibnamefont
  {Werner}}, \bibinfo {author} {\bibfnamefont {D.}~\bibnamefont {Belic}},
  \bibinfo {author} {\bibfnamefont {P.}~\bibnamefont {{von Brentano}}},
  \bibinfo {author} {\bibfnamefont {C.}~\bibnamefont {Fransen}}, \bibinfo
  {author} {\bibfnamefont {A.}~\bibnamefont {Gade}}, \bibinfo {author}
  {\bibfnamefont {H.}~\bibnamefont {{von Garrel}}}, \bibinfo {author}
  {\bibfnamefont {J.}~\bibnamefont {Jolie}}, \bibinfo {author} {\bibfnamefont
  {U.}~\bibnamefont {Kneissl}}, \bibinfo {author} {\bibfnamefont
  {C.}~\bibnamefont {Kohstall}}, \bibinfo {author} {\bibfnamefont
  {A.}~\bibnamefont {Linnemann}}, \bibinfo {author} {\bibfnamefont {A.~F.}\
  \bibnamefont {Lisetskiy}}, \bibinfo {author} {\bibfnamefont {N.}~\bibnamefont
  {Pietralla}}, \bibinfo {author} {\bibfnamefont {H.~H.}\ \bibnamefont {Pitz}},
  \bibinfo {author} {\bibfnamefont {M.}~\bibnamefont {Scheck}}, \bibinfo
  {author} {\bibfnamefont {K.-H.}\ \bibnamefont {Speidel}}, \bibinfo {author}
  {\bibfnamefont {F.}~\bibnamefont {Stedile}},\ and\ \bibinfo {author}
  {\bibfnamefont {S.~W.}\ \bibnamefont {Yates}},\ }\bibfield  {title} {\bibinfo
  {title} {Proton–neutron structure of the ${N}=52$ nucleus $^{92}${Z}r},\
  }\href {https://doi.org/https://doi.org/10.1016/S0370-2693(02)02961-1}
  {\bibfield  {journal} {\bibinfo  {journal} {Phys. Lett. B}\ }\textbf
  {\bibinfo {volume} {550}},\ \bibinfo {pages} {140} (\bibinfo {year}
  {2002})}\BibitemShut {NoStop}%
\bibitem [{\citenamefont {Holt}\ \emph {et~al.}(2007)\citenamefont {Holt},
  \citenamefont {Pietralla}, \citenamefont {Holt}, \citenamefont {Kuo},\ and\
  \citenamefont {Rainovski}}]{holt2007}%
  \BibitemOpen
  \bibfield  {author} {\bibinfo {author} {\bibfnamefont {J.~D.}\ \bibnamefont
  {Holt}}, \bibinfo {author} {\bibfnamefont {N.}~\bibnamefont {Pietralla}},
  \bibinfo {author} {\bibfnamefont {J.~W.}\ \bibnamefont {Holt}}, \bibinfo
  {author} {\bibfnamefont {T.~T.~S.}\ \bibnamefont {Kuo}},\ and\ \bibinfo
  {author} {\bibfnamefont {G.}~\bibnamefont {Rainovski}},\ }\bibfield  {title}
  {\bibinfo {title} {Microscopic restoration of proton-neutron mixed symmetry
  in weakly collective nuclei},\ }\href
  {https://doi.org/10.1103/PhysRevC.76.034325} {\bibfield  {journal} {\bibinfo
  {journal} {Phys. Rev. C}\ }\textbf {\bibinfo {volume} {76}},\ \bibinfo
  {pages} {034325} (\bibinfo {year} {2007})}\BibitemShut {NoStop}%
\bibitem [{\citenamefont {Sieja}\ \emph {et~al.}(2009)\citenamefont {Sieja},
  \citenamefont {Mart\'{\i}nez-Pinedo}, \citenamefont {Coquard},\ and\
  \citenamefont {Pietralla}}]{sieja2009}%
  \BibitemOpen
  \bibfield  {author} {\bibinfo {author} {\bibfnamefont {K.}~\bibnamefont
  {Sieja}}, \bibinfo {author} {\bibfnamefont {G.}~\bibnamefont
  {Mart\'{\i}nez-Pinedo}}, \bibinfo {author} {\bibfnamefont {L.}~\bibnamefont
  {Coquard}},\ and\ \bibinfo {author} {\bibfnamefont {N.}~\bibnamefont
  {Pietralla}},\ }\bibfield  {title} {\bibinfo {title} {{D}escription of
  proton-neutron mixed-symmetry states near $^{132}\mathrm{Sn}$ within a
  realistic large scale shell model},\ }\href
  {https://doi.org/10.1103/PhysRevC.80.054311} {\bibfield  {journal} {\bibinfo
  {journal} {Phys. Rev. C}\ }\textbf {\bibinfo {volume} {80}},\ \bibinfo
  {pages} {054311} (\bibinfo {year} {2009})}\BibitemShut {NoStop}%
\bibitem [{\citenamefont {Soloviev}(1992)}]{soloviev1992}%
  \BibitemOpen
  \bibfield  {author} {\bibinfo {author} {\bibfnamefont {V.~G.}\ \bibnamefont
  {Soloviev}},\ }\href@noop {} {\emph {\bibinfo {title} {Theory of Atomic
  Nuclei: Quasiparticles and Phonons}}}\ (\bibinfo  {publisher} {IOP
  Publishing},\ \bibinfo {address} {Bristol},\ \bibinfo {year}
  {1992})\BibitemShut {NoStop}%
\bibitem [{\citenamefont {Pignanelli}\ \emph {et~al.}(1993)\citenamefont
  {Pignanelli}, \citenamefont {Blasi}, \citenamefont {Bordewijk}, \citenamefont
  {{De Leo}}, \citenamefont {Harakeh}, \citenamefont {Hofstee}, \citenamefont
  {Micheletti}, \citenamefont {Perrino}, \citenamefont {Ponomarev},
  \citenamefont {Soloviev}, \citenamefont {Sushkov},\ and\ \citenamefont {{van
  der Werf}}}]{pignanelli1993}%
  \BibitemOpen
  \bibfield  {author} {\bibinfo {author} {\bibfnamefont {M.}~\bibnamefont
  {Pignanelli}}, \bibinfo {author} {\bibfnamefont {N.}~\bibnamefont {Blasi}},
  \bibinfo {author} {\bibfnamefont {J.~A.}\ \bibnamefont {Bordewijk}}, \bibinfo
  {author} {\bibfnamefont {R.}~\bibnamefont {{De Leo}}}, \bibinfo {author}
  {\bibfnamefont {M.~N.}\ \bibnamefont {Harakeh}}, \bibinfo {author}
  {\bibfnamefont {M.~A.}\ \bibnamefont {Hofstee}}, \bibinfo {author}
  {\bibfnamefont {S.}~\bibnamefont {Micheletti}}, \bibinfo {author}
  {\bibfnamefont {R.}~\bibnamefont {Perrino}}, \bibinfo {author} {\bibfnamefont
  {V.~Y.}\ \bibnamefont {Ponomarev}}, \bibinfo {author} {\bibfnamefont {V.~G.}\
  \bibnamefont {Soloviev}}, \bibinfo {author} {\bibfnamefont {A.~V.}\
  \bibnamefont {Sushkov}},\ and\ \bibinfo {author} {\bibfnamefont {S.~Y.}\
  \bibnamefont {{van der Werf}}},\ }\bibfield  {title} {\bibinfo {title}
  {Strength distributions in neodymium isotopes: A test of collective nuclear
  models},\ }\href
  {https://doi.org/https://doi.org/10.1016/0375-9474(93)90178-Z} {\bibfield
  {journal} {\bibinfo  {journal} {Nucl. Phys. A}\ }\textbf {\bibinfo {volume}
  {559}},\ \bibinfo {pages} {1} (\bibinfo {year} {1993})}\BibitemShut {NoStop}%
\bibitem [{\citenamefont {Ponomarev}\ and\ \citenamefont {von
  Neumann-Cosel}(1999)}]{ponomarev1999}%
  \BibitemOpen
  \bibfield  {author} {\bibinfo {author} {\bibfnamefont {V.~Y.}\ \bibnamefont
  {Ponomarev}}\ and\ \bibinfo {author} {\bibfnamefont {P.}~\bibnamefont {von
  Neumann-Cosel}},\ }\bibfield  {title} {\bibinfo {title} {Fragmentation of the
  two-octupole phonon multiplet in ${}^{208}\mathrm{Pb}$},\ }\href
  {https://doi.org/10.1103/PhysRevLett.82.501} {\bibfield  {journal} {\bibinfo
  {journal} {Phys. Rev. Lett.}\ }\textbf {\bibinfo {volume} {82}},\ \bibinfo
  {pages} {501} (\bibinfo {year} {1999})}\BibitemShut {NoStop}%
\bibitem [{\citenamefont {Ryezayeva}\ \emph {et~al.}(2002)\citenamefont
  {Ryezayeva}, \citenamefont {Hartmann}, \citenamefont {Kalmykov},
  \citenamefont {Lenske}, \citenamefont {von Neumann-Cosel}, \citenamefont
  {Ponomarev}, \citenamefont {Richter}, \citenamefont {Shevchenko},
  \citenamefont {Volz},\ and\ \citenamefont {Wambach}}]{ryezayeva2002}%
  \BibitemOpen
  \bibfield  {author} {\bibinfo {author} {\bibfnamefont {N.}~\bibnamefont
  {Ryezayeva}}, \bibinfo {author} {\bibfnamefont {T.}~\bibnamefont {Hartmann}},
  \bibinfo {author} {\bibfnamefont {Y.}~\bibnamefont {Kalmykov}}, \bibinfo
  {author} {\bibfnamefont {H.}~\bibnamefont {Lenske}}, \bibinfo {author}
  {\bibfnamefont {P.}~\bibnamefont {von Neumann-Cosel}}, \bibinfo {author}
  {\bibfnamefont {V.~Y.}\ \bibnamefont {Ponomarev}}, \bibinfo {author}
  {\bibfnamefont {A.}~\bibnamefont {Richter}}, \bibinfo {author} {\bibfnamefont
  {A.}~\bibnamefont {Shevchenko}}, \bibinfo {author} {\bibfnamefont
  {S.}~\bibnamefont {Volz}},\ and\ \bibinfo {author} {\bibfnamefont
  {J.}~\bibnamefont {Wambach}},\ }\bibfield  {title} {\bibinfo {title} {Nature
  of low-energy dipole strength in nuclei: The case of a resonance at particle
  threshold in $^{\mathrm{208}}\mathrm{P}\mathrm{b}$},\ }\href
  {https://doi.org/10.1103/PhysRevLett.89.272502} {\bibfield  {journal}
  {\bibinfo  {journal} {Phys. Rev. Lett.}\ }\textbf {\bibinfo {volume} {89}},\
  \bibinfo {pages} {272502} (\bibinfo {year} {2002})}\BibitemShut {NoStop}%
\bibitem [{\citenamefont {Savran}\ \emph {et~al.}(2018)\citenamefont {Savran},
  \citenamefont {Derya}, \citenamefont {Bagchi}, \citenamefont {Endres},
  \citenamefont {Harakeh}, \citenamefont {Isaak}, \citenamefont
  {Kalantar-Nayestanaki}, \citenamefont {Lanza}, \citenamefont {L\"oher},
  \citenamefont {Najafi}, \citenamefont {Pascu}, \citenamefont {Pickstone},
  \citenamefont {Pietralla}, \citenamefont {Ponomarev}, \citenamefont
  {Rigollet}, \citenamefont {Romig}, \citenamefont {Spieker}, \citenamefont
  {Vitturi},\ and\ \citenamefont {Zilges}}]{savran2018}%
  \BibitemOpen
  \bibfield  {author} {\bibinfo {author} {\bibfnamefont {D.}~\bibnamefont
  {Savran}}, \bibinfo {author} {\bibfnamefont {V.}~\bibnamefont {Derya}},
  \bibinfo {author} {\bibfnamefont {S.}~\bibnamefont {Bagchi}}, \bibinfo
  {author} {\bibfnamefont {J.}~\bibnamefont {Endres}}, \bibinfo {author}
  {\bibfnamefont {M.~N.}\ \bibnamefont {Harakeh}}, \bibinfo {author}
  {\bibfnamefont {J.}~\bibnamefont {Isaak}}, \bibinfo {author} {\bibfnamefont
  {N.}~\bibnamefont {Kalantar-Nayestanaki}}, \bibinfo {author} {\bibfnamefont
  {E.~G.}\ \bibnamefont {Lanza}}, \bibinfo {author} {\bibfnamefont
  {B.}~\bibnamefont {L\"oher}}, \bibinfo {author} {\bibfnamefont
  {A.}~\bibnamefont {Najafi}}, \bibinfo {author} {\bibfnamefont
  {S.}~\bibnamefont {Pascu}}, \bibinfo {author} {\bibfnamefont {S.~G.}\
  \bibnamefont {Pickstone}}, \bibinfo {author} {\bibfnamefont {N.}~\bibnamefont
  {Pietralla}}, \bibinfo {author} {\bibfnamefont {V.~Y.}\ \bibnamefont
  {Ponomarev}}, \bibinfo {author} {\bibfnamefont {C.}~\bibnamefont {Rigollet}},
  \bibinfo {author} {\bibfnamefont {C.}~\bibnamefont {Romig}}, \bibinfo
  {author} {\bibfnamefont {M.}~\bibnamefont {Spieker}}, \bibinfo {author}
  {\bibfnamefont {A.}~\bibnamefont {Vitturi}},\ and\ \bibinfo {author}
  {\bibfnamefont {A.}~\bibnamefont {Zilges}},\ }\bibfield  {title} {\bibinfo
  {title} {Multi-messenger investigation of the pygmy dipole resonance in
  $^{140}${C}e},\ }\href
  {https://doi.org/https://doi.org/10.1016/j.physletb.2018.09.025} {\bibfield
  {journal} {\bibinfo  {journal} {Phys. Lett. B}\ }\textbf {\bibinfo {volume}
  {786}},\ \bibinfo {pages} {16} (\bibinfo {year} {2018})}\BibitemShut
  {NoStop}%
\bibitem [{\citenamefont {Burda}\ \emph {et~al.}(2007)\citenamefont {Burda},
  \citenamefont {Botha}, \citenamefont {Carter}, \citenamefont {Fearick},
  \citenamefont {F\"ortsch}, \citenamefont {Fransen}, \citenamefont {Fujita},
  \citenamefont {Holt}, \citenamefont {Kuhar}, \citenamefont {Lenhardt},
  \citenamefont {von Neumann-Cosel}, \citenamefont {Neveling}, \citenamefont
  {Pietralla}, \citenamefont {Ponomarev}, \citenamefont {Richter},
  \citenamefont {Scholten}, \citenamefont {Sideras-Haddad}, \citenamefont
  {Smit},\ and\ \citenamefont {Wambach}}]{burda2007}%
  \BibitemOpen
  \bibfield  {author} {\bibinfo {author} {\bibfnamefont {O.}~\bibnamefont
  {Burda}}, \bibinfo {author} {\bibfnamefont {N.}~\bibnamefont {Botha}},
  \bibinfo {author} {\bibfnamefont {J.}~\bibnamefont {Carter}}, \bibinfo
  {author} {\bibfnamefont {R.~W.}\ \bibnamefont {Fearick}}, \bibinfo {author}
  {\bibfnamefont {S.~V.}\ \bibnamefont {F\"ortsch}}, \bibinfo {author}
  {\bibfnamefont {C.}~\bibnamefont {Fransen}}, \bibinfo {author} {\bibfnamefont
  {H.}~\bibnamefont {Fujita}}, \bibinfo {author} {\bibfnamefont {J.~D.}\
  \bibnamefont {Holt}}, \bibinfo {author} {\bibfnamefont {M.}~\bibnamefont
  {Kuhar}}, \bibinfo {author} {\bibfnamefont {A.}~\bibnamefont {Lenhardt}},
  \bibinfo {author} {\bibfnamefont {P.}~\bibnamefont {von Neumann-Cosel}},
  \bibinfo {author} {\bibfnamefont {R.}~\bibnamefont {Neveling}}, \bibinfo
  {author} {\bibfnamefont {N.}~\bibnamefont {Pietralla}}, \bibinfo {author}
  {\bibfnamefont {V.~Y.}\ \bibnamefont {Ponomarev}}, \bibinfo {author}
  {\bibfnamefont {A.}~\bibnamefont {Richter}}, \bibinfo {author} {\bibfnamefont
  {O.}~\bibnamefont {Scholten}}, \bibinfo {author} {\bibfnamefont
  {E.}~\bibnamefont {Sideras-Haddad}}, \bibinfo {author} {\bibfnamefont
  {F.~D.}\ \bibnamefont {Smit}},\ and\ \bibinfo {author} {\bibfnamefont
  {J.}~\bibnamefont {Wambach}},\ }\bibfield  {title} {\bibinfo {title}
  {High-energy-resolution inelastic electron and proton scattering and the
  multiphonon nature of mixed-symmetry ${2}^{+}$ states in
  $^{94}\mathrm{Mo}$},\ }\href {https://doi.org/10.1103/PhysRevLett.99.092503}
  {\bibfield  {journal} {\bibinfo  {journal} {Phys. Rev. Lett.}\ }\textbf
  {\bibinfo {volume} {99}},\ \bibinfo {pages} {092503} (\bibinfo {year}
  {2007})}\BibitemShut {NoStop}%
\bibitem [{\citenamefont {Walz}\ \emph {et~al.}(2011)\citenamefont {Walz},
  \citenamefont {Fujita}, \citenamefont {Krugmann}, \citenamefont {von
  Neumann-Cosel}, \citenamefont {Pietralla}, \citenamefont {Ponomarev},
  \citenamefont {Scheikh-Obeid},\ and\ \citenamefont {Wambach}}]{walz2011}%
  \BibitemOpen
  \bibfield  {author} {\bibinfo {author} {\bibfnamefont {C.}~\bibnamefont
  {Walz}}, \bibinfo {author} {\bibfnamefont {H.}~\bibnamefont {Fujita}},
  \bibinfo {author} {\bibfnamefont {A.}~\bibnamefont {Krugmann}}, \bibinfo
  {author} {\bibfnamefont {P.}~\bibnamefont {von Neumann-Cosel}}, \bibinfo
  {author} {\bibfnamefont {N.}~\bibnamefont {Pietralla}}, \bibinfo {author}
  {\bibfnamefont {V.~Y.}\ \bibnamefont {Ponomarev}}, \bibinfo {author}
  {\bibfnamefont {A.}~\bibnamefont {Scheikh-Obeid}},\ and\ \bibinfo {author}
  {\bibfnamefont {J.}~\bibnamefont {Wambach}},\ }\bibfield  {title} {\bibinfo
  {title} {Origin of low-energy quadrupole collectivity in vibrational
  nuclei},\ }\href {https://doi.org/10.1103/PhysRevLett.106.062501} {\bibfield
  {journal} {\bibinfo  {journal} {Phys. Rev. Lett.}\ }\textbf {\bibinfo
  {volume} {106}},\ \bibinfo {pages} {062501} (\bibinfo {year}
  {2011})}\BibitemShut {NoStop}%
\bibitem [{\citenamefont {Smirnova}\ \emph {et~al.}(2000)\citenamefont
  {Smirnova}, \citenamefont {Pietralla}, \citenamefont {Mizusaki},\ and\
  \citenamefont {{Van Isacker}}}]{smirnova2000}%
  \BibitemOpen
  \bibfield  {author} {\bibinfo {author} {\bibfnamefont {N.~A.}\ \bibnamefont
  {Smirnova}}, \bibinfo {author} {\bibfnamefont {N.}~\bibnamefont {Pietralla}},
  \bibinfo {author} {\bibfnamefont {T.}~\bibnamefont {Mizusaki}},\ and\
  \bibinfo {author} {\bibfnamefont {P.}~\bibnamefont {{Van Isacker}}},\
  }\bibfield  {title} {\bibinfo {title} {Interrelation between the isoscalar
  octupole phonon and the proton–neutron mixed-symmetry quadrupole phonon in
  near-spherical nuclei},\ }\href
  {https://doi.org/https://doi.org/10.1016/S0375-9474(00)00331-6} {\bibfield
  {journal} {\bibinfo  {journal} {Nucl. Phys. A}\ }\textbf {\bibinfo {volume}
  {678}},\ \bibinfo {pages} {235} (\bibinfo {year} {2000})}\BibitemShut
  {NoStop}%
\bibitem [{\citenamefont {Fransen}\ \emph {et~al.}(2003)\citenamefont
  {Fransen}, \citenamefont {Pietralla}, \citenamefont {Ammar}, \citenamefont
  {Bandyopadhyay}, \citenamefont {Boukharouba}, \citenamefont {von Brentano},
  \citenamefont {Dewald}, \citenamefont {Gableske}, \citenamefont {Gade},
  \citenamefont {Jolie}, \citenamefont {Kneissl}, \citenamefont {Lesher},
  \citenamefont {Lisetskiy}, \citenamefont {McEllistrem}, \citenamefont
  {Merrick}, \citenamefont {Pitz}, \citenamefont {Warr}, \citenamefont
  {Werner},\ and\ \citenamefont {Yates}}]{fransen2003}%
  \BibitemOpen
  \bibfield  {author} {\bibinfo {author} {\bibfnamefont {C.}~\bibnamefont
  {Fransen}}, \bibinfo {author} {\bibfnamefont {N.}~\bibnamefont {Pietralla}},
  \bibinfo {author} {\bibfnamefont {Z.}~\bibnamefont {Ammar}}, \bibinfo
  {author} {\bibfnamefont {D.}~\bibnamefont {Bandyopadhyay}}, \bibinfo {author}
  {\bibfnamefont {N.}~\bibnamefont {Boukharouba}}, \bibinfo {author}
  {\bibfnamefont {P.}~\bibnamefont {von Brentano}}, \bibinfo {author}
  {\bibfnamefont {A.}~\bibnamefont {Dewald}}, \bibinfo {author} {\bibfnamefont
  {J.}~\bibnamefont {Gableske}}, \bibinfo {author} {\bibfnamefont
  {A.}~\bibnamefont {Gade}}, \bibinfo {author} {\bibfnamefont {J.}~\bibnamefont
  {Jolie}}, \bibinfo {author} {\bibfnamefont {U.}~\bibnamefont {Kneissl}},
  \bibinfo {author} {\bibfnamefont {S.~R.}\ \bibnamefont {Lesher}}, \bibinfo
  {author} {\bibfnamefont {A.~F.}\ \bibnamefont {Lisetskiy}}, \bibinfo {author}
  {\bibfnamefont {M.~T.}\ \bibnamefont {McEllistrem}}, \bibinfo {author}
  {\bibfnamefont {M.}~\bibnamefont {Merrick}}, \bibinfo {author} {\bibfnamefont
  {H.~H.}\ \bibnamefont {Pitz}}, \bibinfo {author} {\bibfnamefont
  {N.}~\bibnamefont {Warr}}, \bibinfo {author} {\bibfnamefont {V.}~\bibnamefont
  {Werner}},\ and\ \bibinfo {author} {\bibfnamefont {S.~W.}\ \bibnamefont
  {Yates}},\ }\bibfield  {title} {\bibinfo {title} {Comprehensive studies of
  low-spin collective excitations in ${}^{94}\mathrm{Mo}$},\ }\href
  {https://doi.org/10.1103/PhysRevC.67.024307} {\bibfield  {journal} {\bibinfo
  {journal} {Phys. Rev. C}\ }\textbf {\bibinfo {volume} {67}},\ \bibinfo
  {pages} {024307} (\bibinfo {year} {2003})}\BibitemShut {NoStop}%
\bibitem [{\citenamefont {Scheck}\ \emph {et~al.}(2010)\citenamefont {Scheck},
  \citenamefont {Butler}, \citenamefont {Fransen}, \citenamefont {Werner},\
  and\ \citenamefont {Yates}}]{scheck2010}%
  \BibitemOpen
  \bibfield  {author} {\bibinfo {author} {\bibfnamefont {M.}~\bibnamefont
  {Scheck}}, \bibinfo {author} {\bibfnamefont {P.~A.}\ \bibnamefont {Butler}},
  \bibinfo {author} {\bibfnamefont {C.}~\bibnamefont {Fransen}}, \bibinfo
  {author} {\bibfnamefont {V.}~\bibnamefont {Werner}},\ and\ \bibinfo {author}
  {\bibfnamefont {S.~W.}\ \bibnamefont {Yates}},\ }\bibfield  {title} {\bibinfo
  {title} {Strong ${M}1$ components in
  ${3}_{i}^{\ensuremath{-}}\ensuremath{\rightarrow}{3}_{1}^{\ensuremath{-}}$
  transitions in nearly spherical nuclei: Evidence for isovector-octupole
  excitations},\ }\href {https://doi.org/10.1103/PhysRevC.81.064305} {\bibfield
   {journal} {\bibinfo  {journal} {Phys. Rev. C}\ }\textbf {\bibinfo {volume}
  {81}},\ \bibinfo {pages} {064305} (\bibinfo {year} {2010})}\BibitemShut
  {NoStop}%
\bibitem [{\citenamefont {Hennig}\ \emph {et~al.}(2015)\citenamefont {Hennig},
  \citenamefont {Ahn}, \citenamefont {Anagnostatou}, \citenamefont {Blazhev},
  \citenamefont {Cooper}, \citenamefont {Derya}, \citenamefont {Elvers},
  \citenamefont {Endres}, \citenamefont {Goddard}, \citenamefont {Heinz},
  \citenamefont {Hughes}, \citenamefont {Ilie}, \citenamefont {Mineva},
  \citenamefont {Petkov}, \citenamefont {Pickstone}, \citenamefont {Pietralla},
  \citenamefont {Radeck}, \citenamefont {Ross}, \citenamefont {Savran},
  \citenamefont {Spieker}, \citenamefont {Werner},\ and\ \citenamefont
  {Zilges}}]{hennig2015}%
  \BibitemOpen
  \bibfield  {author} {\bibinfo {author} {\bibfnamefont {A.}~\bibnamefont
  {Hennig}}, \bibinfo {author} {\bibfnamefont {T.}~\bibnamefont {Ahn}},
  \bibinfo {author} {\bibfnamefont {V.}~\bibnamefont {Anagnostatou}}, \bibinfo
  {author} {\bibfnamefont {A.}~\bibnamefont {Blazhev}}, \bibinfo {author}
  {\bibfnamefont {N.}~\bibnamefont {Cooper}}, \bibinfo {author} {\bibfnamefont
  {V.}~\bibnamefont {Derya}}, \bibinfo {author} {\bibfnamefont
  {M.}~\bibnamefont {Elvers}}, \bibinfo {author} {\bibfnamefont
  {J.}~\bibnamefont {Endres}}, \bibinfo {author} {\bibfnamefont
  {P.}~\bibnamefont {Goddard}}, \bibinfo {author} {\bibfnamefont
  {A.}~\bibnamefont {Heinz}}, \bibinfo {author} {\bibfnamefont {R.~O.}\
  \bibnamefont {Hughes}}, \bibinfo {author} {\bibfnamefont {G.}~\bibnamefont
  {Ilie}}, \bibinfo {author} {\bibfnamefont {M.~N.}\ \bibnamefont {Mineva}},
  \bibinfo {author} {\bibfnamefont {P.}~\bibnamefont {Petkov}}, \bibinfo
  {author} {\bibfnamefont {S.~G.}\ \bibnamefont {Pickstone}}, \bibinfo {author}
  {\bibfnamefont {N.}~\bibnamefont {Pietralla}}, \bibinfo {author}
  {\bibfnamefont {D.}~\bibnamefont {Radeck}}, \bibinfo {author} {\bibfnamefont
  {T.~J.}\ \bibnamefont {Ross}}, \bibinfo {author} {\bibfnamefont
  {D.}~\bibnamefont {Savran}}, \bibinfo {author} {\bibfnamefont
  {M.}~\bibnamefont {Spieker}}, \bibinfo {author} {\bibfnamefont
  {V.}~\bibnamefont {Werner}},\ and\ \bibinfo {author} {\bibfnamefont
  {A.}~\bibnamefont {Zilges}},\ }\bibfield  {title} {\bibinfo {title}
  {Collective excitations of $^{96}\mathrm{Ru}$ by means of $(p,p\prime\gamma)$
  experiments},\ }\href {https://doi.org/10.1103/PhysRevC.92.064317} {\bibfield
   {journal} {\bibinfo  {journal} {Phys. Rev. C}\ }\textbf {\bibinfo {volume}
  {92}},\ \bibinfo {pages} {064317} (\bibinfo {year} {2015})}\BibitemShut
  {NoStop}%
\bibitem [{\citenamefont {Gregor}\ \emph {et~al.}(2017)\citenamefont {Gregor},
  \citenamefont {Scheck}, \citenamefont {Chapman}, \citenamefont {Gaffney},
  \citenamefont {Keatings}, \citenamefont {Mashtakov}, \citenamefont
  {O'Donnell}, \citenamefont {Smith}, \citenamefont {Spagnoletti},
  \citenamefont {Th{\"u}rauf}, \citenamefont {Werner},\ and\ \citenamefont
  {Wiseman}}]{gregor2017}%
  \BibitemOpen
  \bibfield  {author} {\bibinfo {author} {\bibfnamefont {E.~T.}\ \bibnamefont
  {Gregor}}, \bibinfo {author} {\bibfnamefont {M.}~\bibnamefont {Scheck}},
  \bibinfo {author} {\bibfnamefont {R.}~\bibnamefont {Chapman}}, \bibinfo
  {author} {\bibfnamefont {L.~P.}\ \bibnamefont {Gaffney}}, \bibinfo {author}
  {\bibfnamefont {J.}~\bibnamefont {Keatings}}, \bibinfo {author}
  {\bibfnamefont {K.~R.}\ \bibnamefont {Mashtakov}}, \bibinfo {author}
  {\bibfnamefont {D.}~\bibnamefont {O'Donnell}}, \bibinfo {author}
  {\bibfnamefont {J.~F.}\ \bibnamefont {Smith}}, \bibinfo {author}
  {\bibfnamefont {P.}~\bibnamefont {Spagnoletti}}, \bibinfo {author}
  {\bibfnamefont {M.}~\bibnamefont {Th{\"u}rauf}}, \bibinfo {author}
  {\bibfnamefont {V.}~\bibnamefont {Werner}},\ and\ \bibinfo {author}
  {\bibfnamefont {C.}~\bibnamefont {Wiseman}},\ }\bibfield  {title} {\bibinfo
  {title} {Shell evolution of stable ${N} = 50-56$ {Z}r and {M}o nuclei with
  respect to low-lying octupole excitations},\ }\href
  {https://doi.org/10.1140/epja/i2017-12224-7} {\bibfield  {journal} {\bibinfo
  {journal} {Eur. Phys. J. A}\ }\textbf {\bibinfo {volume} {53}},\ \bibinfo
  {pages} {50} (\bibinfo {year} {2017})}\BibitemShut {NoStop}%
\bibitem [{\citenamefont {Th\"urauf}\ \emph {et~al.}(2019)\citenamefont
  {Th\"urauf}, \citenamefont {Stoyanov}, \citenamefont {Scheck}, \citenamefont
  {Jentschel}, \citenamefont {Bernards}, \citenamefont {Blanc}, \citenamefont
  {Cooper}, \citenamefont {De~France}, \citenamefont {Gregor}, \citenamefont
  {Henrich}, \citenamefont {Hicks}, \citenamefont {Jolie}, \citenamefont
  {Kaleja}, \citenamefont {K\"oster}, \citenamefont {Kr\"oll}, \citenamefont
  {Leguillon}, \citenamefont {Mutti}, \citenamefont {O'Donnell}, \citenamefont
  {Petrache}, \citenamefont {Simpson}, \citenamefont {Smith}, \citenamefont
  {Soldner}, \citenamefont {Tezgel}, \citenamefont {Urban}, \citenamefont
  {Vanhoy}, \citenamefont {Werner}, \citenamefont {Werner}, \citenamefont
  {Zell},\ and\ \citenamefont {Zerrouki}}]{thurauf2019}%
  \BibitemOpen
  \bibfield  {author} {\bibinfo {author} {\bibfnamefont {M.}~\bibnamefont
  {Th\"urauf}}, \bibinfo {author} {\bibfnamefont {C.}~\bibnamefont {Stoyanov}},
  \bibinfo {author} {\bibfnamefont {M.}~\bibnamefont {Scheck}}, \bibinfo
  {author} {\bibfnamefont {M.}~\bibnamefont {Jentschel}}, \bibinfo {author}
  {\bibfnamefont {C.}~\bibnamefont {Bernards}}, \bibinfo {author}
  {\bibfnamefont {A.}~\bibnamefont {Blanc}}, \bibinfo {author} {\bibfnamefont
  {N.}~\bibnamefont {Cooper}}, \bibinfo {author} {\bibfnamefont
  {G.}~\bibnamefont {De~France}}, \bibinfo {author} {\bibfnamefont {E.~T.}\
  \bibnamefont {Gregor}}, \bibinfo {author} {\bibfnamefont {C.}~\bibnamefont
  {Henrich}}, \bibinfo {author} {\bibfnamefont {S.~F.}\ \bibnamefont {Hicks}},
  \bibinfo {author} {\bibfnamefont {J.}~\bibnamefont {Jolie}}, \bibinfo
  {author} {\bibfnamefont {O.}~\bibnamefont {Kaleja}}, \bibinfo {author}
  {\bibfnamefont {U.}~\bibnamefont {K\"oster}}, \bibinfo {author}
  {\bibfnamefont {T.}~\bibnamefont {Kr\"oll}}, \bibinfo {author} {\bibfnamefont
  {R.}~\bibnamefont {Leguillon}}, \bibinfo {author} {\bibfnamefont
  {P.}~\bibnamefont {Mutti}}, \bibinfo {author} {\bibfnamefont
  {D.}~\bibnamefont {O'Donnell}}, \bibinfo {author} {\bibfnamefont {C.~M.}\
  \bibnamefont {Petrache}}, \bibinfo {author} {\bibfnamefont {G.~S.}\
  \bibnamefont {Simpson}}, \bibinfo {author} {\bibfnamefont {J.~F.}\
  \bibnamefont {Smith}}, \bibinfo {author} {\bibfnamefont {T.}~\bibnamefont
  {Soldner}}, \bibinfo {author} {\bibfnamefont {M.}~\bibnamefont {Tezgel}},
  \bibinfo {author} {\bibfnamefont {W.}~\bibnamefont {Urban}}, \bibinfo
  {author} {\bibfnamefont {J.}~\bibnamefont {Vanhoy}}, \bibinfo {author}
  {\bibfnamefont {M.}~\bibnamefont {Werner}}, \bibinfo {author} {\bibfnamefont
  {V.}~\bibnamefont {Werner}}, \bibinfo {author} {\bibfnamefont {K.~O.}\
  \bibnamefont {Zell}},\ and\ \bibinfo {author} {\bibfnamefont
  {T.}~\bibnamefont {Zerrouki}},\ }\bibfield  {title} {\bibinfo {title}
  {Low-lying octupole isovector excitation in $^{144}${N}d},\ }\href
  {https://doi.org/10.1103/PhysRevC.99.011304} {\bibfield  {journal} {\bibinfo
  {journal} {Phys. Rev. C}\ }\textbf {\bibinfo {volume} {99}},\ \bibinfo
  {pages} {011304} (\bibinfo {year} {2019})}\BibitemShut {NoStop}%
\bibitem [{\citenamefont {Fransen}\ \emph {et~al.}(2005)\citenamefont
  {Fransen}, \citenamefont {Werner}, \citenamefont {Bandyopadhyay},
  \citenamefont {Boukharouba}, \citenamefont {Lesher}, \citenamefont
  {McEllistrem}, \citenamefont {Jolie}, \citenamefont {Pietralla},
  \citenamefont {{von Brentano}},\ and\ \citenamefont {Yates}}]{fransen2005}%
  \BibitemOpen
  \bibfield  {author} {\bibinfo {author} {\bibfnamefont {C.}~\bibnamefont
  {Fransen}}, \bibinfo {author} {\bibfnamefont {V.}~\bibnamefont {Werner}},
  \bibinfo {author} {\bibfnamefont {D.}~\bibnamefont {Bandyopadhyay}}, \bibinfo
  {author} {\bibfnamefont {N.}~\bibnamefont {Boukharouba}}, \bibinfo {author}
  {\bibfnamefont {S.~R.}\ \bibnamefont {Lesher}}, \bibinfo {author}
  {\bibfnamefont {M.~T.}\ \bibnamefont {McEllistrem}}, \bibinfo {author}
  {\bibfnamefont {J.}~\bibnamefont {Jolie}}, \bibinfo {author} {\bibfnamefont
  {N.}~\bibnamefont {Pietralla}}, \bibinfo {author} {\bibfnamefont
  {P.}~\bibnamefont {{von Brentano}}},\ and\ \bibinfo {author} {\bibfnamefont
  {S.~W.}\ \bibnamefont {Yates}},\ }\bibfield  {title} {\bibinfo {title}
  {Investigation of low-spin states in $^{92}${Z}r with the
  ($n,n^\prime\gamma$) reaction},\ }\href
  {https://doi.org/10.1103/PhysRevC.71.054304} {\bibfield  {journal} {\bibinfo
  {journal} {Phys. Rev. C}\ }\textbf {\bibinfo {volume} {71}},\ \bibinfo
  {pages} {054304} (\bibinfo {year} {2005})}\BibitemShut {NoStop}%
\bibitem [{\citenamefont {Casperson}\ \emph {et~al.}(2013)\citenamefont
  {Casperson}, \citenamefont {Werner},\ and\ \citenamefont
  {Heinze}}]{casperson2013}%
  \BibitemOpen
  \bibfield  {author} {\bibinfo {author} {\bibfnamefont {R.~J.}\ \bibnamefont
  {Casperson}}, \bibinfo {author} {\bibfnamefont {V.}~\bibnamefont {Werner}},\
  and\ \bibinfo {author} {\bibfnamefont {S.}~\bibnamefont {Heinze}},\
  }\bibfield  {title} {\bibinfo {title} {Hexadecapole degree of freedom in
  $^{94}${M}o},\ }\href
  {https://doi.org/https://doi.org/10.1016/j.physletb.2013.02.042} {\bibfield
  {journal} {\bibinfo  {journal} {Phys. Lett. B}\ }\textbf {\bibinfo {volume}
  {721}},\ \bibinfo {pages} {51} (\bibinfo {year} {2013})}\BibitemShut
  {NoStop}%
\bibitem [{\citenamefont {Pietralla}(2018)}]{pietralla2018}%
  \BibitemOpen
  \bibfield  {author} {\bibinfo {author} {\bibfnamefont {N.}~\bibnamefont
  {Pietralla}},\ }\bibfield  {title} {\bibinfo {title} {The {I}nstitute of
  {N}uclear {P}hysics at the {TU} {D}armstadt},\ }\href
  {https://doi.org/10.1080/10619127.2018.1463013} {\bibfield  {journal}
  {\bibinfo  {journal} {Nucl. Phys. News}\ }\textbf {\bibinfo {volume} {28
  (2)}},\ \bibinfo {pages} {4} (\bibinfo {year} {2018})}\BibitemShut {NoStop}%
\bibitem [{\citenamefont {Lenhardt}\ \emph {et~al.}(2006)\citenamefont
  {Lenhardt}, \citenamefont {Bonnes}, \citenamefont {Burda}, \citenamefont
  {{von Neumann-Cosel}}, \citenamefont {Platz}, \citenamefont {Richter},\ and\
  \citenamefont {Watzlawik}}]{lenhardt2006}%
  \BibitemOpen
  \bibfield  {author} {\bibinfo {author} {\bibfnamefont {A.}~\bibnamefont
  {Lenhardt}}, \bibinfo {author} {\bibfnamefont {U.}~\bibnamefont {Bonnes}},
  \bibinfo {author} {\bibfnamefont {O.}~\bibnamefont {Burda}}, \bibinfo
  {author} {\bibfnamefont {P.}~\bibnamefont {{von Neumann-Cosel}}}, \bibinfo
  {author} {\bibfnamefont {M.}~\bibnamefont {Platz}}, \bibinfo {author}
  {\bibfnamefont {A.}~\bibnamefont {Richter}},\ and\ \bibinfo {author}
  {\bibfnamefont {S.}~\bibnamefont {Watzlawik}},\ }\bibfield  {title} {\bibinfo
  {title} {A silicon microstrip detector in a magnetic spectrometer for
  high-resolution electron scattering experiments at the {S}-{DALINAC}},\
  }\href {https://doi.org/https://doi.org/10.1016/j.nima.2006.03.003}
  {\bibfield  {journal} {\bibinfo  {journal} {Nucl. Instrum. Methods Phys.
  Research, Sect. A}\ }\textbf {\bibinfo {volume} {562}},\ \bibinfo {pages}
  {320} (\bibinfo {year} {2006})}\BibitemShut {NoStop}%
\bibitem [{\citenamefont {Burda}\ \emph {et~al.}(2010)\citenamefont {Burda},
  \citenamefont {von Neumann-Cosel}, \citenamefont {Richter}, \citenamefont
  {Forss\'en},\ and\ \citenamefont {Brown}}]{burda2010}%
  \BibitemOpen
  \bibfield  {author} {\bibinfo {author} {\bibfnamefont {O.}~\bibnamefont
  {Burda}}, \bibinfo {author} {\bibfnamefont {P.}~\bibnamefont {von
  Neumann-Cosel}}, \bibinfo {author} {\bibfnamefont {A.}~\bibnamefont
  {Richter}}, \bibinfo {author} {\bibfnamefont {C.}~\bibnamefont {Forss\'en}},\
  and\ \bibinfo {author} {\bibfnamefont {B.~A.}\ \bibnamefont {Brown}},\
  }\bibfield  {title} {\bibinfo {title} {Resonance parameters of the first
  $1/{2}^{+}$ state in $^{9}\mathrm{Be}$ and astrophysical implications},\
  }\href {https://doi.org/10.1103/PhysRevC.82.015808} {\bibfield  {journal}
  {\bibinfo  {journal} {Phys. Rev. C}\ }\textbf {\bibinfo {volume} {82}},\
  \bibinfo {pages} {015808} (\bibinfo {year} {2010})}\BibitemShut {NoStop}%
\bibitem [{\citenamefont {{Scheik Obeid}}\ \emph {et~al.}(2013)\citenamefont
  {{Scheik Obeid}}, \citenamefont {Burda}, \citenamefont {Chernykh},
  \citenamefont {Krugmann}, \citenamefont {von Neumann-Cosel}, \citenamefont
  {Pietralla}, \citenamefont {Poltoratska}, \citenamefont {Ponomarev},\ and\
  \citenamefont {Walz}}]{scheikh2013}%
  \BibitemOpen
  \bibfield  {author} {\bibinfo {author} {\bibfnamefont {A.}~\bibnamefont
  {{Scheik Obeid}}}, \bibinfo {author} {\bibfnamefont {O.}~\bibnamefont
  {Burda}}, \bibinfo {author} {\bibfnamefont {M.}~\bibnamefont {Chernykh}},
  \bibinfo {author} {\bibfnamefont {A.}~\bibnamefont {Krugmann}}, \bibinfo
  {author} {\bibfnamefont {P.}~\bibnamefont {von Neumann-Cosel}}, \bibinfo
  {author} {\bibfnamefont {N.}~\bibnamefont {Pietralla}}, \bibinfo {author}
  {\bibfnamefont {I.}~\bibnamefont {Poltoratska}}, \bibinfo {author}
  {\bibfnamefont {V.~Y.}\ \bibnamefont {Ponomarev}},\ and\ \bibinfo {author}
  {\bibfnamefont {C.}~\bibnamefont {Walz}},\ }\bibfield  {title} {\bibinfo
  {title} {${E}2$ strengths and transition radii difference of one-phonon
  ${2}^{+}$ states of $^{92}${Z}r from electron scattering at low momentum
  transfer},\ }\href {https://doi.org/10.1103/PhysRevC.87.014337} {\bibfield
  {journal} {\bibinfo  {journal} {Phys. Rev. C}\ }\textbf {\bibinfo {volume}
  {87}},\ \bibinfo {pages} {014337} (\bibinfo {year} {2013})}\BibitemShut
  {NoStop}%
\bibitem [{\citenamefont {{Scheikh Obeid}}\ \emph {et~al.}(2014)\citenamefont
  {{Scheikh Obeid}}, \citenamefont {Aslanidou}, \citenamefont {Birkhan},
  \citenamefont {Krugmann}, \citenamefont {von Neumann-Cosel}, \citenamefont
  {Pietralla}, \citenamefont {Poltoratska},\ and\ \citenamefont
  {Ponomarev}}]{scheikh2014}%
  \BibitemOpen
  \bibfield  {author} {\bibinfo {author} {\bibfnamefont {A.}~\bibnamefont
  {{Scheikh Obeid}}}, \bibinfo {author} {\bibfnamefont {S.}~\bibnamefont
  {Aslanidou}}, \bibinfo {author} {\bibfnamefont {J.}~\bibnamefont {Birkhan}},
  \bibinfo {author} {\bibfnamefont {A.}~\bibnamefont {Krugmann}}, \bibinfo
  {author} {\bibfnamefont {P.}~\bibnamefont {von Neumann-Cosel}}, \bibinfo
  {author} {\bibfnamefont {N.}~\bibnamefont {Pietralla}}, \bibinfo {author}
  {\bibfnamefont {I.}~\bibnamefont {Poltoratska}},\ and\ \bibinfo {author}
  {\bibfnamefont {V.~Y.}\ \bibnamefont {Ponomarev}},\ }\bibfield  {title}
  {\bibinfo {title} {${B}({E}2)$ strength ratio of one-phonon ${2}^{+}$ states
  of $^{94}${Z}r from electron scattering at low momentum transfer},\ }\href
  {https://doi.org/10.1103/PhysRevC.89.037301} {\bibfield  {journal} {\bibinfo
  {journal} {Phys. Rev. C}\ }\textbf {\bibinfo {volume} {89}},\ \bibinfo
  {pages} {037301} (\bibinfo {year} {2014})}\BibitemShut {NoStop}%
\bibitem [{\citenamefont {Kremer}\ \emph {et~al.}(2016)\citenamefont {Kremer},
  \citenamefont {Aslanidou}, \citenamefont {Bassauer}, \citenamefont {Hilcker},
  \citenamefont {Krugmann}, \citenamefont {von Neumann-Cosel}, \citenamefont
  {Otsuka}, \citenamefont {Pietralla}, \citenamefont {Ponomarev}, \citenamefont
  {Shimizu}, \citenamefont {Singer}, \citenamefont {Steinhilber}, \citenamefont
  {Togashi}, \citenamefont {Tsunoda}, \citenamefont {Werner},\ and\
  \citenamefont {Zweidinger}}]{kremer2016}%
  \BibitemOpen
  \bibfield  {author} {\bibinfo {author} {\bibfnamefont {C.}~\bibnamefont
  {Kremer}}, \bibinfo {author} {\bibfnamefont {S.}~\bibnamefont {Aslanidou}},
  \bibinfo {author} {\bibfnamefont {S.}~\bibnamefont {Bassauer}}, \bibinfo
  {author} {\bibfnamefont {M.}~\bibnamefont {Hilcker}}, \bibinfo {author}
  {\bibfnamefont {A.}~\bibnamefont {Krugmann}}, \bibinfo {author}
  {\bibfnamefont {P.}~\bibnamefont {von Neumann-Cosel}}, \bibinfo {author}
  {\bibfnamefont {T.}~\bibnamefont {Otsuka}}, \bibinfo {author} {\bibfnamefont
  {N.}~\bibnamefont {Pietralla}}, \bibinfo {author} {\bibfnamefont {V.~Y.}\
  \bibnamefont {Ponomarev}}, \bibinfo {author} {\bibfnamefont {N.}~\bibnamefont
  {Shimizu}}, \bibinfo {author} {\bibfnamefont {M.}~\bibnamefont {Singer}},
  \bibinfo {author} {\bibfnamefont {G.}~\bibnamefont {Steinhilber}}, \bibinfo
  {author} {\bibfnamefont {T.}~\bibnamefont {Togashi}}, \bibinfo {author}
  {\bibfnamefont {Y.}~\bibnamefont {Tsunoda}}, \bibinfo {author} {\bibfnamefont
  {V.}~\bibnamefont {Werner}},\ and\ \bibinfo {author} {\bibfnamefont
  {M.}~\bibnamefont {Zweidinger}},\ }\bibfield  {title} {\bibinfo {title}
  {First measurement of collectivity of coexisting shapes based on type {II}
  shell evolution: The case of $^{96}${Z}r},\ }\href
  {https://doi.org/10.1103/PhysRevLett.117.172503} {\bibfield  {journal}
  {\bibinfo  {journal} {Phys. Rev. Lett.}\ }\textbf {\bibinfo {volume} {117}},\
  \bibinfo {pages} {172503} (\bibinfo {year} {2016})}\BibitemShut {NoStop}%
\bibitem [{\citenamefont {Hofmann}\ \emph {et~al.}(2002)\citenamefont
  {Hofmann}, \citenamefont {von Neumann-Cosel}, \citenamefont {Neumeyer},
  \citenamefont {Rangacharyulu}, \citenamefont {Reitz}, \citenamefont
  {Richter}, \citenamefont {Schrieder}, \citenamefont {Sober}, \citenamefont
  {Fagg},\ and\ \citenamefont {Brown}}]{hofmann2002}%
  \BibitemOpen
  \bibfield  {author} {\bibinfo {author} {\bibfnamefont {F.}~\bibnamefont
  {Hofmann}}, \bibinfo {author} {\bibfnamefont {P.}~\bibnamefont {von
  Neumann-Cosel}}, \bibinfo {author} {\bibfnamefont {F.}~\bibnamefont
  {Neumeyer}}, \bibinfo {author} {\bibfnamefont {C.}~\bibnamefont
  {Rangacharyulu}}, \bibinfo {author} {\bibfnamefont {B.}~\bibnamefont
  {Reitz}}, \bibinfo {author} {\bibfnamefont {A.}~\bibnamefont {Richter}},
  \bibinfo {author} {\bibfnamefont {G.}~\bibnamefont {Schrieder}}, \bibinfo
  {author} {\bibfnamefont {D.~I.}\ \bibnamefont {Sober}}, \bibinfo {author}
  {\bibfnamefont {L.~W.}\ \bibnamefont {Fagg}},\ and\ \bibinfo {author}
  {\bibfnamefont {B.~A.}\ \bibnamefont {Brown}},\ }\bibfield  {title} {\bibinfo
  {title} {Magnetic dipole transitions in $^{32}${S} from electron scattering
  at 180$^\circ$},\ }\href {https://doi.org/10.1103/PhysRevC.65.024311}
  {\bibfield  {journal} {\bibinfo  {journal} {Phys. Rev. C}\ }\textbf {\bibinfo
  {volume} {65}},\ \bibinfo {pages} {024311} (\bibinfo {year}
  {2002})}\BibitemShut {NoStop}%
\bibitem [{\citenamefont {Burda}(2008)}]{burda2007a}%
  \BibitemOpen
  \bibfield  {author} {\bibinfo {author} {\bibfnamefont {O.}~\bibnamefont
  {Burda}},\ }\emph {\bibinfo {title} {Nature of Mixed-Symmetry $2^+$ States in
  $^{94}${M}o from High-Resolution Electron and Proton Scattering and Line
  Shape of the First Excited $1/2^+$ State in $^9${B}e}},\ \href
  {http://tubiblio.ulb.tu-darmstadt.de/38394/} {Ph.D. thesis},\ \bibinfo
  {school} {Technische Universit{\"a}t Darmstadt} (\bibinfo {year}
  {2008})\BibitemShut {NoStop}%
\bibitem [{\citenamefont {Neveling}\ \emph {et~al.}(2011)\citenamefont
  {Neveling}, \citenamefont {Fujita}, \citenamefont {Smit}, \citenamefont
  {Adachi}, \citenamefont {Berg}, \citenamefont {Buthelezi}, \citenamefont
  {Carter}, \citenamefont {Conradie}, \citenamefont {Couder}, \citenamefont
  {Fearick}, \citenamefont {F\"ortsch}, \citenamefont {T.}, \citenamefont
  {Fujita}, \citenamefont {G\"orres}, \citenamefont {Hatanaka}, \citenamefont
  {Jingo}, \citenamefont {Krumbholz}, \citenamefont {Kureba}, \citenamefont
  {Mira}, \citenamefont {Murray}, \citenamefont {{von Neumann-Cosel}},
  \citenamefont {O'Brien}, \citenamefont {Papka}, \citenamefont {Poltoratska},
  \citenamefont {Richter}, \citenamefont {Sideras-Haddad}, \citenamefont
  {Swartz}, \citenamefont {Tamii}, \citenamefont {Usman},\ and\ \citenamefont
  {{van Zyl}}}]{neveling2011}%
  \BibitemOpen
  \bibfield  {author} {\bibinfo {author} {\bibfnamefont {R.}~\bibnamefont
  {Neveling}}, \bibinfo {author} {\bibfnamefont {H.}~\bibnamefont {Fujita}},
  \bibinfo {author} {\bibfnamefont {F.~D.}\ \bibnamefont {Smit}}, \bibinfo
  {author} {\bibfnamefont {T.}~\bibnamefont {Adachi}}, \bibinfo {author}
  {\bibfnamefont {G.~P.~A.}\ \bibnamefont {Berg}}, \bibinfo {author}
  {\bibfnamefont {E.~Z.}\ \bibnamefont {Buthelezi}}, \bibinfo {author}
  {\bibfnamefont {J.}~\bibnamefont {Carter}}, \bibinfo {author} {\bibfnamefont
  {J.~L.}\ \bibnamefont {Conradie}}, \bibinfo {author} {\bibfnamefont
  {M.}~\bibnamefont {Couder}}, \bibinfo {author} {\bibfnamefont {R.~W.}\
  \bibnamefont {Fearick}}, \bibinfo {author} {\bibfnamefont {S.~V.}\
  \bibnamefont {F\"ortsch}}, \bibinfo {author} {\bibfnamefont {F.~D.}\
  \bibnamefont {T.}}, \bibinfo {author} {\bibfnamefont {Y.}~\bibnamefont
  {Fujita}}, \bibinfo {author} {\bibfnamefont {J.}~\bibnamefont {G\"orres}},
  \bibinfo {author} {\bibfnamefont {K.}~\bibnamefont {Hatanaka}}, \bibinfo
  {author} {\bibfnamefont {M.}~\bibnamefont {Jingo}}, \bibinfo {author}
  {\bibfnamefont {A.~M.}\ \bibnamefont {Krumbholz}}, \bibinfo {author}
  {\bibfnamefont {C.~O.}\ \bibnamefont {Kureba}}, \bibinfo {author}
  {\bibfnamefont {J.~P.}\ \bibnamefont {Mira}}, \bibinfo {author}
  {\bibfnamefont {S.~H.~T.}\ \bibnamefont {Murray}}, \bibinfo {author}
  {\bibfnamefont {P.}~\bibnamefont {{von Neumann-Cosel}}}, \bibinfo {author}
  {\bibfnamefont {S.}~\bibnamefont {O'Brien}}, \bibinfo {author} {\bibfnamefont
  {P.}~\bibnamefont {Papka}}, \bibinfo {author} {\bibfnamefont
  {I.}~\bibnamefont {Poltoratska}}, \bibinfo {author} {\bibfnamefont
  {A.}~\bibnamefont {Richter}}, \bibinfo {author} {\bibfnamefont
  {E.}~\bibnamefont {Sideras-Haddad}}, \bibinfo {author} {\bibfnamefont
  {J.~A.}\ \bibnamefont {Swartz}}, \bibinfo {author} {\bibfnamefont
  {A.}~\bibnamefont {Tamii}}, \bibinfo {author} {\bibfnamefont {I.~T.}\
  \bibnamefont {Usman}},\ and\ \bibinfo {author} {\bibfnamefont {J.~J.}\
  \bibnamefont {{van Zyl}}},\ }\bibfield  {title} {\bibinfo {title} {High
  energy-resolution zero-degree facility for light-ion scattering and reactions
  at i{T}hemba {LABS}},\ }\href
  {https://doi.org/https://doi.org/10.1016/j.nima.2011.06.077} {\bibfield
  {journal} {\bibinfo  {journal} {Nucl. Instrum. Methods Phys. Research, Sect.
  A}\ }\textbf {\bibinfo {volume} {654}},\ \bibinfo {pages} {29} (\bibinfo
  {year} {2011})}\BibitemShut {NoStop}%
\bibitem [{\citenamefont {Walz}(2014)}]{walz2014}%
  \BibitemOpen
  \bibfield  {author} {\bibinfo {author} {\bibfnamefont {C.}~\bibnamefont
  {Walz}},\ }\emph {\bibinfo {title} {The two-photon decay of the $11/2^-$
  isomer of $^{137}${B}a and mixed-symmetry states of $^{92,94}${Z}r and
  $^{94}${M}o}},\ \href {http://tubiblio.ulb.tu-darmstadt.de/65733/} {Ph.D.
  thesis},\ \bibinfo  {school} {Technische Universit{\"a}t Darmstadt} (\bibinfo
  {year} {2014})\BibitemShut {NoStop}%
\bibitem [{\citenamefont {Martin}\ \emph {et~al.}(2017)\citenamefont {Martin},
  \citenamefont {von Neumann-Cosel}, \citenamefont {Tamii}, \citenamefont
  {Aoi}, \citenamefont {Bassauer}, \citenamefont {Bertulani}, \citenamefont
  {Carter}, \citenamefont {Donaldson}, \citenamefont {Fujita}, \citenamefont
  {Fujita}, \citenamefont {Hashimoto}, \citenamefont {Hatanaka}, \citenamefont
  {Ito}, \citenamefont {Krugmann}, \citenamefont {Liu}, \citenamefont {Maeda},
  \citenamefont {Miki}, \citenamefont {Neveling}, \citenamefont {Pietralla},
  \citenamefont {Poltoratska}, \citenamefont {Ponomarev}, \citenamefont
  {Richter}, \citenamefont {Shima}, \citenamefont {Yamamoto},\ and\
  \citenamefont {Zweidinger}}]{martin2017}%
  \BibitemOpen
  \bibfield  {author} {\bibinfo {author} {\bibfnamefont {D.}~\bibnamefont
  {Martin}}, \bibinfo {author} {\bibfnamefont {P.}~\bibnamefont {von
  Neumann-Cosel}}, \bibinfo {author} {\bibfnamefont {A.}~\bibnamefont {Tamii}},
  \bibinfo {author} {\bibfnamefont {N.}~\bibnamefont {Aoi}}, \bibinfo {author}
  {\bibfnamefont {S.}~\bibnamefont {Bassauer}}, \bibinfo {author}
  {\bibfnamefont {C.~A.}\ \bibnamefont {Bertulani}}, \bibinfo {author}
  {\bibfnamefont {J.}~\bibnamefont {Carter}}, \bibinfo {author} {\bibfnamefont
  {L.}~\bibnamefont {Donaldson}}, \bibinfo {author} {\bibfnamefont
  {H.}~\bibnamefont {Fujita}}, \bibinfo {author} {\bibfnamefont
  {Y.}~\bibnamefont {Fujita}}, \bibinfo {author} {\bibfnamefont
  {T.}~\bibnamefont {Hashimoto}}, \bibinfo {author} {\bibfnamefont
  {K.}~\bibnamefont {Hatanaka}}, \bibinfo {author} {\bibfnamefont
  {T.}~\bibnamefont {Ito}}, \bibinfo {author} {\bibfnamefont {A.}~\bibnamefont
  {Krugmann}}, \bibinfo {author} {\bibfnamefont {B.}~\bibnamefont {Liu}},
  \bibinfo {author} {\bibfnamefont {Y.}~\bibnamefont {Maeda}}, \bibinfo
  {author} {\bibfnamefont {K.}~\bibnamefont {Miki}}, \bibinfo {author}
  {\bibfnamefont {R.}~\bibnamefont {Neveling}}, \bibinfo {author}
  {\bibfnamefont {N.}~\bibnamefont {Pietralla}}, \bibinfo {author}
  {\bibfnamefont {I.}~\bibnamefont {Poltoratska}}, \bibinfo {author}
  {\bibfnamefont {V.~Y.}\ \bibnamefont {Ponomarev}}, \bibinfo {author}
  {\bibfnamefont {A.}~\bibnamefont {Richter}}, \bibinfo {author} {\bibfnamefont
  {T.}~\bibnamefont {Shima}}, \bibinfo {author} {\bibfnamefont
  {T.}~\bibnamefont {Yamamoto}},\ and\ \bibinfo {author} {\bibfnamefont
  {M.}~\bibnamefont {Zweidinger}},\ }\bibfield  {title} {\bibinfo {title} {Test
  of the {B}rink-{A}xel hypothesis for the pygmy dipole resonance},\ }\href
  {https://doi.org/10.1103/PhysRevLett.119.182503} {\bibfield  {journal}
  {\bibinfo  {journal} {Phys. Rev. Lett.}\ }\textbf {\bibinfo {volume} {119}},\
  \bibinfo {pages} {182503} (\bibinfo {year} {2017})}\BibitemShut {NoStop}%
\bibitem [{\citenamefont {Shevchenko}\ \emph {et~al.}(2004)\citenamefont
  {Shevchenko}, \citenamefont {Carter}, \citenamefont {Fearick}, \citenamefont
  {F\"ortsch}, \citenamefont {Fujita}, \citenamefont {Fujita}, \citenamefont
  {Kalmykov}, \citenamefont {Lacroix}, \citenamefont {Lawrie}, \citenamefont
  {von Neumann-Cosel}, \citenamefont {Neveling}, \citenamefont {Ponomarev},
  \citenamefont {Richter}, \citenamefont {Sideras-Haddad}, \citenamefont
  {Smit},\ and\ \citenamefont {Wambach}}]{shevchenko2004}%
  \BibitemOpen
  \bibfield  {author} {\bibinfo {author} {\bibfnamefont {A.}~\bibnamefont
  {Shevchenko}}, \bibinfo {author} {\bibfnamefont {J.}~\bibnamefont {Carter}},
  \bibinfo {author} {\bibfnamefont {R.~W.}\ \bibnamefont {Fearick}}, \bibinfo
  {author} {\bibfnamefont {S.~V.}\ \bibnamefont {F\"ortsch}}, \bibinfo {author}
  {\bibfnamefont {H.}~\bibnamefont {Fujita}}, \bibinfo {author} {\bibfnamefont
  {Y.}~\bibnamefont {Fujita}}, \bibinfo {author} {\bibfnamefont
  {Y.}~\bibnamefont {Kalmykov}}, \bibinfo {author} {\bibfnamefont
  {D.}~\bibnamefont {Lacroix}}, \bibinfo {author} {\bibfnamefont {J.~J.}\
  \bibnamefont {Lawrie}}, \bibinfo {author} {\bibfnamefont {P.}~\bibnamefont
  {von Neumann-Cosel}}, \bibinfo {author} {\bibfnamefont {R.}~\bibnamefont
  {Neveling}}, \bibinfo {author} {\bibfnamefont {V.~Y.}\ \bibnamefont
  {Ponomarev}}, \bibinfo {author} {\bibfnamefont {A.}~\bibnamefont {Richter}},
  \bibinfo {author} {\bibfnamefont {E.}~\bibnamefont {Sideras-Haddad}},
  \bibinfo {author} {\bibfnamefont {F.~D.}\ \bibnamefont {Smit}},\ and\
  \bibinfo {author} {\bibfnamefont {J.}~\bibnamefont {Wambach}},\ }\bibfield
  {title} {\bibinfo {title} {Fine structure in the energy region of the
  isoscalar giant quadrupole resonance: Characteristic scales from a wavelet
  analysis},\ }\href {https://doi.org/10.1103/PhysRevLett.93.122501} {\bibfield
   {journal} {\bibinfo  {journal} {Phys. Rev. Lett.}\ }\textbf {\bibinfo
  {volume} {93}},\ \bibinfo {pages} {122501} (\bibinfo {year}
  {2004})}\BibitemShut {NoStop}%
\bibitem [{\citenamefont {von Neumann-Cosel}\ \emph {et~al.}(2019)\citenamefont
  {von Neumann-Cosel}, \citenamefont {Ponomarev}, \citenamefont {Richter},\
  and\ \citenamefont {Wambach}}]{vonneumanncosel2019}%
  \BibitemOpen
  \bibfield  {author} {\bibinfo {author} {\bibfnamefont {P.}~\bibnamefont {von
  Neumann-Cosel}}, \bibinfo {author} {\bibfnamefont {V.~Y.}\ \bibnamefont
  {Ponomarev}}, \bibinfo {author} {\bibfnamefont {A.}~\bibnamefont {Richter}},\
  and\ \bibinfo {author} {\bibfnamefont {J.}~\bibnamefont {Wambach}},\
  }\bibfield  {title} {\bibinfo {title} {Gross, intermediate and fine structure
  of nuclear giant resonances: Evidence for doorway states},\ }\href
  {https://doi.org/10.1140/epja/i2019-12795-1} {\bibfield  {journal} {\bibinfo
  {journal} {Eur. Phys. J. A}\ }\textbf {\bibinfo {volume} {55}},\ \bibinfo
  {pages} {224} (\bibinfo {year} {2019})}\BibitemShut {NoStop}%
\bibitem [{\citenamefont {Love}\ and\ \citenamefont {Franey}(1981)}]{love1981}%
  \BibitemOpen
  \bibfield  {author} {\bibinfo {author} {\bibfnamefont {W.~G.}\ \bibnamefont
  {Love}}\ and\ \bibinfo {author} {\bibfnamefont {M.~A.}\ \bibnamefont
  {Franey}},\ }\bibfield  {title} {\bibinfo {title} {{E}ffective
  nucleon-nucleon interaction for scattering at intermediate energies},\ }\href
  {https://doi.org/10.1103/PhysRevC.24.1073} {\bibfield  {journal} {\bibinfo
  {journal} {Phys. Rev. C}\ }\textbf {\bibinfo {volume} {24}},\ \bibinfo
  {pages} {1073} (\bibinfo {year} {1981})}\BibitemShut {NoStop}%
\bibitem [{\citenamefont {Kunz}()}]{CHUCK3}%
  \BibitemOpen
  \bibfield  {author} {\bibinfo {author} {\bibfnamefont {P.~D.}\ \bibnamefont
  {Kunz}},\ }\href@noop {} {\bibinfo {title} {code {CHUCK}3,
  unpublished}}\BibitemShut {NoStop}%
\bibitem [{\citenamefont {Schwandt}\ \emph {et~al.}(1982)\citenamefont
  {Schwandt}, \citenamefont {Meyer}, \citenamefont {Jacobs}, \citenamefont
  {Bacher}, \citenamefont {Vigdor}, \citenamefont {Kaitchuck},\ and\
  \citenamefont {Donoghue}}]{schwandt1982}%
  \BibitemOpen
  \bibfield  {author} {\bibinfo {author} {\bibfnamefont {P.}~\bibnamefont
  {Schwandt}}, \bibinfo {author} {\bibfnamefont {H.~O.}\ \bibnamefont {Meyer}},
  \bibinfo {author} {\bibfnamefont {W.~W.}\ \bibnamefont {Jacobs}}, \bibinfo
  {author} {\bibfnamefont {A.~D.}\ \bibnamefont {Bacher}}, \bibinfo {author}
  {\bibfnamefont {S.~E.}\ \bibnamefont {Vigdor}}, \bibinfo {author}
  {\bibfnamefont {M.~D.}\ \bibnamefont {Kaitchuck}},\ and\ \bibinfo {author}
  {\bibfnamefont {T.~R.}\ \bibnamefont {Donoghue}},\ }\bibfield  {title}
  {\bibinfo {title} {Analyzing power of proton-nucleus elastic scattering
  between 80 and 180 {M}e{V}},\ }\href {https://doi.org/10.1103/PhysRevC.26.55}
  {\bibfield  {journal} {\bibinfo  {journal} {Phys. Rev. C}\ }\textbf {\bibinfo
  {volume} {26}},\ \bibinfo {pages} {55} (\bibinfo {year} {1982})}\BibitemShut
  {NoStop}%
\bibitem [{\citenamefont {Singh}\ \emph {et~al.}(1986)\citenamefont {Singh},
  \citenamefont {Rychel}, \citenamefont {Gyufko}, \citenamefont {{Van
  Krüchten}}, \citenamefont {Lahanas},\ and\ \citenamefont
  {Wiedner}}]{aa94zr}%
  \BibitemOpen
  \bibfield  {author} {\bibinfo {author} {\bibfnamefont {P.}~\bibnamefont
  {Singh}}, \bibinfo {author} {\bibfnamefont {D.}~\bibnamefont {Rychel}},
  \bibinfo {author} {\bibfnamefont {R.}~\bibnamefont {Gyufko}}, \bibinfo
  {author} {\bibfnamefont {B.}~\bibnamefont {{Van Krüchten}}}, \bibinfo
  {author} {\bibfnamefont {M.}~\bibnamefont {Lahanas}},\ and\ \bibinfo {author}
  {\bibfnamefont {C.~A.}\ \bibnamefont {Wiedner}},\ }\bibfield  {title}
  {\bibinfo {title} {Alpha scattering from $^{92}${Z}r and $^{94}${Z}r},\
  }\href {https://doi.org/https://doi.org/10.1016/0375-9474(86)90279-4}
  {\bibfield  {journal} {\bibinfo  {journal} {Nucl. Phys. A}\ }\textbf
  {\bibinfo {volume} {458}},\ \bibinfo {pages} {1} (\bibinfo {year}
  {1986})}\BibitemShut {NoStop}%
\bibitem [{\citenamefont {Wang}\ \emph {et~al.}(2014)\citenamefont {Wang},
  \citenamefont {Liu}, \citenamefont {Zhang}, \citenamefont {Zhou},
  \citenamefont {Guo}, \citenamefont {Wang}, \citenamefont {Lei}, \citenamefont
  {Zheng}, \citenamefont {Fang}, \citenamefont {Qiang}, \citenamefont {Zhang},
  \citenamefont {Ding}, \citenamefont {Li}, \citenamefont {Ma}, \citenamefont
  {Yan}, \citenamefont {Wang}, \citenamefont {Gao}, \citenamefont {Fang},
  \citenamefont {Hu}, \citenamefont {Wu}, \citenamefont {He},\ and\
  \citenamefont {Zheng}}]{wang2014}%
  \BibitemOpen
  \bibfield  {author} {\bibinfo {author} {\bibfnamefont {Z.~G.}\ \bibnamefont
  {Wang}}, \bibinfo {author} {\bibfnamefont {M.~L.}\ \bibnamefont {Liu}},
  \bibinfo {author} {\bibfnamefont {Y.~H.}\ \bibnamefont {Zhang}}, \bibinfo
  {author} {\bibfnamefont {X.~H.}\ \bibnamefont {Zhou}}, \bibinfo {author}
  {\bibfnamefont {S.}~\bibnamefont {Guo}}, \bibinfo {author} {\bibfnamefont
  {J.~G.}\ \bibnamefont {Wang}}, \bibinfo {author} {\bibfnamefont {X.~G.}\
  \bibnamefont {Lei}}, \bibinfo {author} {\bibfnamefont {Y.}~\bibnamefont
  {Zheng}}, \bibinfo {author} {\bibfnamefont {Y.~D.}\ \bibnamefont {Fang}},
  \bibinfo {author} {\bibfnamefont {Y.~H.}\ \bibnamefont {Qiang}}, \bibinfo
  {author} {\bibfnamefont {N.~T.}\ \bibnamefont {Zhang}}, \bibinfo {author}
  {\bibfnamefont {B.}~\bibnamefont {Ding}}, \bibinfo {author} {\bibfnamefont
  {G.~S.}\ \bibnamefont {Li}}, \bibinfo {author} {\bibfnamefont
  {F.}~\bibnamefont {Ma}}, \bibinfo {author} {\bibfnamefont {X.~L.}\
  \bibnamefont {Yan}}, \bibinfo {author} {\bibfnamefont {S.~C.}\ \bibnamefont
  {Wang}}, \bibinfo {author} {\bibfnamefont {B.~S.}\ \bibnamefont {Gao}},
  \bibinfo {author} {\bibfnamefont {F.}~\bibnamefont {Fang}}, \bibinfo {author}
  {\bibfnamefont {B.~T.}\ \bibnamefont {Hu}}, \bibinfo {author} {\bibfnamefont
  {X.~G.}\ \bibnamefont {Wu}}, \bibinfo {author} {\bibfnamefont {C.~Y.}\
  \bibnamefont {He}},\ and\ \bibinfo {author} {\bibfnamefont {Y.}~\bibnamefont
  {Zheng}},\ }\bibfield  {title} {\bibinfo {title} {High-spin level structures
  of the near-spherical nuclei $^{91,92}${Z}r},\ }\href
  {https://doi.org/10.1103/PhysRevC.89.044308} {\bibfield  {journal} {\bibinfo
  {journal} {Phys. Rev. C}\ }\textbf {\bibinfo {volume} {89}},\ \bibinfo
  {pages} {044308} (\bibinfo {year} {2014})}\BibitemShut {NoStop}%
\bibitem [{\citenamefont {Elhami}\ \emph {et~al.}(2008)\citenamefont {Elhami},
  \citenamefont {Orce}, \citenamefont {Scheck}, \citenamefont {Mukhopadhyay},
  \citenamefont {Choudry}, \citenamefont {McEllistrem}, \citenamefont {Yates},
  \citenamefont {Angell}, \citenamefont {Boswell}, \citenamefont {Fallin},
  \citenamefont {Howell}, \citenamefont {Hutcheson}, \citenamefont {Karwowski},
  \citenamefont {Kelley}, \citenamefont {Parpottas}, \citenamefont {Tonchev},\
  and\ \citenamefont {Tornow}}]{elhami2008}%
  \BibitemOpen
  \bibfield  {author} {\bibinfo {author} {\bibfnamefont {E.}~\bibnamefont
  {Elhami}}, \bibinfo {author} {\bibfnamefont {J.~N.}\ \bibnamefont {Orce}},
  \bibinfo {author} {\bibfnamefont {M.}~\bibnamefont {Scheck}}, \bibinfo
  {author} {\bibfnamefont {S.}~\bibnamefont {Mukhopadhyay}}, \bibinfo {author}
  {\bibfnamefont {S.~N.}\ \bibnamefont {Choudry}}, \bibinfo {author}
  {\bibfnamefont {M.~T.}\ \bibnamefont {McEllistrem}}, \bibinfo {author}
  {\bibfnamefont {S.~W.}\ \bibnamefont {Yates}}, \bibinfo {author}
  {\bibfnamefont {C.}~\bibnamefont {Angell}}, \bibinfo {author} {\bibfnamefont
  {M.}~\bibnamefont {Boswell}}, \bibinfo {author} {\bibfnamefont
  {B.}~\bibnamefont {Fallin}}, \bibinfo {author} {\bibfnamefont {C.~R.}\
  \bibnamefont {Howell}}, \bibinfo {author} {\bibfnamefont {A.}~\bibnamefont
  {Hutcheson}}, \bibinfo {author} {\bibfnamefont {H.~J.}\ \bibnamefont
  {Karwowski}}, \bibinfo {author} {\bibfnamefont {J.~H.}\ \bibnamefont
  {Kelley}}, \bibinfo {author} {\bibfnamefont {Y.}~\bibnamefont {Parpottas}},
  \bibinfo {author} {\bibfnamefont {A.~P.}\ \bibnamefont {Tonchev}},\ and\
  \bibinfo {author} {\bibfnamefont {W.}~\bibnamefont {Tornow}},\ }\bibfield
  {title} {\bibinfo {title} {Experimental study of the low-lying structure of
  $^{94}${Z}r with the $(n,n^\prime\gamma)$ reaction},\ }\href
  {https://doi.org/10.1103/PhysRevC.78.064303} {\bibfield  {journal} {\bibinfo
  {journal} {Phys. Rev. C}\ }\textbf {\bibinfo {volume} {78}},\ \bibinfo
  {pages} {064303} (\bibinfo {year} {2008})}\BibitemShut {NoStop}%
\bibitem [{\citenamefont {Gavrielov}\ \emph {et~al.}(2022)\citenamefont
  {Gavrielov}, \citenamefont {Leviatan},\ and\ \citenamefont
  {Iachello}}]{gavrielov2022}%
  \BibitemOpen
  \bibfield  {author} {\bibinfo {author} {\bibfnamefont {N.}~\bibnamefont
  {Gavrielov}}, \bibinfo {author} {\bibfnamefont {A.}~\bibnamefont
  {Leviatan}},\ and\ \bibinfo {author} {\bibfnamefont {F.}~\bibnamefont
  {Iachello}},\ }\bibfield  {title} {\bibinfo {title} {Zr isotopes as a region
  of intertwined quantum phase transitions},\ }\href
  {https://doi.org/10.1103/PhysRevC.105.014305} {\bibfield  {journal} {\bibinfo
   {journal} {Phys. Rev. C}\ }\textbf {\bibinfo {volume} {105}},\ \bibinfo
  {pages} {014305} (\bibinfo {year} {2022})}\BibitemShut {NoStop}%
\bibitem [{\citenamefont {Fotiades}\ \emph {et~al.}(2002)\citenamefont
  {Fotiades}, \citenamefont {Cizewski}, \citenamefont {Becker}, \citenamefont
  {Bernstein}, \citenamefont {McNabb}, \citenamefont {Younes}, \citenamefont
  {Clark}, \citenamefont {Fallon}, \citenamefont {Lee}, \citenamefont
  {Macchiavelli}, \citenamefont {Holt},\ and\ \citenamefont
  {Hjorth-Jensen}}]{fotiades2002}%
  \BibitemOpen
  \bibfield  {author} {\bibinfo {author} {\bibfnamefont {N.}~\bibnamefont
  {Fotiades}}, \bibinfo {author} {\bibfnamefont {J.~A.}\ \bibnamefont
  {Cizewski}}, \bibinfo {author} {\bibfnamefont {J.~A.}\ \bibnamefont
  {Becker}}, \bibinfo {author} {\bibfnamefont {L.~A.}\ \bibnamefont
  {Bernstein}}, \bibinfo {author} {\bibfnamefont {D.~P.}\ \bibnamefont
  {McNabb}}, \bibinfo {author} {\bibfnamefont {W.}~\bibnamefont {Younes}},
  \bibinfo {author} {\bibfnamefont {R.~M.}\ \bibnamefont {Clark}}, \bibinfo
  {author} {\bibfnamefont {P.}~\bibnamefont {Fallon}}, \bibinfo {author}
  {\bibfnamefont {I.~Y.}\ \bibnamefont {Lee}}, \bibinfo {author} {\bibfnamefont
  {A.~O.}\ \bibnamefont {Macchiavelli}}, \bibinfo {author} {\bibfnamefont
  {A.}~\bibnamefont {Holt}},\ and\ \bibinfo {author} {\bibfnamefont
  {M.}~\bibnamefont {Hjorth-Jensen}},\ }\bibfield  {title} {\bibinfo {title}
  {High-spin excitations in $^{92,93,94,95}${Z}r},\ }\href
  {https://doi.org/10.1103/PhysRevC.65.044303} {\bibfield  {journal} {\bibinfo
  {journal} {Phys. Rev. C}\ }\textbf {\bibinfo {volume} {65}},\ \bibinfo
  {pages} {044303} (\bibinfo {year} {2002})}\BibitemShut {NoStop}%
\bibitem [{\citenamefont {Pignanelli}\ \emph {et~al.}(1992)\citenamefont
  {Pignanelli}, \citenamefont {Blasi}, \citenamefont {Micheletti},
  \citenamefont {{De Leo}}, \citenamefont {LaGamba}, \citenamefont {Perrino},
  \citenamefont {Bordewijk}, \citenamefont {Hofstee}, \citenamefont
  {Schippers}, \citenamefont {{van der Werf}}, \citenamefont {Wesseling},\ and\
  \citenamefont {Harakeh}}]{pp94mo}%
  \BibitemOpen
  \bibfield  {author} {\bibinfo {author} {\bibfnamefont {M.}~\bibnamefont
  {Pignanelli}}, \bibinfo {author} {\bibfnamefont {N.}~\bibnamefont {Blasi}},
  \bibinfo {author} {\bibfnamefont {S.}~\bibnamefont {Micheletti}}, \bibinfo
  {author} {\bibfnamefont {R.}~\bibnamefont {{De Leo}}}, \bibinfo {author}
  {\bibfnamefont {L.}~\bibnamefont {LaGamba}}, \bibinfo {author} {\bibfnamefont
  {R.}~\bibnamefont {Perrino}}, \bibinfo {author} {\bibfnamefont {J.~A.}\
  \bibnamefont {Bordewijk}}, \bibinfo {author} {\bibfnamefont {M.~A.}\
  \bibnamefont {Hofstee}}, \bibinfo {author} {\bibfnamefont {J.~M.}\
  \bibnamefont {Schippers}}, \bibinfo {author} {\bibfnamefont {S.~Y.}\
  \bibnamefont {{van der Werf}}}, \bibinfo {author} {\bibfnamefont
  {J.}~\bibnamefont {Wesseling}},\ and\ \bibinfo {author} {\bibfnamefont
  {M.~N.}\ \bibnamefont {Harakeh}},\ }\bibfield  {title} {\bibinfo {title}
  {Hexadecapole strength distributions of vibrational nuclei in the ${A} = 100$
  mass region},\ }\href
  {https://doi.org/https://doi.org/10.1016/0375-9474(92)90192-M} {\bibfield
  {journal} {\bibinfo  {journal} {Nucl. Phys. A}\ }\textbf {\bibinfo {volume}
  {540}},\ \bibinfo {pages} {27} (\bibinfo {year} {1992})}\BibitemShut
  {NoStop}%
\bibitem [{\citenamefont {Fretwurst}\ \emph {et~al.}(1987)\citenamefont
  {Fretwurst}, \citenamefont {Lindstr\"om}, \citenamefont {{von Reden}},
  \citenamefont {Riech}, \citenamefont {Vasiljev}, \citenamefont {Zarubin},
  \citenamefont {Knyazkov},\ and\ \citenamefont {Kuchtina}}]{fretwurst1987}%
  \BibitemOpen
  \bibfield  {author} {\bibinfo {author} {\bibfnamefont {E.}~\bibnamefont
  {Fretwurst}}, \bibinfo {author} {\bibfnamefont {G.}~\bibnamefont
  {Lindstr\"om}}, \bibinfo {author} {\bibfnamefont {K.}~\bibnamefont {{von
  Reden}}}, \bibinfo {author} {\bibfnamefont {V.}~\bibnamefont {Riech}},
  \bibinfo {author} {\bibfnamefont {S.}~\bibnamefont {Vasiljev}}, \bibinfo
  {author} {\bibfnamefont {P.}~\bibnamefont {Zarubin}}, \bibinfo {author}
  {\bibfnamefont {O.}~\bibnamefont {Knyazkov}},\ and\ \bibinfo {author}
  {\bibfnamefont {I.}~\bibnamefont {Kuchtina}},\ }\bibfield  {title} {\bibinfo
  {title} {Scattering of 25.6 {M}e{V} protons on $^{94}${M}o, $^{96}${M}o and
  $^{100}${M}o},\ }\href
  {https://doi.org/https://doi.org/10.1016/0375-9474(87)90517-3} {\bibfield
  {journal} {\bibinfo  {journal} {Nucl. Phys. A}\ }\textbf {\bibinfo {volume}
  {468}},\ \bibinfo {pages} {247} (\bibinfo {year} {1987})}\BibitemShut
  {NoStop}%
\bibitem [{\citenamefont {Lesher}\ \emph {et~al.}(2007)\citenamefont {Lesher},
  \citenamefont {McKay}, \citenamefont {Mynk}, \citenamefont {Bandyopadhyay},
  \citenamefont {Boukharouba}, \citenamefont {Fransen}, \citenamefont {Orce},
  \citenamefont {McEllistrem},\ and\ \citenamefont {Yates}}]{lesher2007}%
  \BibitemOpen
  \bibfield  {author} {\bibinfo {author} {\bibfnamefont {S.~R.}\ \bibnamefont
  {Lesher}}, \bibinfo {author} {\bibfnamefont {C.~J.}\ \bibnamefont {McKay}},
  \bibinfo {author} {\bibfnamefont {M.}~\bibnamefont {Mynk}}, \bibinfo {author}
  {\bibfnamefont {D.}~\bibnamefont {Bandyopadhyay}}, \bibinfo {author}
  {\bibfnamefont {N.}~\bibnamefont {Boukharouba}}, \bibinfo {author}
  {\bibfnamefont {C.}~\bibnamefont {Fransen}}, \bibinfo {author} {\bibfnamefont
  {J.~N.}\ \bibnamefont {Orce}}, \bibinfo {author} {\bibfnamefont {M.~T.}\
  \bibnamefont {McEllistrem}},\ and\ \bibinfo {author} {\bibfnamefont {S.~W.}\
  \bibnamefont {Yates}},\ }\bibfield  {title} {\bibinfo {title} {Low-spin
  structure of $^{96}\mathrm{Mo}$ studied with the ($n,n^\prime\gamma$)
  reaction},\ }\href {https://doi.org/10.1103/PhysRevC.75.034318} {\bibfield
  {journal} {\bibinfo  {journal} {Phys. Rev. C}\ }\textbf {\bibinfo {volume}
  {75}},\ \bibinfo {pages} {034318} (\bibinfo {year} {2007})}\BibitemShut
  {NoStop}%
\bibitem [{\citenamefont {Jabbour}\ \emph {et~al.}(1987)\citenamefont
  {Jabbour}, \citenamefont {Rosier}, \citenamefont {Ramstein}, \citenamefont
  {Tamisier},\ and\ \citenamefont {Avignon}}]{jabbour1987}%
  \BibitemOpen
  \bibfield  {author} {\bibinfo {author} {\bibfnamefont {J.}~\bibnamefont
  {Jabbour}}, \bibinfo {author} {\bibfnamefont {L.~H.}\ \bibnamefont {Rosier}},
  \bibinfo {author} {\bibfnamefont {B.}~\bibnamefont {Ramstein}}, \bibinfo
  {author} {\bibfnamefont {R.}~\bibnamefont {Tamisier}},\ and\ \bibinfo
  {author} {\bibfnamefont {P.}~\bibnamefont {Avignon}},\ }\bibfield  {title}
  {\bibinfo {title} {Elastic and inelastic scattering of 22 {M}e{V} protons
  from natural even-even {Z}n and {G}e isotopes: (i). spectroscopy of the
  $^{64,66}${Z}n isotopes},\ }\href
  {https://doi.org/https://doi.org/10.1016/0375-9474(87)90338-1} {\bibfield
  {journal} {\bibinfo  {journal} {Nucl. Phys. A}\ }\textbf {\bibinfo {volume}
  {464}},\ \bibinfo {pages} {260} (\bibinfo {year} {1987})}\BibitemShut
  {NoStop}%
\bibitem [{\citenamefont {Hudson}\ and\ \citenamefont
  {Glover}(1972)}]{hudson1972}%
  \BibitemOpen
  \bibfield  {author} {\bibinfo {author} {\bibfnamefont {F.~R.}\ \bibnamefont
  {Hudson}}\ and\ \bibinfo {author} {\bibfnamefont {R.~N.}\ \bibnamefont
  {Glover}},\ }\bibfield  {title} {\bibinfo {title} {The $(t,p)$ reaction on
  the zinc isotopes},\ }\href
  {https://doi.org/https://doi.org/10.1016/0375-9474(72)90295-3} {\bibfield
  {journal} {\bibinfo  {journal} {Nucl. Phys. A}\ }\textbf {\bibinfo {volume}
  {189}},\ \bibinfo {pages} {264} (\bibinfo {year} {1972})}\BibitemShut
  {NoStop}%
\bibitem [{\citenamefont {M\"ucher}(2009)}]{muecher2009}%
  \BibitemOpen
  \bibfield  {author} {\bibinfo {author} {\bibfnamefont {D.}~\bibnamefont
  {M\"ucher}},\ }\href {http://kups.ub.uni-koeln.de/id/eprint/2868} {\bibinfo
  {title} {Dynamische {S}ymmetrien von {A}tomkernen an
  {U}nterschalenabschl\"ussen, {D}octoral thesis, {U}niversit\"at zu {K}\"oln}}
  (\bibinfo {year} {2009})\BibitemShut {NoStop}%
\bibitem [{\citenamefont {M\"ucher}\ \emph {et~al.}(2009)\citenamefont
  {M\"ucher}, \citenamefont {G\"urdal}, \citenamefont {Speidel}, \citenamefont
  {Kumbartzki}, \citenamefont {Benczer-Koller}, \citenamefont {Robinson},
  \citenamefont {Sharon}, \citenamefont {Zamick}, \citenamefont {Lisetskiy},
  \citenamefont {Casperson}, \citenamefont {Heinz}, \citenamefont {Krieger},
  \citenamefont {Leske}, \citenamefont {Maier-Komor}, \citenamefont {Werner},
  \citenamefont {Williams},\ and\ \citenamefont {Winkler}}]{muecher2009a}%
  \BibitemOpen
  \bibfield  {author} {\bibinfo {author} {\bibfnamefont {D.}~\bibnamefont
  {M\"ucher}}, \bibinfo {author} {\bibfnamefont {G.}~\bibnamefont {G\"urdal}},
  \bibinfo {author} {\bibfnamefont {K.-H.}\ \bibnamefont {Speidel}}, \bibinfo
  {author} {\bibfnamefont {G.~J.}\ \bibnamefont {Kumbartzki}}, \bibinfo
  {author} {\bibfnamefont {N.}~\bibnamefont {Benczer-Koller}}, \bibinfo
  {author} {\bibfnamefont {S.~J.~Q.}\ \bibnamefont {Robinson}}, \bibinfo
  {author} {\bibfnamefont {Y.~Y.}\ \bibnamefont {Sharon}}, \bibinfo {author}
  {\bibfnamefont {L.}~\bibnamefont {Zamick}}, \bibinfo {author} {\bibfnamefont
  {A.~F.}\ \bibnamefont {Lisetskiy}}, \bibinfo {author} {\bibfnamefont {R.~J.}\
  \bibnamefont {Casperson}}, \bibinfo {author} {\bibfnamefont {A.}~\bibnamefont
  {Heinz}}, \bibinfo {author} {\bibfnamefont {B.}~\bibnamefont {Krieger}},
  \bibinfo {author} {\bibfnamefont {J.}~\bibnamefont {Leske}}, \bibinfo
  {author} {\bibfnamefont {P.}~\bibnamefont {Maier-Komor}}, \bibinfo {author}
  {\bibfnamefont {V.}~\bibnamefont {Werner}}, \bibinfo {author} {\bibfnamefont
  {E.}~\bibnamefont {Williams}},\ and\ \bibinfo {author} {\bibfnamefont
  {R.}~\bibnamefont {Winkler}},\ }\bibfield  {title} {\bibinfo {title} {Nuclear
  structure studies of $^{70}\mathrm{Zn}$ from $g$-factor and lifetime
  measurements},\ }\href {https://doi.org/10.1103/PhysRevC.79.054310}
  {\bibfield  {journal} {\bibinfo  {journal} {Phys. Rev. C}\ }\textbf {\bibinfo
  {volume} {79}},\ \bibinfo {pages} {054310} (\bibinfo {year}
  {2009})}\BibitemShut {NoStop}%
\bibitem [{\citenamefont {Tuli}(2004)}]{tuli2004}%
  \BibitemOpen
  \bibfield  {author} {\bibinfo {author} {\bibfnamefont {J.}~\bibnamefont
  {Tuli}},\ }\bibfield  {title} {\bibinfo {title} {Nuclear {D}ata {S}heets for
  ${A} = 70$},\ }\href
  {https://doi.org/https://doi.org/10.1016/j.nds.2004.11.005} {\bibfield
  {journal} {\bibinfo  {journal} {Nucl. Data Sheets}\ }\textbf {\bibinfo
  {volume} {103}},\ \bibinfo {pages} {389} (\bibinfo {year}
  {2004})}\BibitemShut {NoStop}%
\end{thebibliography}%

\end{document}